\documentclass[journal=jctcce,manuscript=article]{achemso}

\usepackage[version=3]{mhchem} 
\usepackage[T1]{fontenc}       
\usepackage{amsmath}
\usepackage{amssymb}
\usepackage{dcolumn}
\usepackage{graphicx}
\usepackage{float}

\author{Francesco Nattino}
\affiliation{Theory and Simulations of Materials (THEOS) and National Centre for Computational Design and Discovery of Novel Materials (MARVEL), \'{E}cole Polytechnique F\'{e}d\'{e}rale de Lausanne, CH-1015 Lausanne, Switzerland.}
\author{C\'{e}line Dupont}
\affiliation{Laboratoire Interdisciplinaire Carnot de Bourgogne (ICB), UMR 6303, CNRS, Universit\'{e} de Bourgogne Franche-Comt\'{e}, BP 47870, 21078 Dijon Cedex, France.}
\author{Nicola Marzari}
\affiliation{Theory and Simulations of Materials (THEOS) and National Centre for Computational Design and Discovery of Novel Materials (MARVEL), \'{E}cole Polytechnique F\'{e}d\'{e}rale de Lausanne, CH-1015 Lausanne, Switzerland.}
\author{Oliviero Andreussi}
\email{oliviero.andreussi@unt.edu}
\affiliation{Department of Physics, University of North Texas, Denton, TX 76207, USA}

\title{%
Functional extrapolations to tame unbound anions in density-functional theory calculations
}

\begin{document}

%
%
%
%
%

\begin{abstract}
Standard flavors of density-functional theory (DFT) calculations are known to fail in describing anions, due to large self-interaction errors. The problem may be circumvented by using localized basis sets of reduced size, leaving no variational flexibility for the extra electron to delocalize. Alternatively, a recent approach exploiting DFT evaluations of total energies on electronic densities optimized at the Hartree-Fock (HF) level has been reported, showing that the self-interaction-free HF densities are able to lead to an improved description of the additional electron, returning affinities in close agreement with the experiments. Nonetheless, such an approach can fail when the HF densities are too inaccurate. Here, an alternative approach is presented, in which an embedding environment is used to stabilize the anion in a bound configuration. Similarly to the HF case, when computing total energies at the DFT level on these corrected densities, electron affinities in very good agreement with experiments can be recovered. The effect of the environment can be evaluated and removed by an extrapolation of the results to the limit of vanishing embedding. Moreover, the approach can be easily applied to DFT calculations with delocalized basis sets, e.g. plane-waves, for which alternative approaches are either not viable or more computationally demanding. The proposed extrapolation strategy can be thus applied also to extended systems, as often studied in condensed-matter physics and materials science, and we illustrate how the embedding environment can be exploited to determine the energy of an adsorbing anion - here a chloride ion on a  metal surface - whose charge configuration would be incorrectly predicted by standard density functionals.
\end{abstract}

\section{Introduction\label{sec:potential}}

Charge-transfer processes are of paramount importance in many technological and biological processes\cite{Marcus-ReviewModPhys-1993} and they are exploited in many energy-conversion devices, such as solar cells\cite{Bates-SEF-2017} and electro-catalysts\cite{Perez-JACS-2017}. It is thus unfortunate that practical implementations of density functional theory (DFT), which is the most widespread electronic-structure method in materials science, fail in describing a wide range of negatively-charged atomic and molecular species. The issue is linked to the approximate form of the unknown exchange-correlation functional in DFT, which contains a spurious fraction of the electron-electron self-repulsion, also known as self-interaction error (SIE)\cite{Koch-2001}. The SIE is particularly severe for anions of atoms and small molecules, for which a careful inspection of the single-particle eigenvalues often reveals positive HOMO energies\cite{Shore-PRB-1977}. Small anions are thus incorrectly predicted to be unbound by routinely-used density functionals. 

If a localized basis set is employed, the HOMO levels can be artificially confined through the use of a reduced basis. While calculations with a fully-converged basis would result in the extra electron being lost to the continuum\cite{Wasserman-AnnualReviews-2017}, the pragmatic approach that exploits moderate basis-set sizes (MBS) allows to self-consistently optimize anion electron densities using DFT. Despite reasonable concerns\cite{Rosch-JCP-1997}, the MBS approach allows to calculate electron affinities (EAs) as total energy differences between neutral and negatively-charged species\cite{Galbraith-JCP-1996},  with results that are overall rather accurate, with mean-absolute errors (MAEs) of the order of 100-200 meV\cite{Rienstra-ChemRev-2002}. 

Burke and coworkers have shown how the large errors in the HOMO energies of atomic anions can be reconciled with the  accuracy of the corresponding total energies\cite{Lee-JPCL-2010, Lee-MolPhys-2010}. Briefly, the approximate treatment of exchange and correlation gives rise to an almost rigid upshift of the Kohn-Sham (KS) potential with respect to the exact KS reference. The shift is such that a barrier emerges at several angstroms from the nucleus. Standard localized basis-sets only sample the `inner' potential region, so that positive HOMO levels appear as bound even though they are actually resonances in the fully-converged basis-set limit. Nevertheless, the potential shift that strongly affects the HOMO energies has little effect on the self-consistently computed electron densities, therefore enabling the calculation of accurate total energies\cite{Wasserman-AnnualReviews-2017}.  

An approach that has been successfully applied to calculate EAs of atomic \cite{Lee-JPCL-2010, Lee-MolPhys-2010} and molecular \cite{Kim-JCP-2011} systems consists in evaluating DFT total energies from electron densities that are non-self-consistently computed using other electronic-structure methods. In particular, feeding the PBE density functional \cite{Perdew-PRL-1996, Perdew-PRL-1997} with densities optimized at the Hartree-Fock (HF) level allows to calculate EAs with a lower MAE than the MBS one\cite{Kim-JCP-2011}. In contrast with approximate DFT functionals the HF framework is one-electron self-interaction free\cite{Koch-2001}, and it consistently returns negative HOMO energies even for the anionic systems that are metastable in most DFT approximations\cite{Kim-JCP-2011}. However, problems can still arise for systems for which HF densities are not accurate enough. This approach is also not ideal for extended (metallic) systems, for which the computation of the HF exchange is very expensive, and HF provides a poor reference. 

In this work, we propose an alternative scheme that allows stabilizing localized anion configurations within a DFT framework.  By exploiting a continuum embedding environment that favors electron localization, we are able to self-consistently compute anion densities and total energies for properly bound systems. The artificial contribution that derives from the embedding can be removed through extrapolation to zero embedding intensity, allowing to estimate EAs in the absence of the environment. Specifically, we suggests two possible embedding schemes. In the first one, a dielectric embedding, analogous to the one employed in implicit solvation models to mimic the solvent response, provides an electrostatic stabilization of bound states. In the second scheme, an ad-hoc confining potential favors electron localization by providing instead a de-stabilizing contribution for the delocalized states. The two schemes provide very similar results, with the MAE computed for the G2-1 EA data set\cite{Curtiss-JCP-1991} being in the range 0.12-0.15 eV. This approach thus allows one to estimate the intrinsic density functional accuracy within a self-consistent framework. In contrast with the MBS approach, our strategy is not limited to localized basis-set and thus one can reach a well-defined basis-set convergence limit. 

Furthermore, the strategy can be trivially applied to solids and periodic systems, considering the very limited additional cost of the embedding procedure with respect to standard DFT calculations. As an example, we consider here a system constituted by a chloride ion and a Pt(111) surface, modeled as a periodically repeated slab. Standard PBE-DFT predicts Cl$^-$ to be unbound\cite{Lee-JPCL-2010, Lee-MolPhys-2010}, displacing a considerable fraction of the electron beyond charge neutrality on the metallic substrate, even at large distances between the slab and the ion. We show here that the dielectric embedding allows to stabilize the configuration in which the full additional charge is localized on the chlorine atom. Extrapolation to vanishing embedding intensity allows to estimate the correct energy of the Pt(111) + Cl$^-$ system. 

The article is structured as follows. Section \ref{sec:methods} describes the embedding procedure and it presents the computational parameters employed in the calculations. Results on the EA extrapolation for the various embedding schemes investigated, and the comparison with previous approaches is then reported in Section \ref{sub:EA}. The application of the dielectric extrapolation method to extended systems is then illustrated in Section \ref{sub:Pt111-Cl}. Finally, the conclusions are presented in Section \ref{sec:conclusions}.

\section{Methods\label{sec:methods}}

\subsection{Dielectric Embedding\label{sec:dielectric-embedding}}

We first consider a dielectric embedding, as typically employed in continuum solvation models to mimic the electrostatic response of the solvent on the embedded solvated system. Namely, an interface function $s(\textbf{r})$ is defined in terms of some of the system's degrees of freedom and chosen to smoothly vary between the value of 1 in the volume where the embedded system's degrees of freedom are present and the value of 0 in the embedding region. In particular, we have tested here two possible cavity definitions: the electron density-based function from the revised self-consistent continuum solvation (SCCS) model \cite{Fattebert-JIntQuantumChem-2003, Scherlis-JCP-2006, Andreussi-JCP-2012} and a rigid interface function based on atom-centered spheres from the soft-sphere continuum solvation (SSCS) model \cite{Fisicaro-JCTC-2017}. In the former, the cavity function is defined using the following piece-wise definition:
\begin{equation}
s_{\text{SCCS}}\left(\mathbf{r}\right)=\begin{cases}
0 & \rho^{el}\left(\mathbf{r}\right)\le\rho^{min}\\
1-t\left(\frac{\ln\rho^{max}-\ln\rho^{el}\left(\mathbf{r}\right)}{\ln\rho^{max}-\ln\rho^{min}}\right) & \rho^{max}>\rho^{el}\left(\mathbf{r}\right)>\rho^{min}\\
1 & \rho^{el}\left(\mathbf{r}\right)\ge\rho^{max}\label{eq:cavity-sccs}
\end{cases}
\end{equation}
where $t\left(x\right)$ is a smooth step function that goes from $0$ to $1$, with continuous first and second derivatives:
\begin{equation}
t\left(x\right)=x-\frac{\sin\left(2\pi x\right)}{2\pi}.
\end{equation}
The interface is defined in terms of two physically intuitive parameters, $\rho^{max}$ and $\rho^{min}$, that control how close to the embedded system the interface lies: the smaller their values, the further away from the embedded system is the interface. The interface function $s(\textbf{r})$ is used to construct the embedding dielectric function $\varepsilon(\textbf{r})$, which in the SCCS model takes the following form:
\begin{equation}
\varepsilon_{\text{SCCS}}\left(\mathbf{r}\right)=\exp\left(\ln\varepsilon_0\cdot\left(1 - s_{\text{SCCS}}\left(\mathbf{r}\right)\right)\right).\label{eq:epsilon-sccs}
\end{equation}

The second cavity definition exploits instead interlocking spheres centered on the system's nuclei with a smooth error-function profile:
\begin{equation}
s_{\text{SSCS}}\left(\mathbf{r}\right)=1 - \prod_a \frac{1}{2}\left[1 + \text{erf}\left(\frac{\left|\textbf{r}-\textbf{R}\right|-r_a}{\Delta}\right)\right],\label{eq:cavity-ss}
\end{equation}
where the $r_a$ quantity defines the radius of the sphere and the $\Delta$ parameter regulates the smoothness of the transition. Following the original SSCS model, we have set $r_a=\alpha r^{\text{vdW}}_a$, where $r^{\text{vdW}}_a$ is the van der Waals radius of the element corresponding to the atom $a$ as defined in the universal force field library \cite{Rappe-JACS-1992}, and $\alpha$ is a dimensionless scaling parameter. The dielectric function in the SSCS model is defined according to:
\begin{equation}
\varepsilon_{\text{SSCS}}\left(\mathbf{r}\right)=\left(1-\varepsilon_0\right)s_{\text{SSCS}}(\textbf{r}) + \varepsilon_0. \label{eq:epsilon-ss}
\end{equation}
Similarly to Equation \ref{eq:epsilon-sccs}, also Equation \ref{eq:epsilon-ss} allows to recover the vacuum permittivity ($\varepsilon=1$) inside the quantum-mechanical region, where $s(\textbf{r})$ is equal to one, and a constant permittivity $\varepsilon_{0}$ in the surrounding volume, where $s(\textbf{r})$ assumes a value of zero. 

The electrostatic energy of the embedded system will be expressed as 
\begin{equation}
E^{el}=\frac{1}{2}\int\rho\left(\textbf{r}\right)\phi\left(\mathbf{r}\right)\mathrm{d}\mathbf{r} = \underbrace{\frac{1}{2}\int\rho\left(\textbf{r}\right)\phi^{sys}\left(\mathbf{r}\right)\mathrm{d}\mathbf{r}}_{\displaystyle E^{sys}} + \underbrace{\frac{1}{2}\int\rho\left(\textbf{r}\right)\phi^{pol}\left(\mathbf{r}\right)\mathrm{d}\mathbf{r}}_{\displaystyle E^{pol}},\label{eq:electrostatic-energy}
\end{equation}
where $\displaystyle E^{sys}$ and $\displaystyle E^{pol}$ can be seen as the electrostatic contributions that arise from the interactions within the embedded system and between the system and the embedding environment, respectively, $\rho\left(\mathbf{r}\right) = \rho^{el}\left(\mathbf{r}\right)+\sum_{a}\rho_{a}^{ion}\left(\mathbf{r}-\mathbf{R}_{a}\right)$ is the total (electron and nuclear) charge density of the embedded system and the electrostatic potential $\phi\left(\textbf{r}\right)=\phi^{sys}\left(\mathbf{r}\right)+ \phi^{pol}\left(\mathbf{r}\right)$ is the solution of the generalized Poisson equation:
\begin{equation}
\nabla\cdot\varepsilon\left(\mathbf{r}\right)\nabla\phi\left(\mathbf{r}\right)=-4\pi\rho\left(\textbf{r}\right).
\end{equation}
The modified electrostatic potential tends to stabilize localized anions' electronic densities: indeed, it provides an electrostatic stabilization, which is greater for charged and dipolar systems. In the dielectric embedding we can play with the intensity of the embedding (in this case the dielectric permittivity of the environment $\varepsilon_0$) to fictitiously stabilize the electronic density of difficult (unbound) systems and extrapolate their energies for vanishing embedding (i.e. for  $\varepsilon_0\rightarrow1$).

\subsection{Confining potential\label{sec:confinement-potential}}

The second embedding environment is a confining potential contribution to be added to the Kohn-Sham potential. In particular, such contribution may be defined as proportional to the value of the complementary of the interface function. The corresponding energy contribution can be written as:
\begin{equation}
E^{confine}=\int\kappa\left(1-s\left(\mathbf{r}\right)\right)\rho^{el}\left(\mathbf{r}\right)\mathrm{d}\mathbf{r},\label{eq:confine-energy}
\end{equation}
where the confining constant $\kappa$ is a positive tunable parameter, which acts as a destabilization factor for the electronic density that spills out of the interface. The corresponding addition to the Kohn-Sham potential is given by 
\begin{equation}
v^{confine}\left(\mathbf{r}\right)=\kappa\left(1-s\left(\mathbf{r}\right)\right)-\kappa\int\rho^{el}\left(\mathbf{r}'\right)\frac{\delta s\left(\mathbf{r}'\right)}{\delta\rho^{el}\left(\mathbf{r}\right)}\mathrm{d}\mathbf{r}'
=\kappa\left(1-s\left(\mathbf{r}\right)\right)-\kappa\rho^{el}\left(\mathbf{r}\right)\frac{ds}{d\rho^{el}}\left(\mathbf{r}\right).
\end{equation}
The last term naturally vanishes if the interface function is not an explicit function of the system electron density. This is the case, for instance, if the cavity from the SSCS model (Equation \ref{eq:cavity-ss}) is employed. In order to evaluate the electron affinity of a system, we can simulate it in the presence of a fictitious confining potential and look at its total energy after removing the corresponding non-physical energy contribution (Equation \ref{eq:confine-energy}). By extrapolating this energy to vanishing confinement conditions, the energy of a non-embedded system can be estimated, even for those cases where unbound states would make optimization of the electronic density impossible.

\subsection{Electron Affinities Calculations\label{sec:electron-affinities}}

Electron affinities (EAs) have been computed as energy differences between the neutral species and the corresponding anions ($\Delta$SCF approach). Consistently with Ref.\cite{Kim-JCP-2011}, we have computed adiabatic EA values by considering optimized equilibrium geometries for both the neutral and the negatively charged species, and by additionally including (harmonic) zero-point energy (ZPE) corrections:  
\begin{equation}
\text{EA} = (E_0 + \frac{1}{2}\sum_i^N\hbar\omega_{0,i}) - (E_{-1} + \frac{1}{2}\sum_i^N\hbar\omega_{-1,i})=\Delta E + \Delta \text{ZPE}, \label{eq:EA} 
\end{equation}
where $\Delta E = E_0 - E_{-1}$ is the total energy difference between the neutral and the anionic species and $\Delta\text{ZPE} = \frac{1}{2}\sum_i^N\hbar\omega_{0,i} - \frac{1}{2}\sum_i^N\hbar\omega_{-1,i}$ is the corresponding difference between ZPE contributions. Note that the sums extend over the $N=3N_A - 6$ vibrational degrees of freedom of each molecule ($N=3N_A - 5$ for linear molecules), where $N_A$ is the number of atoms in the molecule. 

\subsection{Computational Details\label{sec:comput-details}}

All calculations have been performed with the \textsc{Quantum ESPRESSO} (QE) distribution \cite{Giannozzi-JPhysCondensMatter-2009,Giannozzi-JPCM-2017}. For the simulations involving continuum embeddings, we have used the ENVIRON module \cite{ENVIRON} for QE, where we have also implemented the confining potential. Note that for the dielectric embedding calculations the non-electrostatic terms that are typically employed to estimate cavity, repulsion and dispersion contributions to solvation energies have not been considered here \cite{Scherlis-JCP-2006, Andreussi-JCP-2012}. We have used the PBE generalized-gradient approximation for the exchange-correlation functional \cite{Perdew-PRL-1996, Perdew-PRL-1997} and pseudo-potentials from the Standard Solid-State Pseudopotential library \cite{Prandini-npjCompuMats-2018} (SSSP efficiency 1.0). Plane waves up to a kinetic energy of 40 Ry and 320 Ry have been used for the expansion of the wave-function and of the density, respectively. 

We have performed $\Gamma$-only spin-polarized calculations using a cubic box with a 13 \AA -long side, if not mentioned otherwise. The Martyna-Tuckerman reciprocal-space correction \cite{Martyna-JCP-1999}, opportunely generalized for dielectric embedding \cite{Andreussi-PRB-2014} when necessary, has been employed to remove artifacts from periodic-boundary conditions (PBC). We have verified that the computed EA values are well converged with respect to the cell size and other computational parameters (see Supporting Information). 

Vibrational calculations have been performed using the finite-difference approach as implemented in the relevant tool in the atomic simulation environment (ASE)\cite{HjorthLarsen-JPhysCondensMatter-2017}. The harmonic frequencies have been obtained by diagonalizing the force-constant matrix, constructed from the forces computed for two (opposite) displacements of 0.015 \AA\ per atom and Cartesian coordinate.

\section{Results and Discussion\label{sec:results}}

\subsection{Electron Affinity Extrapolation\label{sub:EA}}

\begin{figure}
\begin{centering}
	\includegraphics[width=0.7\columnwidth]{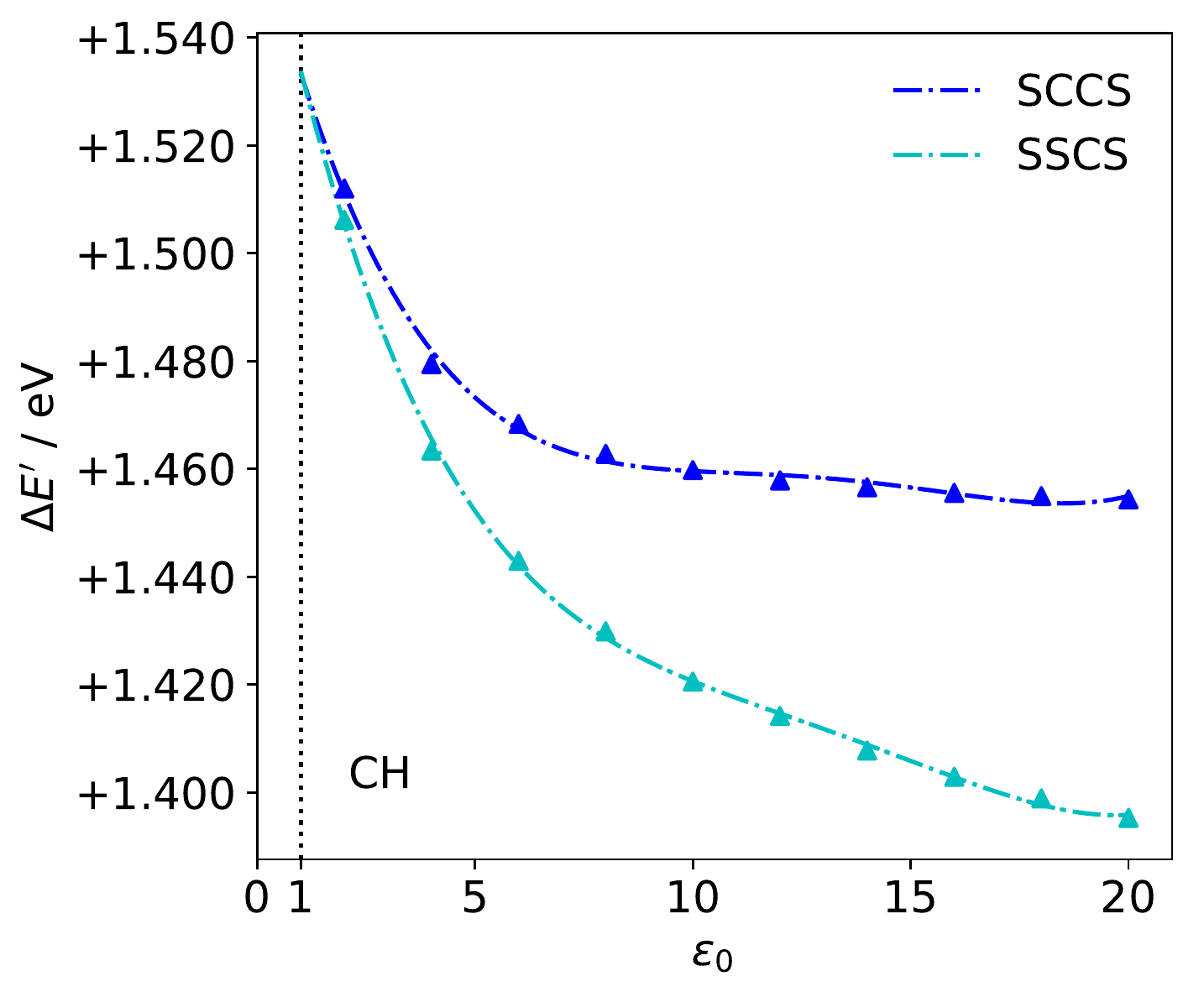}
\end{centering}
\caption{Dielectric extrapolation of $\Delta E'$, using the SCCS cavity (blue curve) or the SSCS cavity (cyan curve) for a representative molecule (CH).}
\label{fig:extrapolation-eps}
\end{figure}

\begin{figure}
\begin{centering}
	\includegraphics[width=0.7\columnwidth]{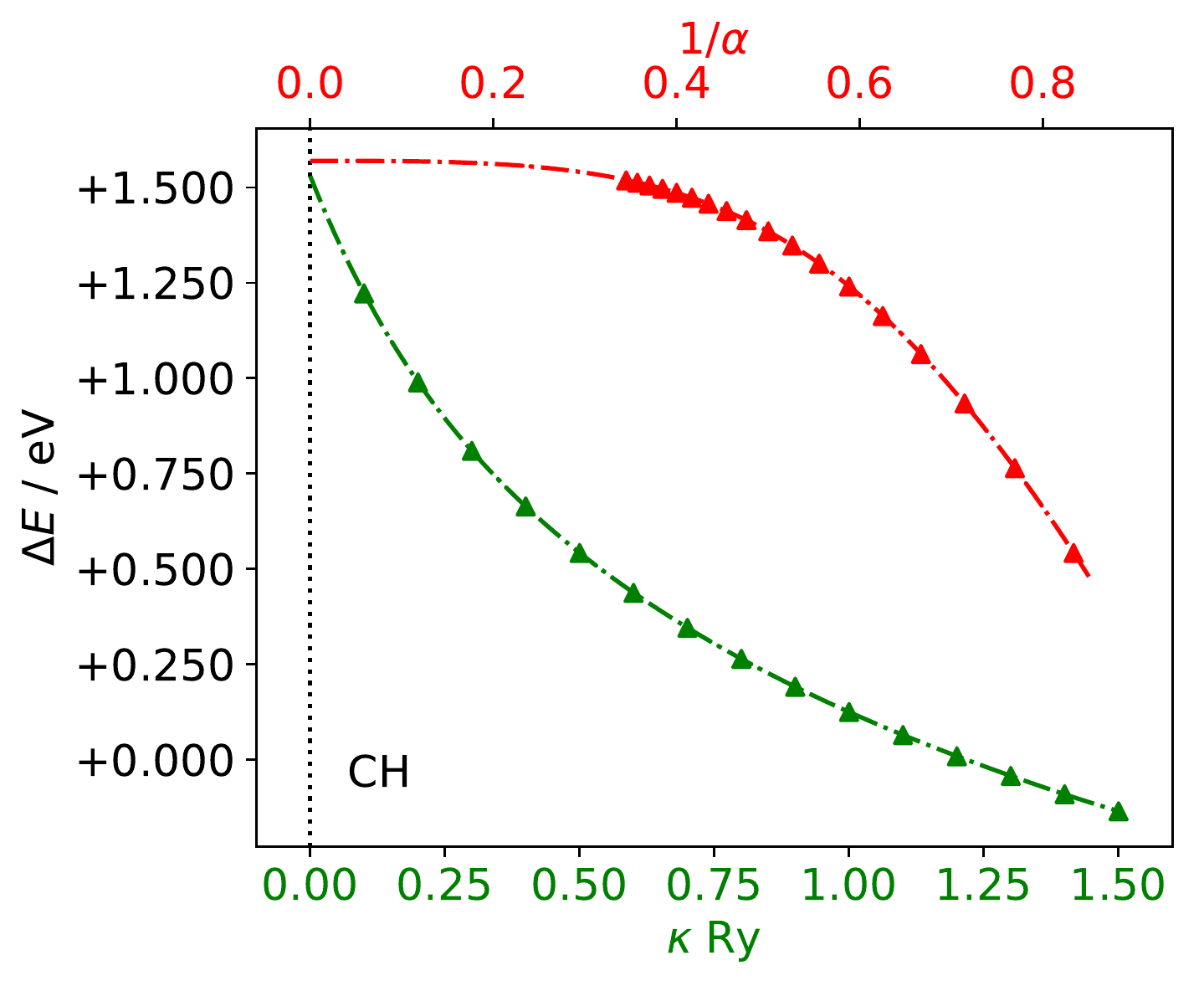}
\end{centering}
\caption{Extrapolation of $\Delta E$ using the confining potential and the SSCS cavity. $\Delta E$ is plotted as a function of the confining potential ($\kappa$, green) and as a function of the scaling factor of the soft-sphere radii ($1/\alpha$, red).}
\label{fig:extrapolation-confinement}
\end{figure}

Figure \ref{fig:extrapolation-eps} and Figure \ref{fig:extrapolation-confinement} illustrate how the two proposed extrapolation procedures can be implemented for a representative molecular species (CH, for all other species see Supporting Information). In Figure \ref{fig:extrapolation-eps}, the vacuum energy difference between the neutral and the anionic form of CH has been obtained by extrapolating results of calculations performed in an embedding dielectric continuum. In particular, we consider the energy difference $\Delta E'\left(\varepsilon_0\right)$: 
\begin{equation}
\Delta E'\left(\varepsilon_0\right) = \left[E_{0}\left(\varepsilon_0\right) - E_{0}^{pol}\left(\varepsilon_0\right)\right] - \left[E_{-1}\left(\varepsilon_0\right) - E_{-1}^{pol}\left(\varepsilon_0\right)\right] = \Delta E\left(\varepsilon_0\right) - \Delta E^{pol}\left(\varepsilon_0\right),
\label{eq:deltaE-electrostatic}
\end{equation}
where $\Delta E\left(\varepsilon_0\right) = E_{0}\left(\varepsilon_0\right) - E_{-1}\left(\varepsilon_0\right)$ is the total energy difference between the charge-neutral and anionic species, respectively, both embedded in a dielectric medium with dielectric constant $\varepsilon_0$. $\Delta E^{pol}\left(\varepsilon_0\right) = E_{0}^{pol}\left(\varepsilon_0\right) - E_{-1}^{pol}\left(\varepsilon_0\right)$ is the corresponding energy difference between the dielectric polarization contributions to the electrostatic energy of the systems (see Equation \ref{eq:electrostatic-energy}). The vacuum energy differences $\Delta E$ is obtained by extrapolating to $\varepsilon_0 = 1$ a polynomial fit of $\Delta E'\left(\varepsilon_0\right)$ as a function of the dielectric permittivity. We note in passing  that vacuum results could be equivalently obtained by taking the  $\varepsilon_0\rightarrow1$ limit of the energy difference $\Delta E\left(\varepsilon_0\right)$, since both $E_{0}^{pol}\left(\varepsilon_0\right)$ and $E_{-1}^{pol}\left(\varepsilon_0\right)$ tend to zero for $\varepsilon_0$ approaching the vacuum dielectric constant. We have found, however, that by subtracting the polarization contributions to the total energies we obtain smoother functions of $\varepsilon_0$, which are thus preferable for numerical extrapolations. 

The two curves in Figure \ref{fig:extrapolation-eps} differ in the choice of the cavity function $s(\textbf{r})$. In particular, we have tested the electron-density-based cavity from the SCCS model (Equation \ref{eq:cavity-sccs}) and the cavity based on atom-centered spheres from the SSCS model (Equation \ref{eq:cavity-ss}). While we observe different trends for the various molecules and cavities (the trend is not always monotonic), we always obtain smooth $\Delta E'$ vs $\varepsilon_0$ curves that allow for a stable polynomial extrapolation to $\varepsilon_0=1$. 

Figure \ref{fig:extrapolation-confinement} shows how the vacuum $\Delta E$ value for unbound anions can be alternatively extrapolated using the confining potential described in Section \ref{sec:confinement-potential}. Note that the atom-centered interface function from the SSCS model has been employed to construct the confining potential. The following energy difference has been considered in this second type of extrapolation:
\begin{equation}
\Delta E\left(\kappa\right) =  E_{0}\left(\kappa\right) - E_{-1}\left(\kappa\right).
\label{eq:deltaE-confine}
\end{equation}
where $E_{0}\left(\kappa\right)$ and  $E_{-1}\left(\kappa\right)$ are the total energy of the charge-neutral and anionic species, respectively, where we have explicitly indicated the dependence on the confining potential $\kappa$. For the confining potential case, we have found that sufficiently smooth curves can be obtained without the need of subtracting the confining energy contributions from the respective total energies, and we have thus used these for the vanishing-embedding extrapolation.

Figure \ref{fig:extrapolation-confinement} illustrates two equivalent extrapolation approaches based on the use of the confining potential. On the one hand, the vacuum energy difference $\Delta E$ can be obtained for vanishing confining potentials, i.e. by taking the $\kappa\rightarrow 0$ Ry limit of $\Delta E(\kappa)$. On the other hand, the same result is obtained by keeping fixed the magnitude of the potential but systematically increasing the size of the cavity, thereby shifting the confining potential to  larger distances from the anion. Using the SSCS interface function, this is achieved by increasing the value of the $\alpha$ parameter, which is the scaling factor that multiplies the ionic radii of the atom-centered spheres that constitute the cavity  (see Equation \ref{eq:cavity-ss}).  As shown in Figure \ref{fig:extrapolation-confinement}, the vacuum $\Delta E$ value for CH can be obtained by either extrapolating $\Delta E(\kappa)$ to $\kappa = 0$ Ry or to $1/\alpha = 0$. A polynomial fit is employed for the extrapolation in both cases (more details are provided in the Supporting Information). 

\begin{figure}
\begin{centering}
	\includegraphics[width=0.35\columnwidth]{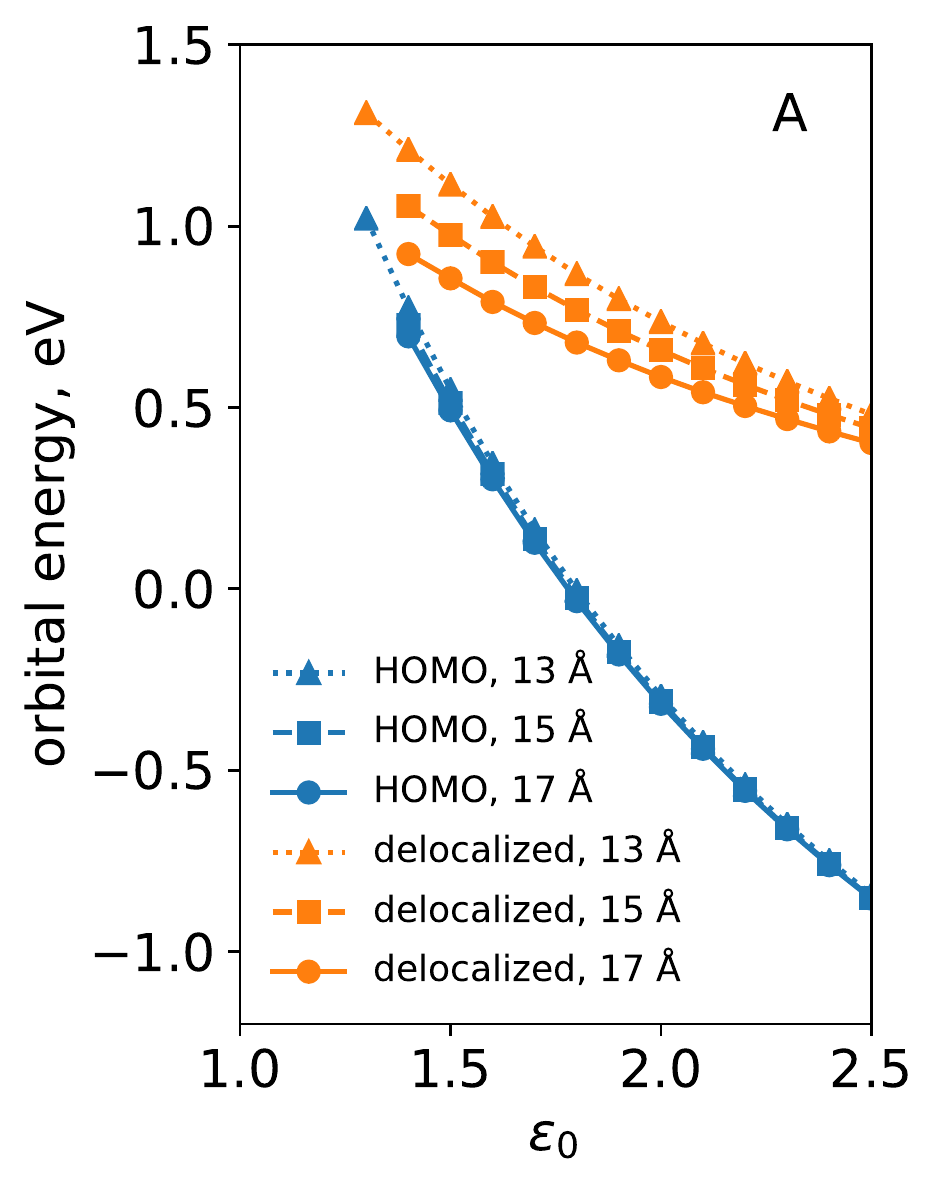}
	\includegraphics[width=0.35\columnwidth]{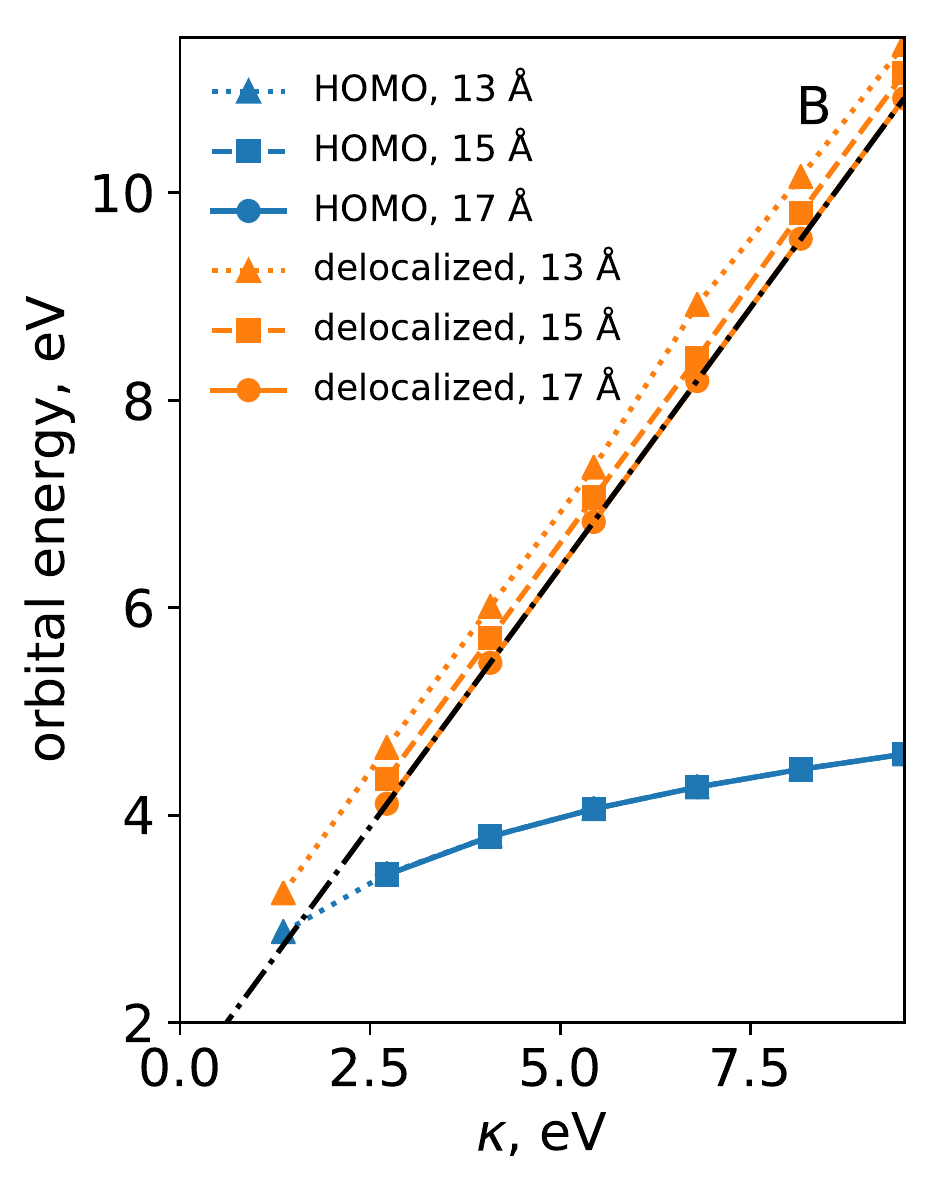}
\end{centering}
\caption{Orbital energies as a function of the dielectric constant of the embedding medium $\varepsilon_0$ (left) and the asymptotic confining potential $\kappa$ (right) for CH$_3^-$. Blue and orange lines illustrate the energies of the anion's HOMO and the lowest-energy delocalized state, respectively. Different line styles and symbols illustrate the various cells sizes employed (the corresponding cubic cell's side is reported in the legend).
As a reference, the black curve illustrates a line with a slope of one.  }
\label{fig:bands}
\end{figure}

Figure \ref{fig:bands} shows how the dielectric embedding and the confining potential stabilize localized electronic states that would otherwise be unbound, using the CH$_3^-$ species as an illustrative example. In particular, Figure \ref{fig:bands} reports the energy dependence of the HOMO and of the lowest-energy delocalized level on the parameters that define the two embedding environments:  the dielectric constant of the medium $\varepsilon_0$ and the confining potential factor $\kappa$. For what concerns the dielectric embedding (Figure \ref{fig:bands}A), both the energy of the HOMO and the one of the delocalized level decrease with increasing  $\varepsilon_0$. Indeed, both states are stabilized by the dielectric embedding, due to the larger electrostatic interaction with the polarization charge density. Localized states like the HOMO, however, undergo larger stabilizations for increasing values of $\varepsilon_0$. This is intuitively understood from the fact that a higher electron localization is linked to larger potential gradients, which give rise to larger polarization densities and, in turn, to more negative electrostatic energy contributions. Thus, the dielectric embedding stabilizes both localized and delocalized states, but it promotes electron localization by providing a larger stabilization to the localized states than to the delocalized ones.

Figure \ref{fig:bands}B illustrates corresponding trends for the confining potential embedding. This second embedding approach introduces a destabilizing term that affects both the HOMO and the delocalized states. This is clearly visible from the energy increase of the two levels for increasing values of $\kappa$. The latter, however, undergoes a larger destabilization, following a linear trend. In particular, the upward potential shift of the delocalized level coincides with the value of the confining potential applied (cf. line with unitary slope in Figure \ref{fig:bands}B). In contrast, the HOMO energy follows a milder dependence on $\kappa$, since the corresponding state is mostly localized in the region of space where the confining potential is zero. Therefore, the confining potential embedding fosters electron localization through a destabilizing contribution, which is larger for delocalized states than for the localized ones.   

Figure \ref{fig:bands} also illustrates how the delocalized energy levels are affected by the volume of the cell employed in the calculations. While the HOMO energies are not affected by the cell size and they virtually remain constant when increasing the cubic box side from 13 \AA\ to 17 \AA , the delocalized levels are considerably affected by a volume change. This is consistent with a `particle-in-a-box' model, where the minimum-energy  level shifts down for increasing volumes. According to this picture, the minimal embedding conditions for which localized anion states can be effectively stabilized are a function of the cell size. We observe, in fact, that the threshold values of $\varepsilon_0$ and $\kappa$ that allow for converging anion calculations shifts to larger values with increasing cell volumes. These thresholds can be identified with the points where the (localized) HOMO level becomes lower in energy than the corresponding delocalized state, i.e. the points where the blue and orange curves cross in Figure \ref{fig:bands}.

\begin{figure}
\begin{centering}
	\includegraphics[width=\columnwidth]{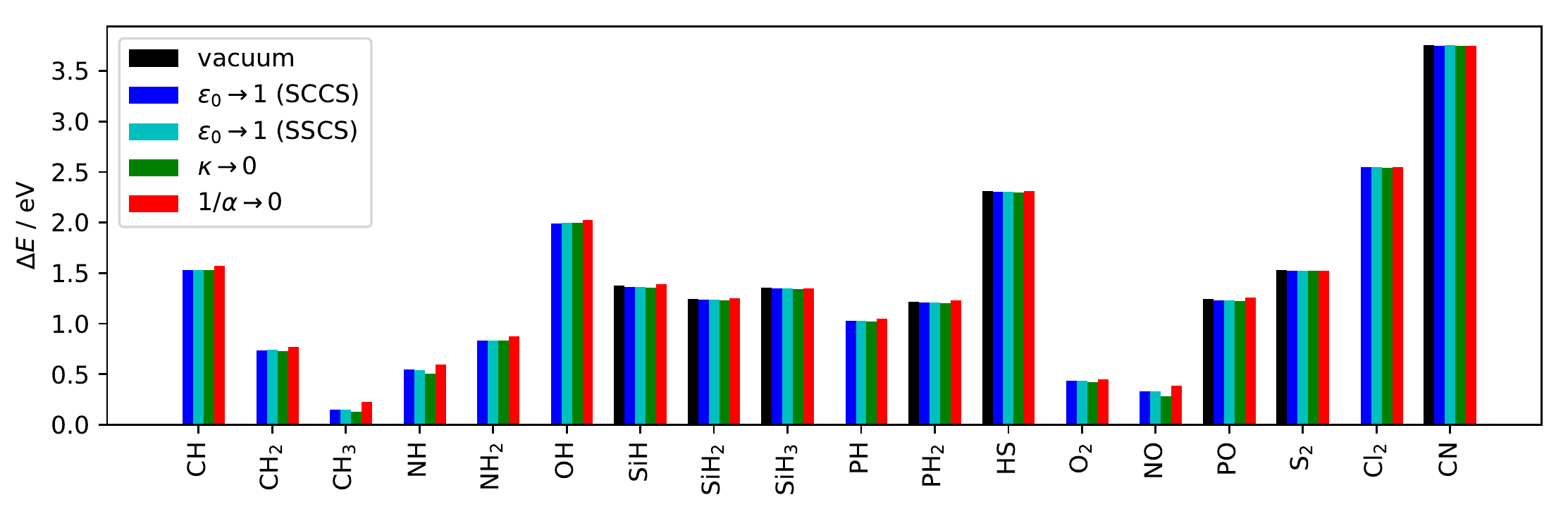}
\end{centering}
\caption{Vacuum extrapolated $\Delta E$ values computed for all the elements of the G2-1 set. The various extrapolation methods are compared to each other: blue and cyan bars are for the dielectric extrapolations using the SCCS cavity and the SSCS cavity, respectively; the green and red bars are for the confining-potential extrapolations to zero potential and large cavity, respectively; the  $\Delta E$ values computed in vacuum are plotted as black bars (no bar is shown if the anion calculation did not converge). }
\label{fig:compare-extrapolations}
\end{figure}

The results of the different extrapolation techniques considered are illustrated in Figure \ref{fig:compare-extrapolations} for all the molecules of the G2-1 set. We also report the results of vacuum calculations for the molecules whose anion calculation converged for the cell size considered (13$\times$13$\times$13 \AA$^3$). For these molecules, all techniques lead to extrapolated values that agree within less than 30 meVs with the reference vacuum results. Overall, the dielectric extrapolation technique, using either the SCCS or the SSCS cavity, and the confining-potential extrapolation approach based on either the zero-potential or the large-cavity limit, give rise to total energy differences between neutral and anion species that are in good agreement with each others. The largest deviations are observed for the confining potential extrapolation scheme based on the cavity size: the infinitely-large-cavity limit, in fact, is approached rather slowly (see Figure \ref{fig:extrapolation-confinement}), introducing the largest error in the extrapolated values. 

EA estimates are obtained by adding ZPE corrections to the extrapolated energy differences between the optimized neutral and anionic species (see Equation \ref{eq:EA}). In order to determine the $\Delta \text{ZPE}$ corrections, we follow an approach that is analogous to the one employed to extrapolate $\Delta E$ values to vacuum conditions, using the frequencies computed for the anion and for the neutral species in different embedding environments. While all described approaches can in principle be employed for this purpose, we determine the $\Delta \text{ZPE}$ corrections using the only dielectric extrapolation technique in combination with the SSCS cavity and apply those to all extrapolation methods. The computed values of $\Delta \text{ZPE}$ (see Supporting Information) agree well with the values determined in Ref. \cite{Kim-JCP-2011}~. Our approach, however, allows to consistently determine ZPE corrections using the same scheme that is employed to calculate the energy differences $\Delta E$. This instead is not possible for the approaches that exploit different electronic-structure method for the optimization of the electron density and for the energy evaluation (e.g. the HF-PBE method from Ref. \cite{Kim-JCP-2011}).
  
\begin{figure}
\begin{centering}
	\includegraphics[width=0.5\columnwidth]{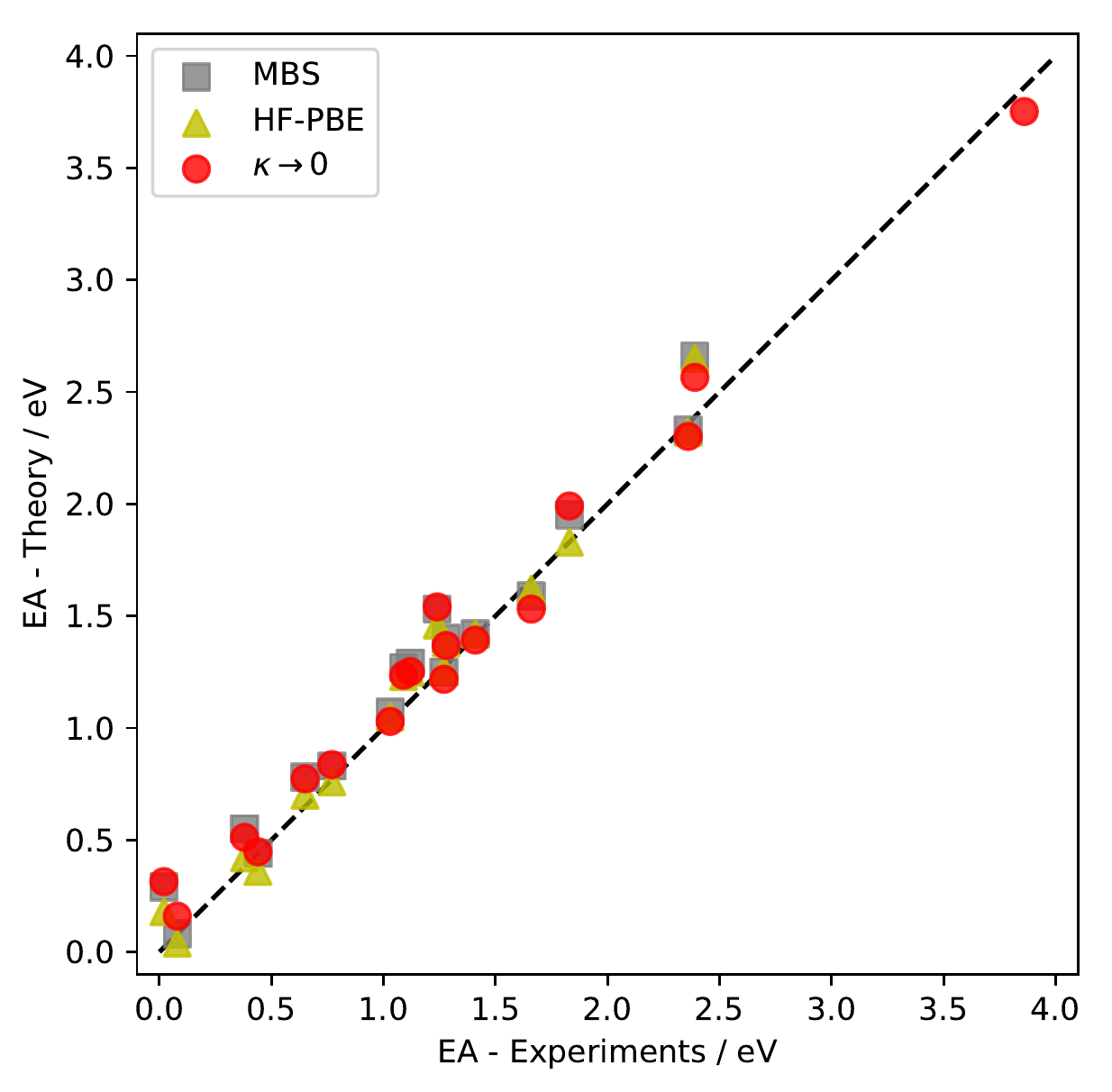}
\end{centering}
\caption{Theoretically-computed EAs versus corresponding experimental values for the G2-1 set. Red symbols corresponds to results obtained from the extrapolation method (vanishing potential limit in the confining-potential embedding), while grey and yellow symbols are for the moderate basis set (MBS) method \cite{Kim-JCP-2011} and for PBE calculations based on Hartree-Fock densities (HF-PBE) \cite{Kim-JCP-2011}, respectively.}
\label{fig:compare-experiments}
\end{figure}

Predicted EA values for the G2-1 set are plotted against experimental data in Figure \ref{fig:compare-experiments}. Results are  compared to the MBS calculations and to EA values obtained by performing PBE-DFT calculations on pre-computed Hartree-Fock electron densities (HF-PBE approach)\cite{Kim-JCP-2011}. The mean absolute errors (MAEs) for all theoretical models considered here are presented in Table \ref{tab:MAEs}. Considering the good agreement across the various extrapolation methods (Figure \ref{fig:compare-extrapolations}), which is reflected in the similarity of the corresponding MAEs (Table \ref{tab:MAEs}), only one set of extrapolated data is plotted in Figure \ref{fig:compare-experiments} (tabulated EA values computed using all the extrapolation approaches are reported in the Supporting Information). The various extrapolation methods generally return similar level of agreement with experimental data, with the MAE across the G2-1 being approximately 0.12 eV. A slightly larger MAE (0.146 eV) is obtained for the confining-potential extrapolation method based on the cavity size. We ascribe this to the larger uncertainty in the determination of the vacuum EA, due to the observed slower convergence towards the large-cavity limit as obtained with this strategy. Overall, the MAE obtained for the various extrapolation methods is very similar to what obtained through the MBS. This is not surprising, as both the embedding extrapolation calculations and the MBS ones are based on the PBE density functional for both the density optimization and the energy evaluation. A lower MAE (0.079 meV) is obtained using the HF-PBE method, which is consistent with the use of higher quality densities; similar results could be expected here using simple self-interaction corrections\cite{Perdew-PRB-1981, DAvezac-PRB-2005}.

It is important to stress the fact that the embedding extrapolation approach described here does not aim at producing highly accurate electron affinities, as the limiting factor of the method's accuracy is the density functional employed in the underlying electronic-structure calculations. However, the proposed scheme provides two main advantages with respect to available methods. First, it provides a framework that is fully self-consistent, and that makes use of a single electronic-structure methods for the electron density optimization and the energy estimate, enabling consistent geometry optimizations and frequency calculations. Most importantly, it can be employed in combination with any basis-set type, including plane waves, and it can be straightforwardly applied in combination with extended systems, as illustrated in the following section. 
 
\begin{table}
\begin{tabular}{cc}
\hline
\hline
Method	& MAE (eV) \tabularnewline
\hline 
MBS	\cite{Kim-JCP-2011} 				&  0.115 \tabularnewline
HF-PBE \cite{Kim-JCP-2011} 			&  0.079 \tabularnewline
Dielectric extrapolation (SCCS)			&  0.123 \tabularnewline
Dielectric extrapolation (SSCS)			&  0.123 \tabularnewline
Confinement extrapolation (energy)		&  0.115 \tabularnewline
Confinement extrapolation (cavity size)	&  0.146 \tabularnewline
\hline
\hline
\end{tabular}
\caption{Mean absolute errors (MAEs) for the various theoretical methods considered. }
\label{tab:MAEs}
\end{table}

\subsection{Application to Extended Systems\label{sub:Pt111-Cl}}

Plane-wave calculations on isolated anions that are predicted to have a positive HOMO by DFT present severe converge issues. This is because the system tends to delocalize the fraction of the additional electron that can not be bound by the nuclei. If the simulation box, however, includes a second subsystem that can accept the unbound charge, like e.g. a metal surface, a different and problematic aspect can emerge. Indeed, under these circumstances, the lowest-energy electron density configuration involves the extra electron to be split between the anion and the metal, regardless of the distance between the two subsystems. 

As a study system, we consider here a chloride ion sitting at 10 \AA\ from a Pt(111) surface. The surface has been modeled as a bulk-like 4-layer slab, constructed using the computed equilibrium lattice constant $a = 3.961$ \AA . A 3$\times$3 multiple of the surface primitive cell has been considered, and the first Brillouin zone has been sampled with a 6$\times$6$\times$1 $\Gamma$-centered k-point grid. A large separation between periodic replicas of the slab (40 \AA ) has been introduced along the surface normal.

\begin{figure}
\begin{centering}
	\includegraphics[width=0.5\columnwidth]{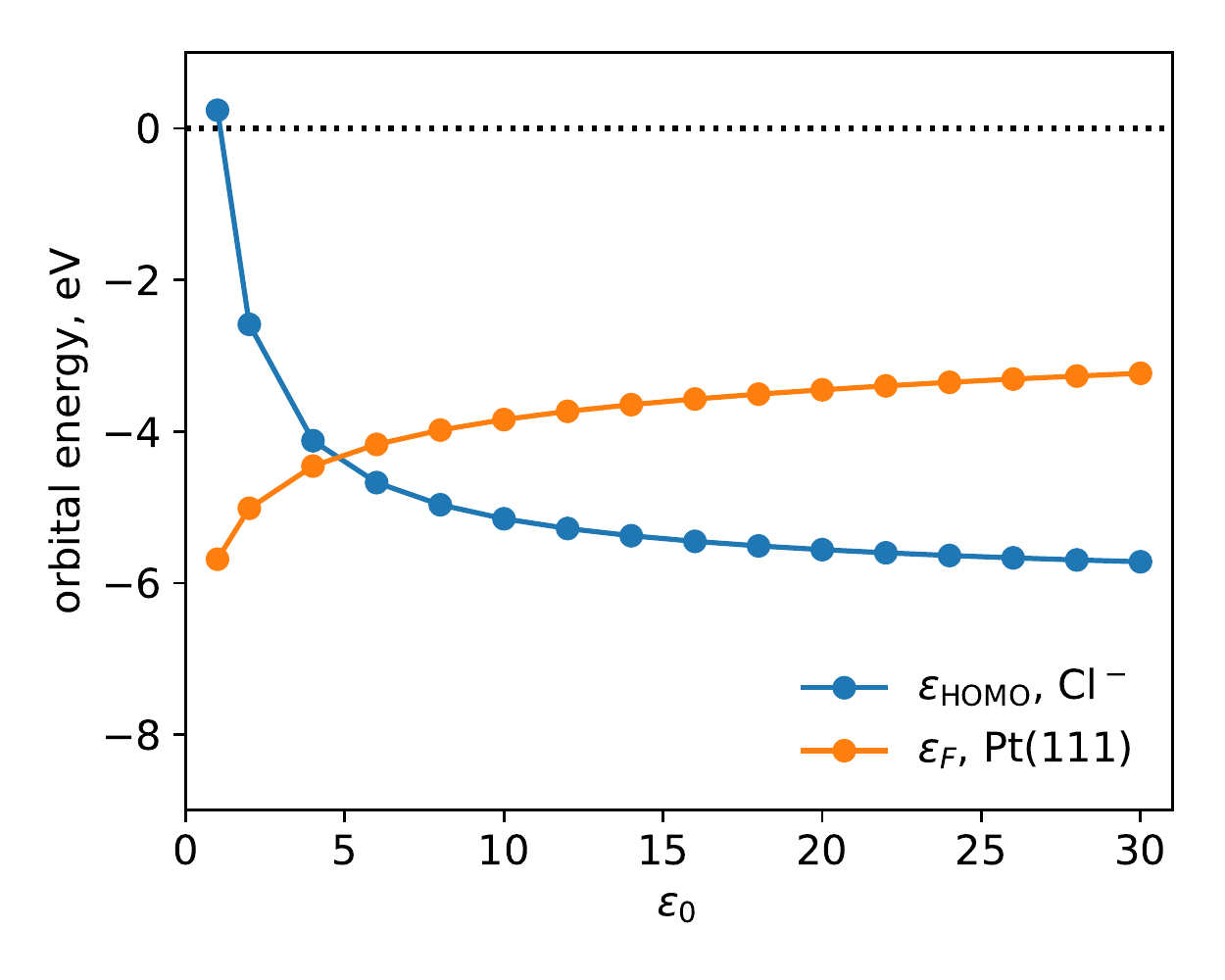}
\end{centering}
\caption{The HOMO energy of an isolated Cl$^-$ anion (blue) is plotted as a function of the dielectric constant of the embedding medium. The Fermi energy of the Pt(111) slab, $\varepsilon_{F}$, is shown in green. A 15$\times$15$\times$15 \AA$^3$ cubic cell has been employed for the isolated anion. The SSCS cavity, with an $\alpha$ parameter of 1.2 has been employed to set the boundary between the quantum-mechanical and the embedding regions.}
\label{fig:Pt111-Cl_orbitals}
\end{figure}

Figure \ref{fig:Pt111-Cl_orbitals} shows the HOMO energy calculated for an isolated Cl$^-$ ions in a continuum embedding environment as a function of $\varepsilon_0$. Similarly to what observed for CH$_3^-$ (see Figure \ref{fig:bands}), the HOMO energy, which is positive (thus unbound) under vacuum conditions ($\varepsilon_0 = 1$), rapidly decreases for increasing values of $\varepsilon_0$. Figure \ref{fig:Pt111-Cl_orbitals} also reports the Fermi energy  of the Pt(111) slab, $\varepsilon_F$, as a function of $\varepsilon_0$. $\varepsilon_F$ increases with increasing  $\varepsilon_0$: the dielectric continuum, indeed, screens the surface dipole, thereby lowering the work function across the interface or, in other words, increasing the Fermi level with respect to the asymptotic electrostatic potential, which is set as the zero. 

The HOMO energy of the chloride ion, as the LUMO energy of the neutral Cl atom, should lie below the Fermi energy of the Pt slab. However, in a vacuum environment, the large SIE that affects Cl$^-$ shifts the HOMO to an energy that is considerably larger than $\varepsilon_F$. A fraction of the extra electron is thus expected to be transferred to the Fermi level of the metal, even if the subsystems lie at very large distance from each other. This is a well-known issue in the context of the so-called ion-unbalance model for electrochemical interfaces\cite{Cucinotta2016}. This model exploits the alignment of single-particle energies to drive the formation of an electrified metal surface and charged electrolyte particles in an overall charge neutral unit cell. The large DFT SIE, unfortunately, prevents the formation of anions with the full expected charge, with consequences on their solvation environment. 

\begin{figure}
\begin{centering}
	\includegraphics[width=0.5\columnwidth]{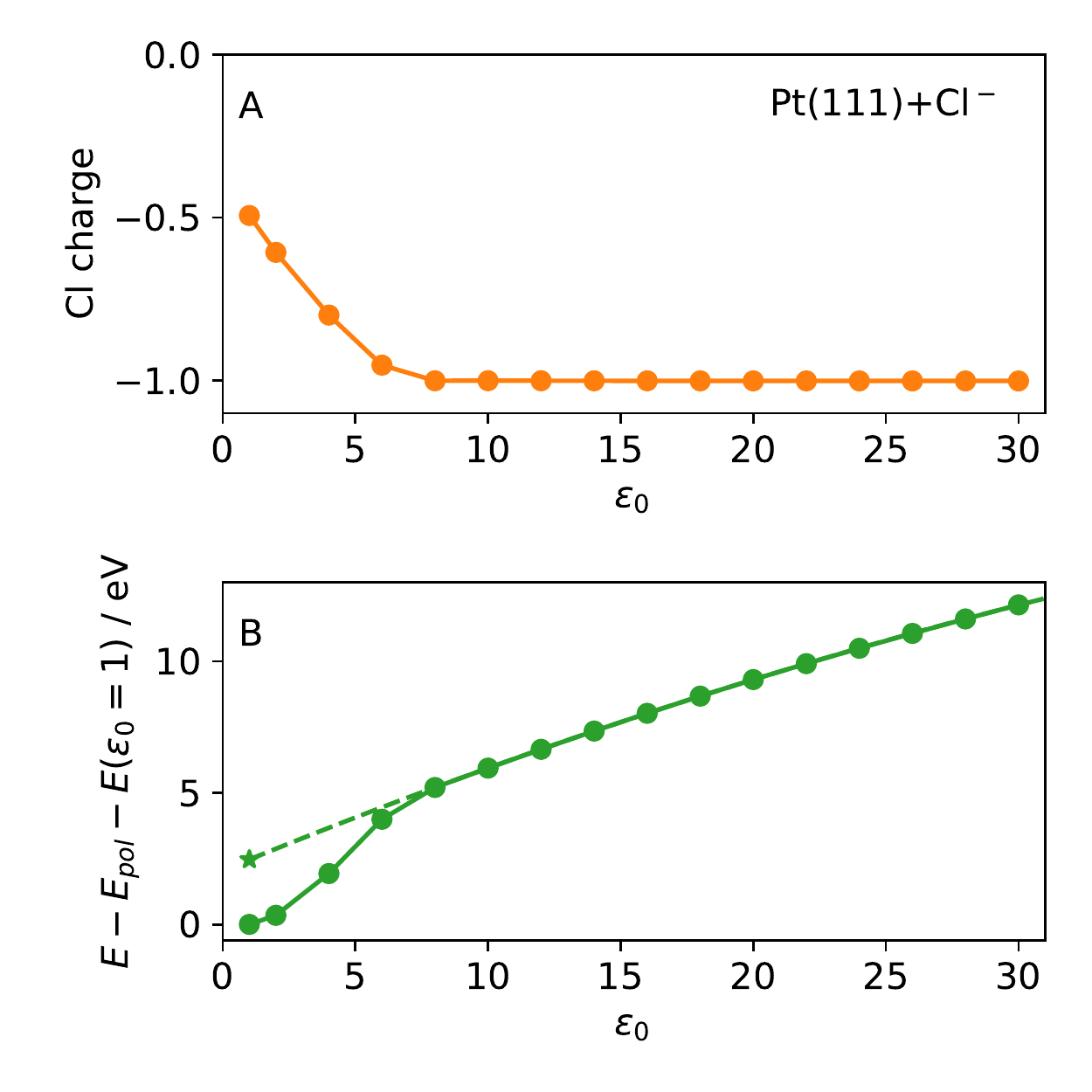}
\end{centering}
\caption{(A): The partial charge on the Cl atom is plotted as a function of the dielectric constant of the embedding medium. (C): total energy of the Pt(111) + Cl$^-$ system as a function of the dielectric constant of the medium. Note that the dielectric polarization contribution has been subtracted from the total energy in order to obtain a smoother curve. The zero has been set as the energy of the system in vacuum. The full dots illustrate the computed energies (the solid line guides the eye). The energies that correspond to a chlorine charge of $\sim -1$ (i.e. the values corresponding to $\varepsilon_0 \ge 8$) have been used for the dielectric extrapolation. The extrapolating curve is illustrated by a dashed line and the star symbol indicates the energy extrapolated to vacuum conditions.}
\label{fig:Pt111-Cl}
\end{figure}

Figure \ref{fig:Pt111-Cl} A illustrates the charge computed for the chlorine atom in the simulations that include the metal surface. In order to determine the net Cl charge we split the simulation cell in two parts using as dividing surface the plane that bisects the vertical distance between the ion and the uppermost Pt layer. The partial charge assigned to each subsystem is then obtained by integrating the charge distributions that reside in the corresponding cell partition. For vacuum conditions, the chlorine charge is approximately $-0.5$, meaning that close to half of an electron actually resides on the platinum slab. The Cl charge gradually decreases for increasing values of $\varepsilon_0$ until it reaches the value of $\sim-1$ for $\varepsilon_0 \sim 5$. This trend can be explained on the basis of the relative difference between the anion HOMO energy and the metal Fermi energy. Indeed, the difference between the two becomes smaller and smaller for increasing values of $\varepsilon_0$ up to $\varepsilon_0\sim5$ (Figure \ref{fig:Pt111-Cl_orbitals}). For larger values of $\varepsilon_0$, the HOMO energy becomes lower than the Pt Fermi energy, and we consistently observe full occupancy of the Cl$^-$ HOMO. 

Figure \ref{fig:Pt111-Cl} B illustrates the total energy of the Pt(111) + Cl$^-$ system, computed as a function of the dielectric constant of the environment. While a smooth energy trend is observed for large values of the dielectric constant of the medium, a sudden drop is observed at $\varepsilon_0 \sim 5$, in correspondence of the $\varepsilon_0$ value at which charge transfer to the surface starts to take place. Nevertheless, we can obtain the energy of the system with the correct charge configuration (i.e. with the electron beyond charge neutrality sitting entirely on the Cl atom) by extrapolating the total energies computed for large $\varepsilon_0$ values, for which the chlorine charge is close to $-1$. We thus extrapolate the value expected for vacuum conditions ($\varepsilon_0$ = 1) from these energies using a polynomial function. The procedure leads to a total energy that differs from the result of the self-consistent vacuum calculation by a considerable amount ($\sim$ 2.5 eV), which is consistent with the significant different charge state of the two subsystems.

\section{Conclusions\label{sec:conclusions}}

Summarizing, we have presented a strategy that allows to stabilize localized anion configurations within a DFT framework. This is achieved by means of a continuum embedding, whose effect on the calculations can be removed through extrapolation to zero intensity. Two embedding schemes, based on a dielectric medium that favors electron localization and a confining potential that penalizes delocalization have been tested and shown to provide virtually identical EAs estimates through $\Delta$SCF calculations.

The proposed strategy allows one to estimate the accuracy of self-consistently-evaluated density functionals without relying on electron densities optimized using other electronic-structure methods. The MAE obtained with the PBE functional for the G2-1 dataset is in line with previous estimates based on the MBS approach, but, in contrast with the latter, our framework presents well-defined basis-set convergence limits and it is not specific to localized basis functions. In addition, the functional extrapolation procedure allows for straightforward force evaluations, which enables self-consistent geometry optimizations and frequency calculations. 

The extrapolation method described here can be similarly applied in the context of periodic calculations for extended systems. As a study case, we have shown how the dielectric embedding can be employed to stabilize the correct charge configuration for the Pt(111)+Cl$^-$ system, and how the corresponding energy for vacuum conditions can be obtained by a suitable extrapolation procedure.  
 
\begin{acknowledgement}
This project has received funding from the European Union's Horizon 2020 research and innovation programme under grant agreements No. 798532. 
N. M. acknowledge support from the MARVEL National Centre of Competence in Research of the Swiss National Science Foundation.
This work was supported by a grant from the Swiss National Supercomputing Centre (CSCS) under project ID s836.
\end{acknowledgement}

\begin{suppinfo}

Supporting Information includes additional data on the EA extrapolation procedure for the molecules of the G2-1 dataset.

\end{suppinfo} 

\bibliography{bibliography}

\end{document}


%
%
%
%
%


\newpage
\section{EA functional extrapolation for the G2-1 dataset\label{sec:extrapolation}}

\begin{figure}
\begin{centering}
\includegraphics[width=0.26\columnwidth]{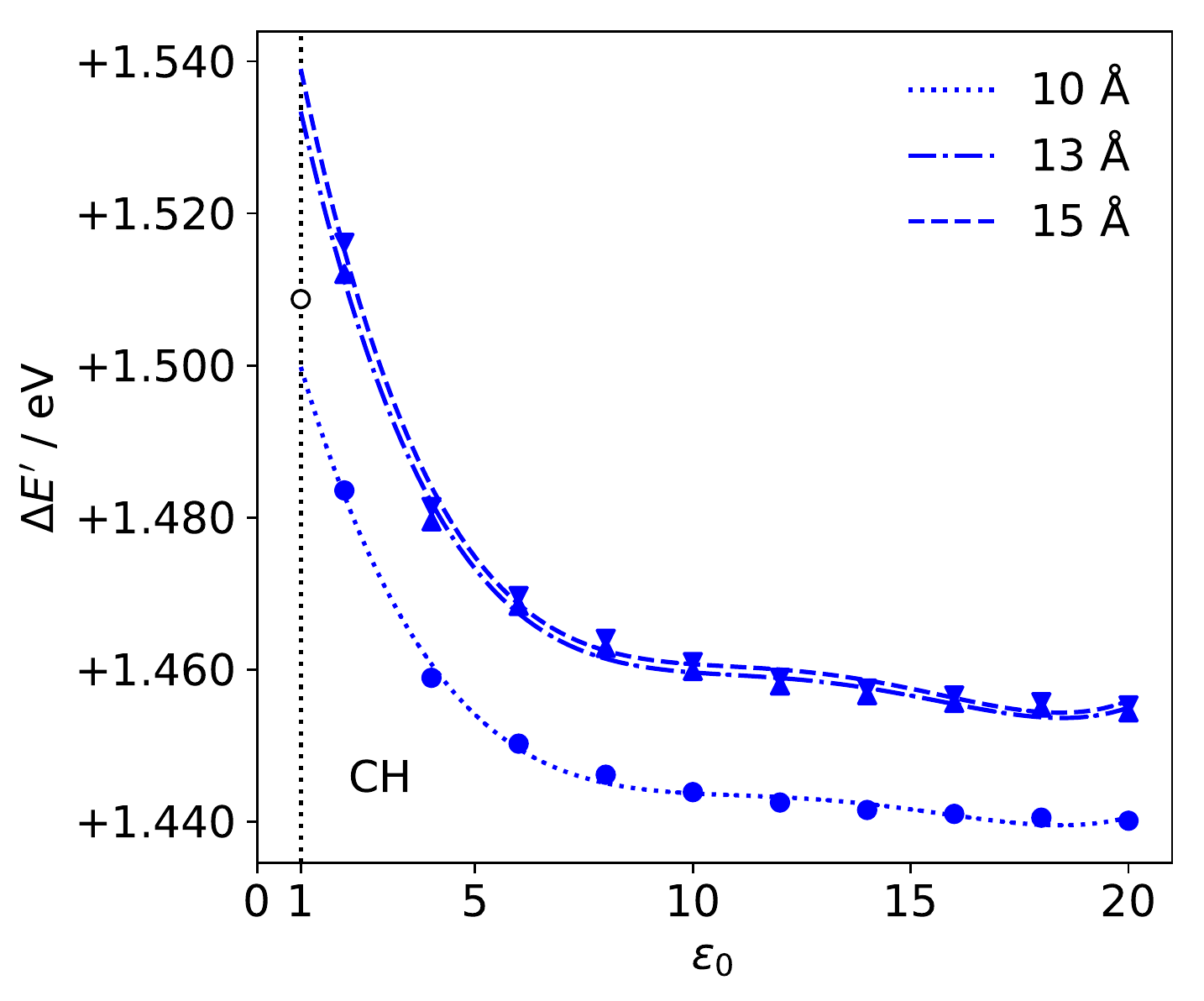}\includegraphics[width=0.26\columnwidth]{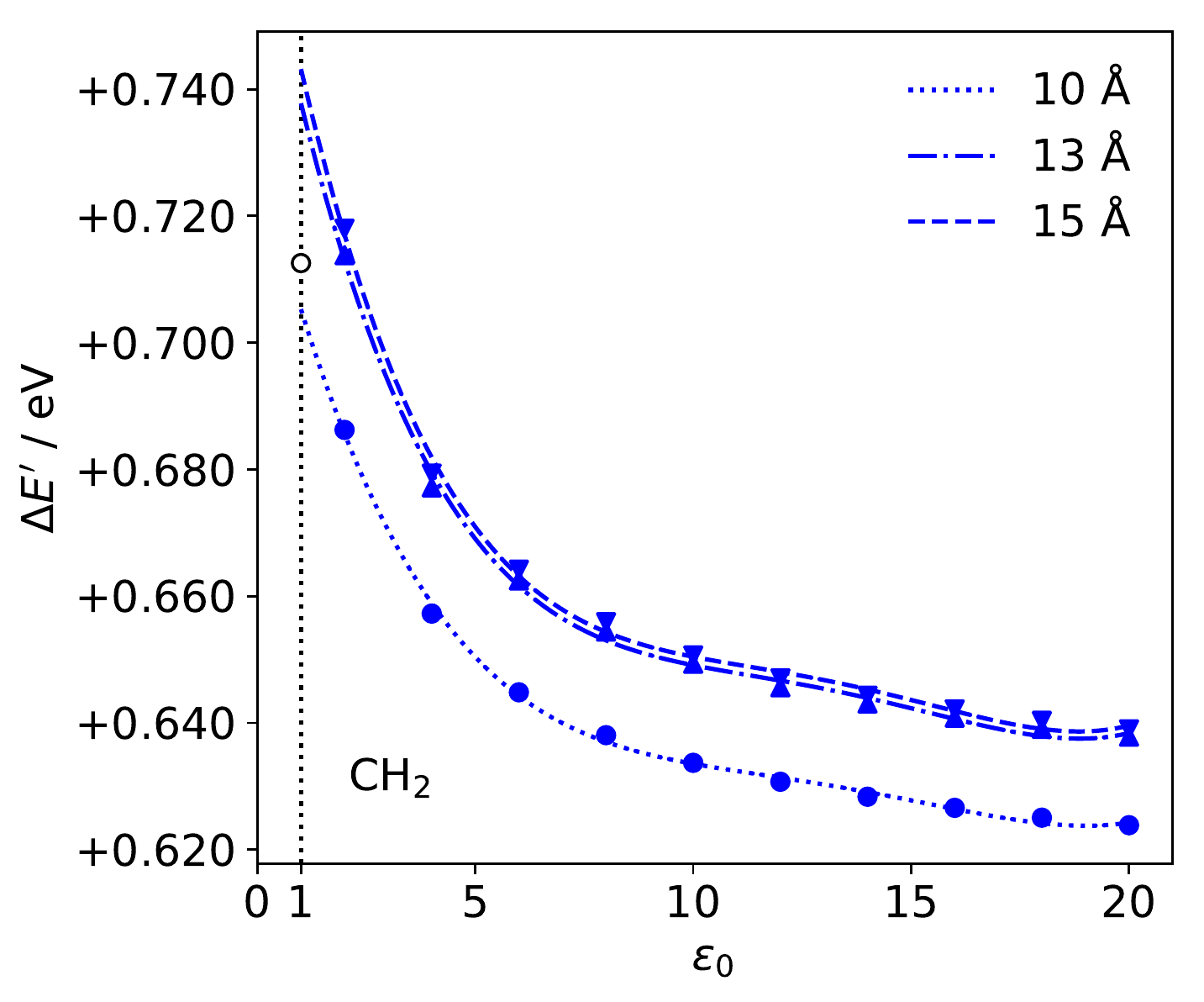}\includegraphics[width=0.26\columnwidth]{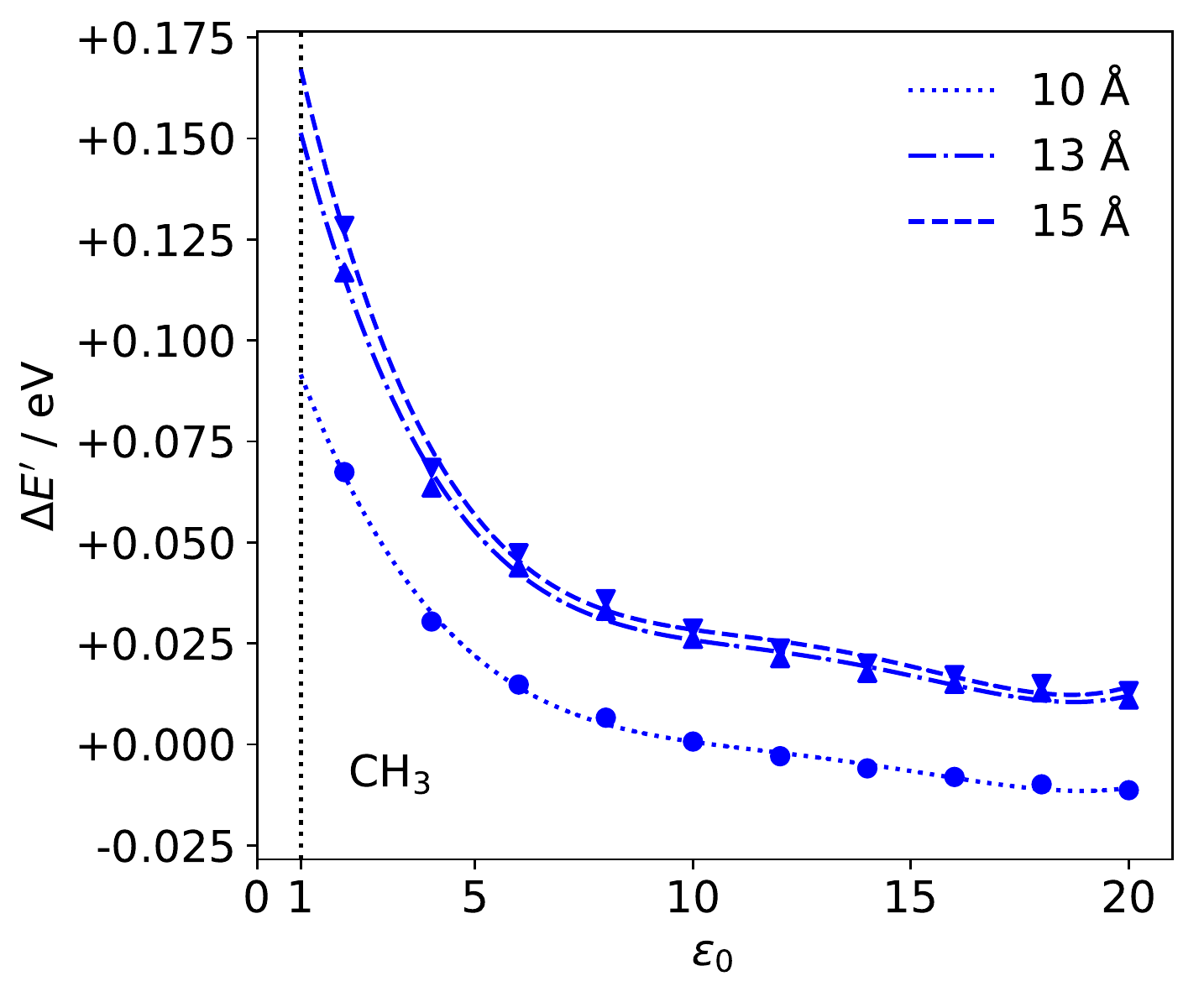} \\
\includegraphics[width=0.26\columnwidth]{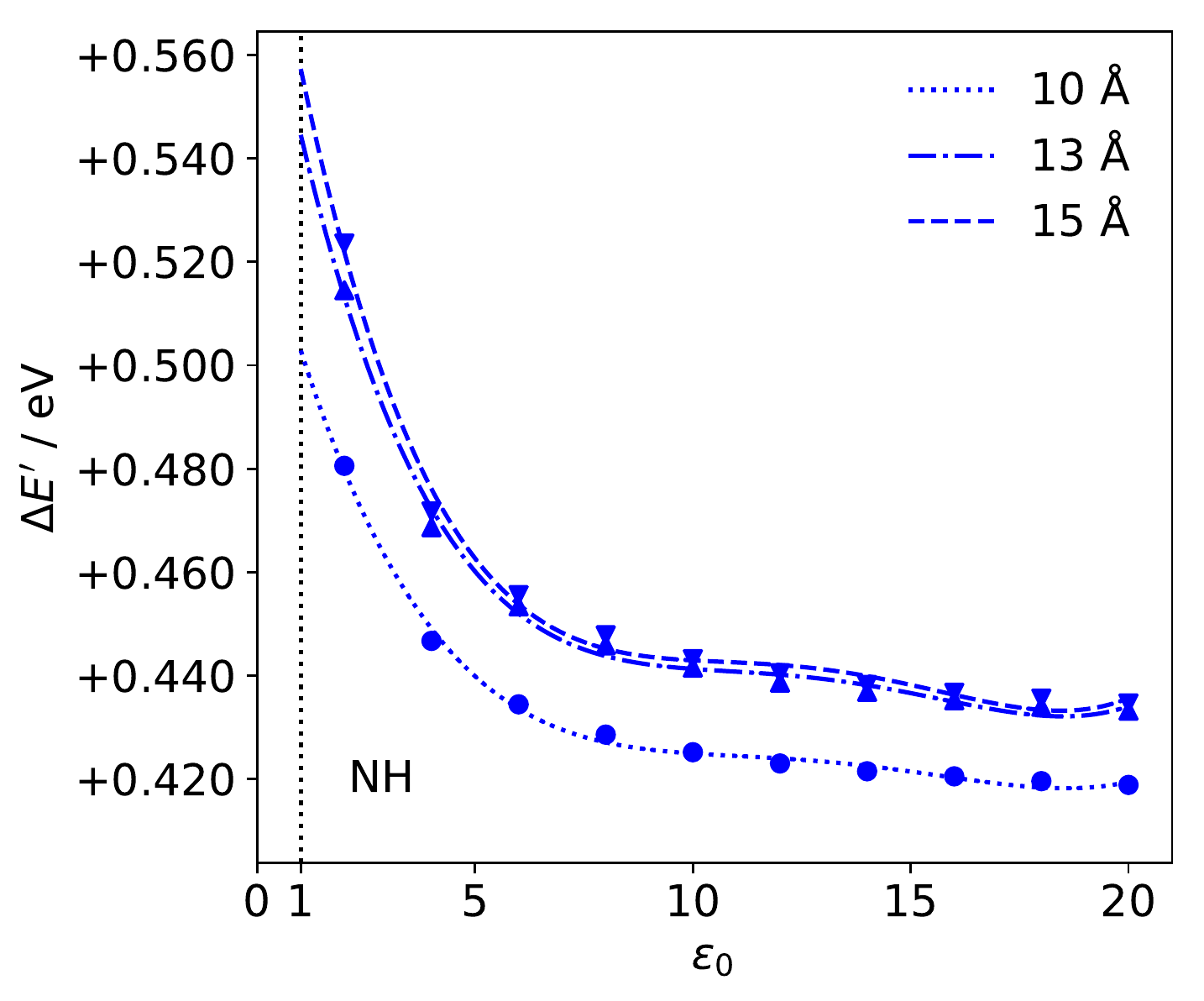}\includegraphics[width=0.26\columnwidth]{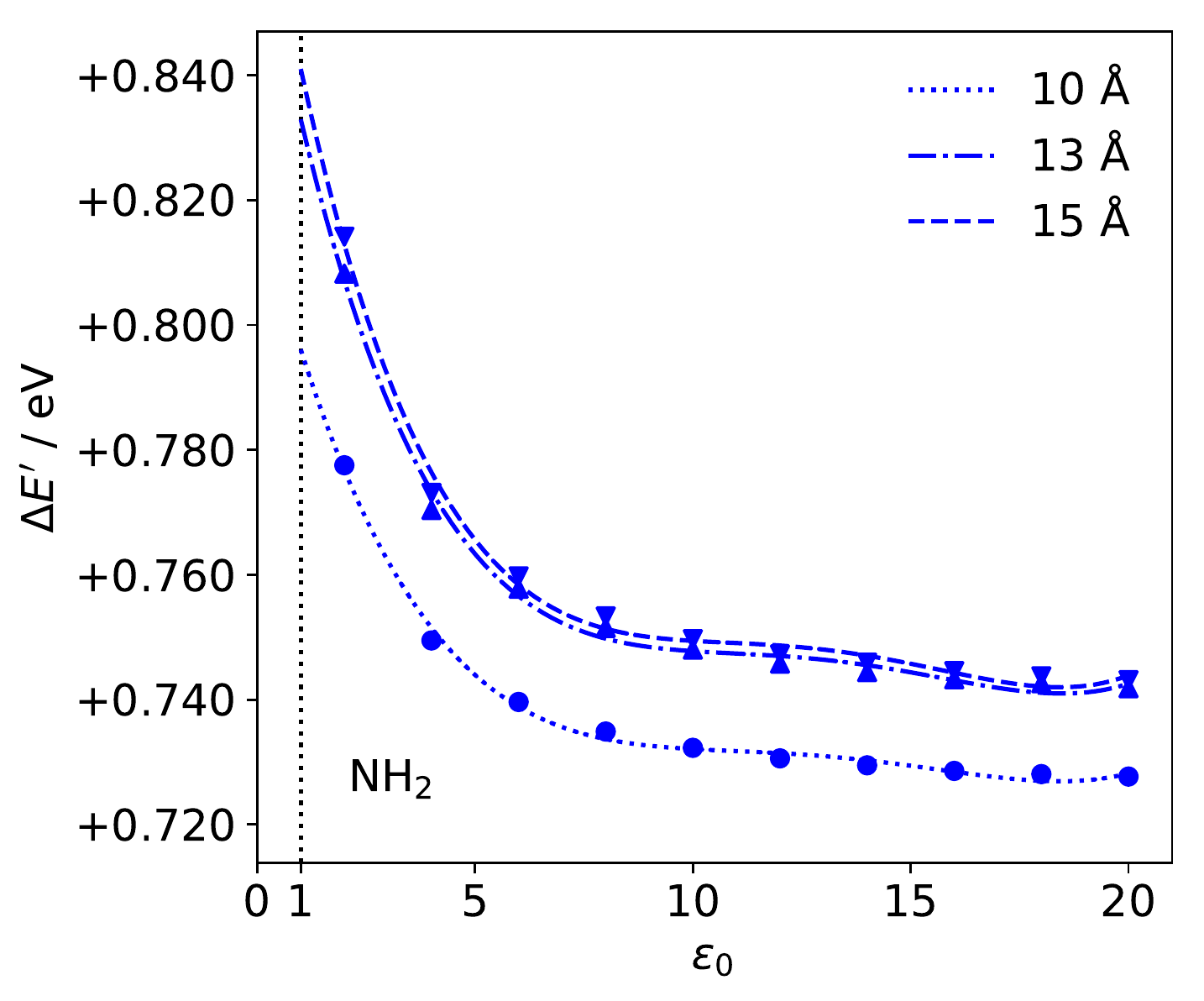}\includegraphics[width=0.26\columnwidth]{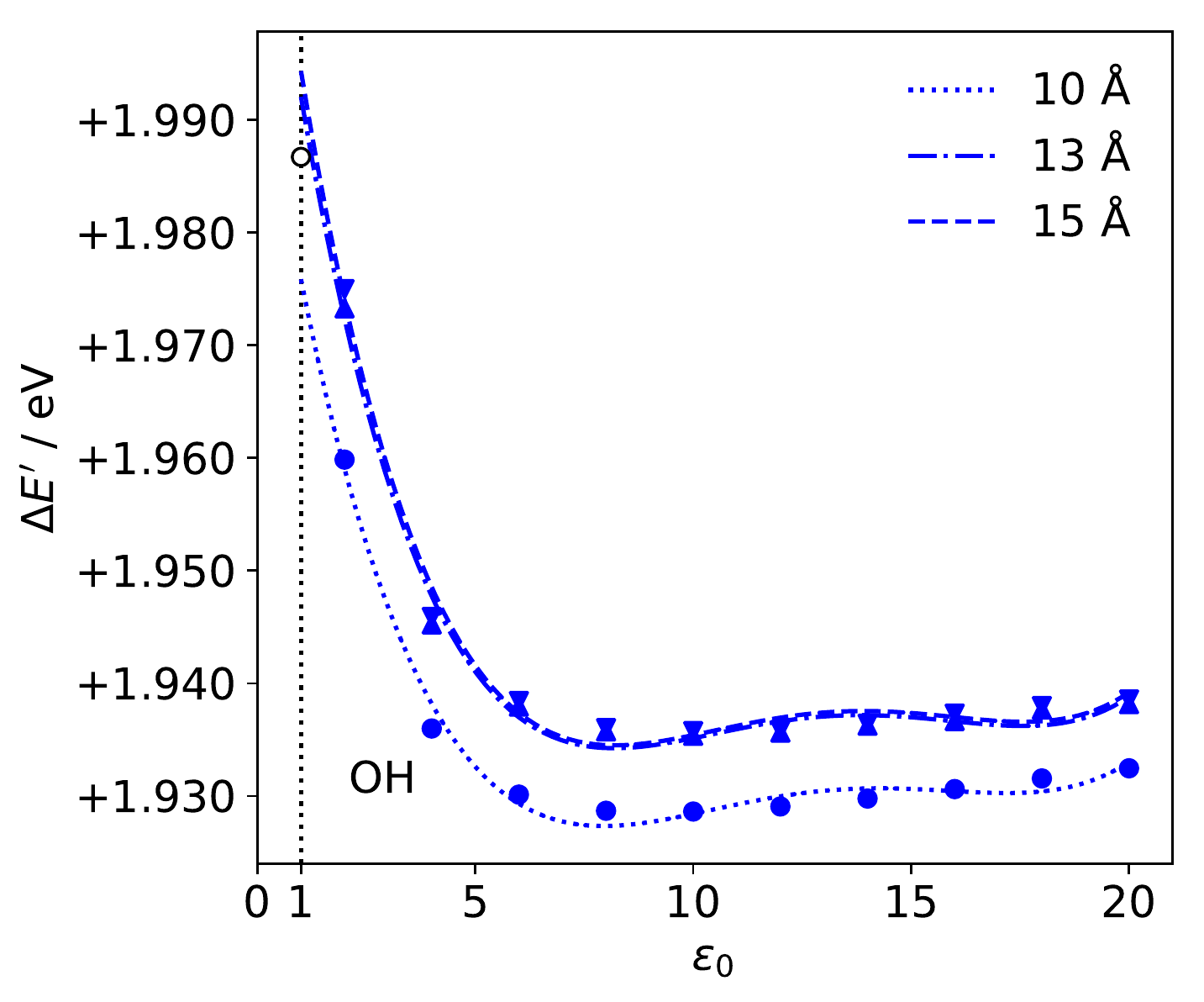} \\
\includegraphics[width=0.26\columnwidth]{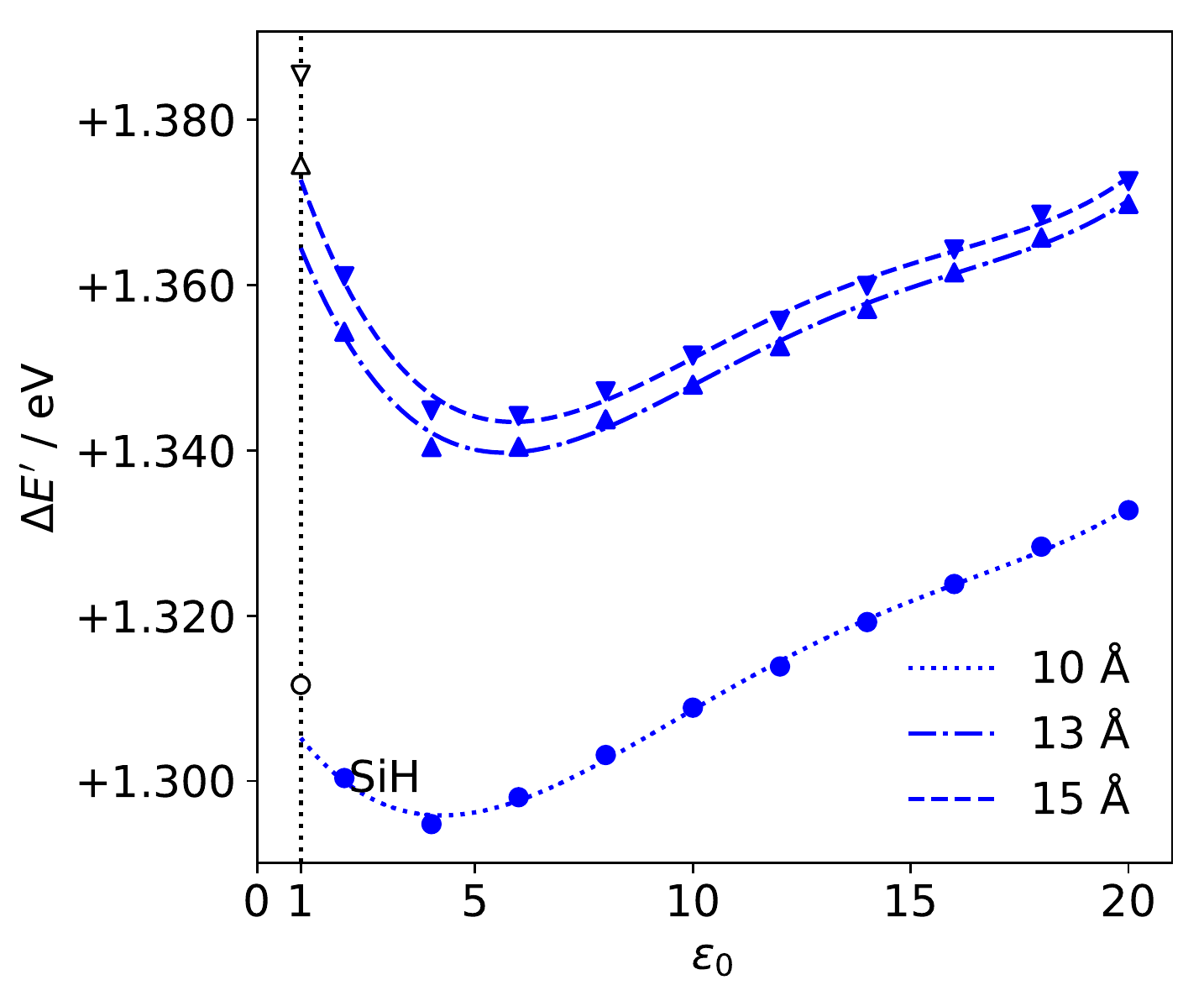}\includegraphics[width=0.26\columnwidth]{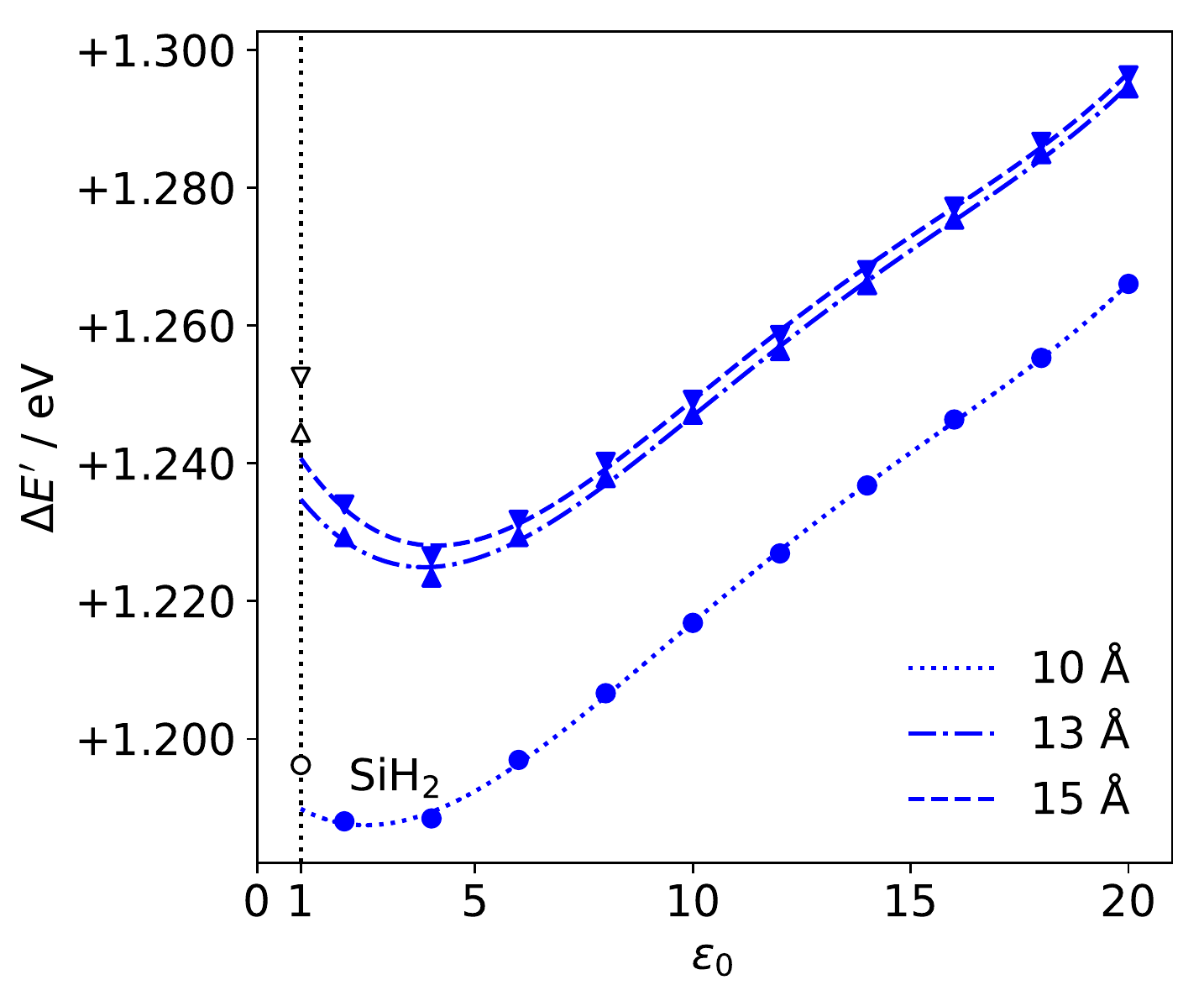}\includegraphics[width=0.26\columnwidth]{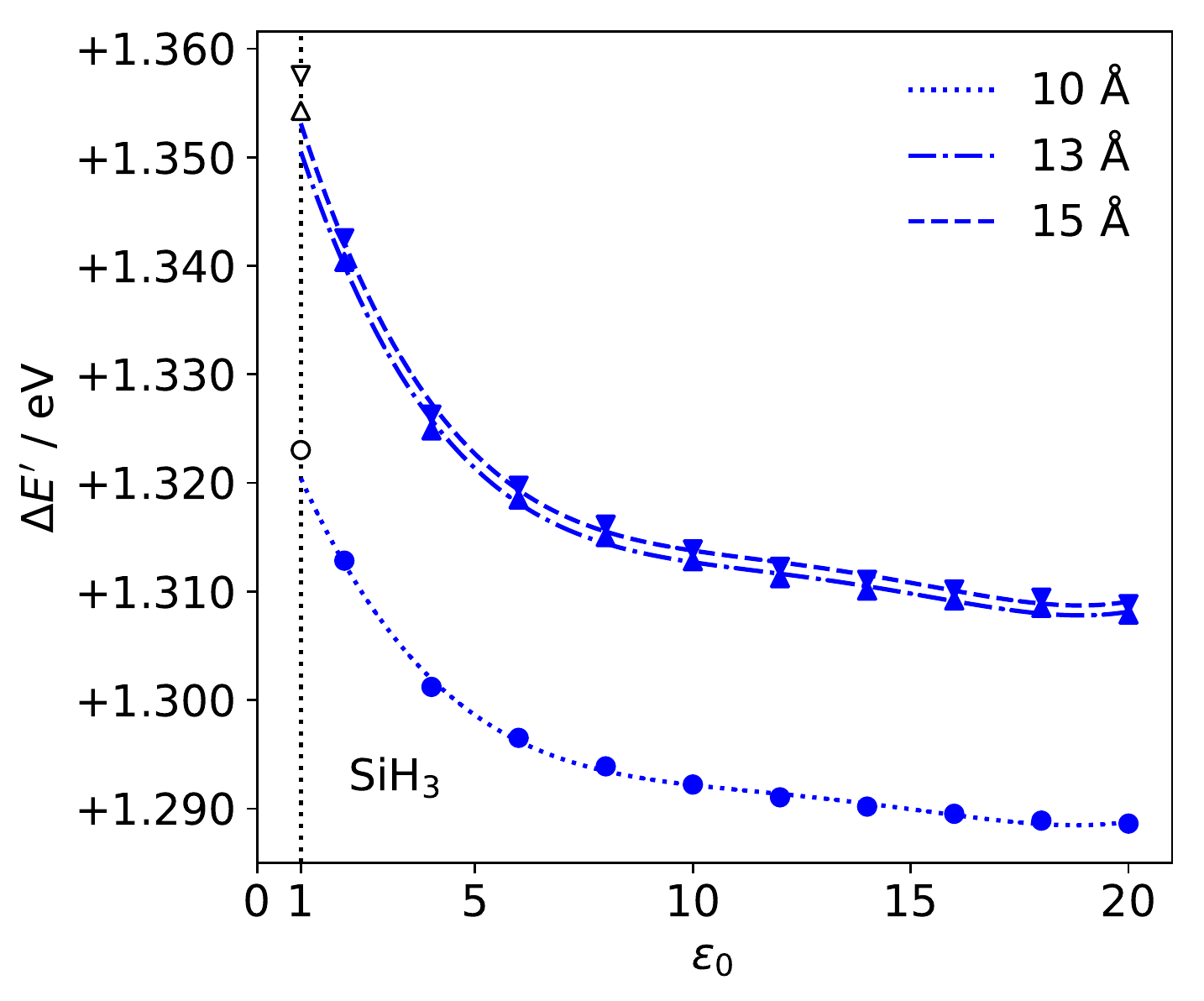} \\
\includegraphics[width=0.26\columnwidth]{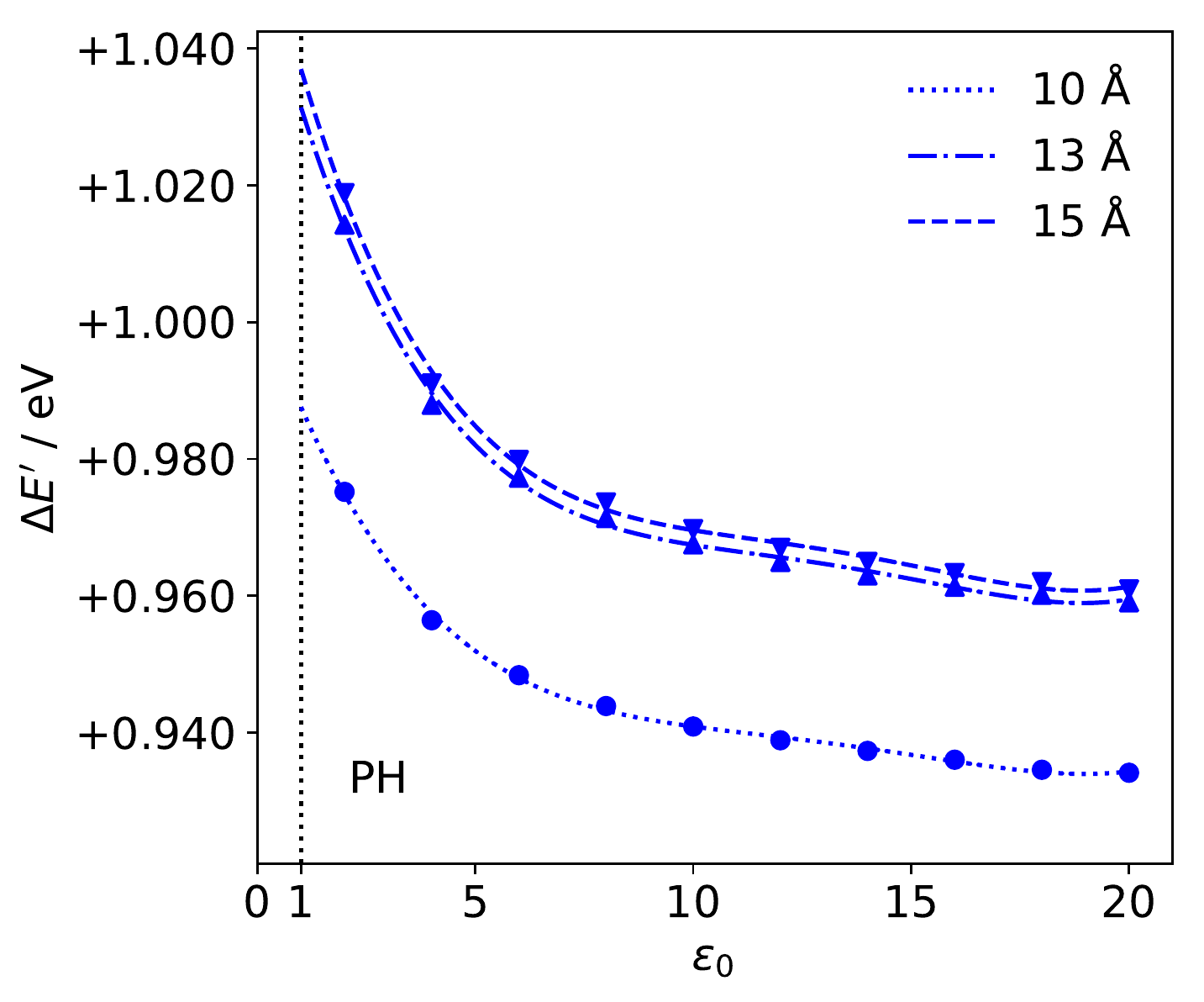}\includegraphics[width=0.26\columnwidth]{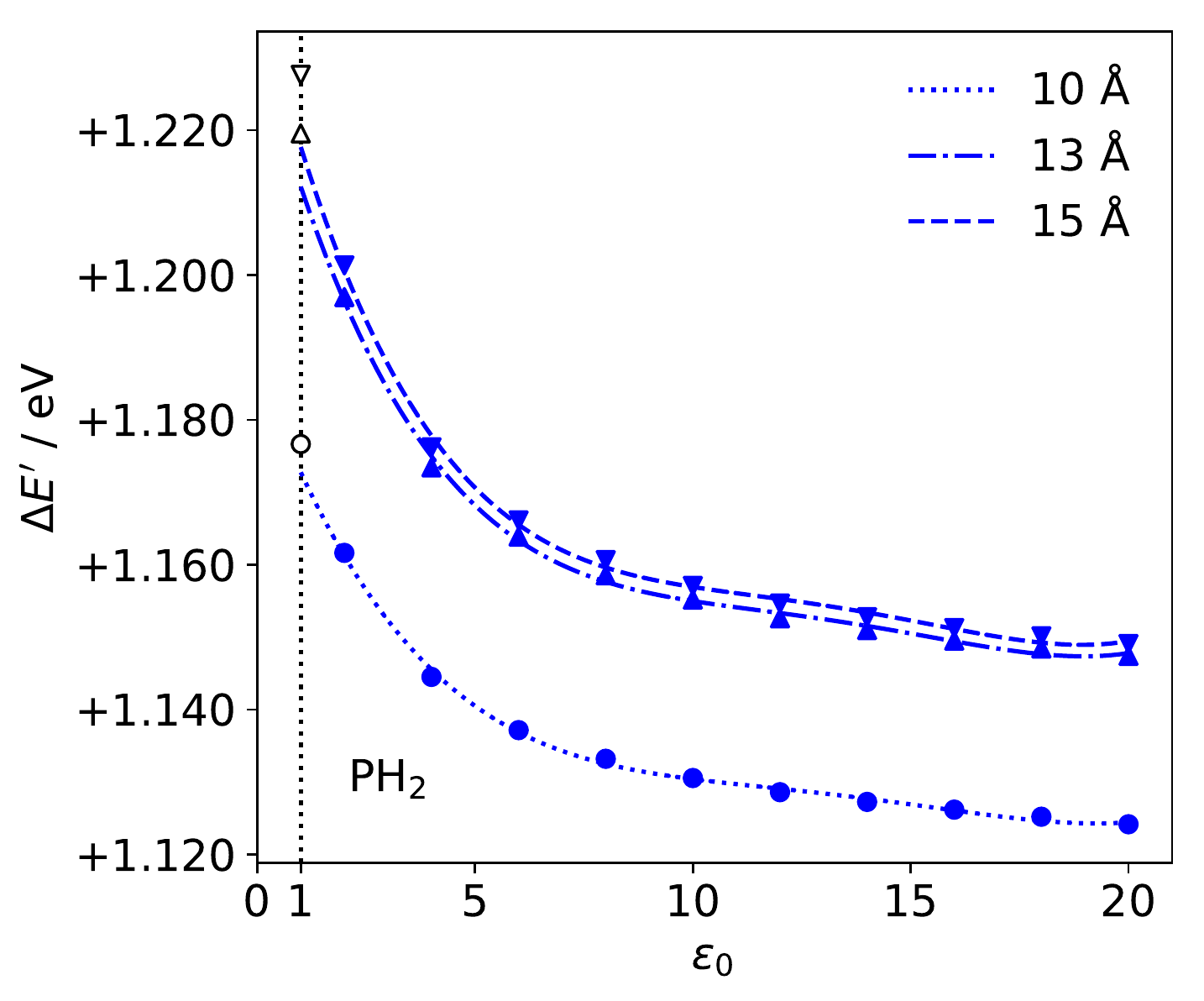}\includegraphics[width=0.26\columnwidth]{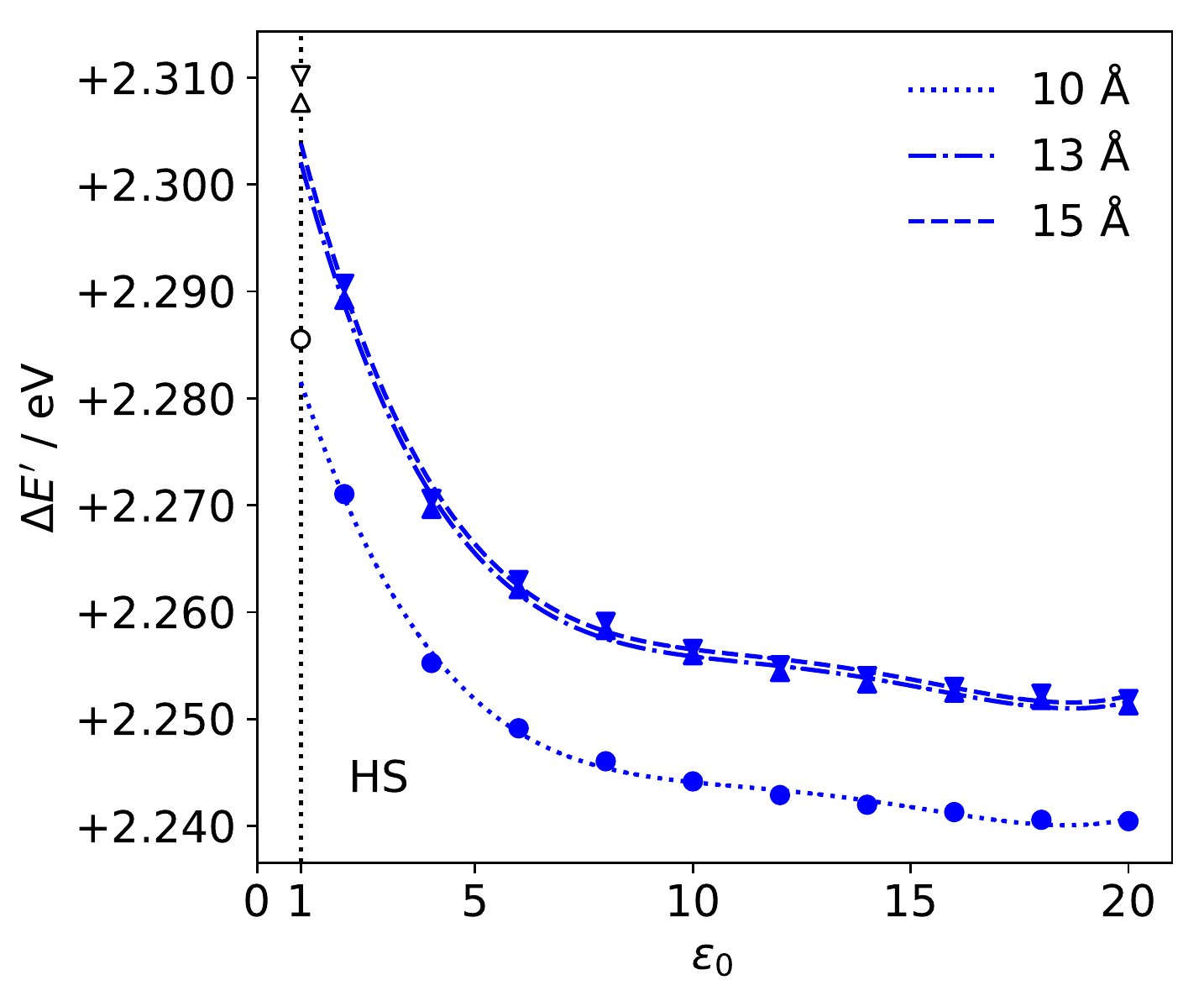} \\
\includegraphics[width=0.26\columnwidth]{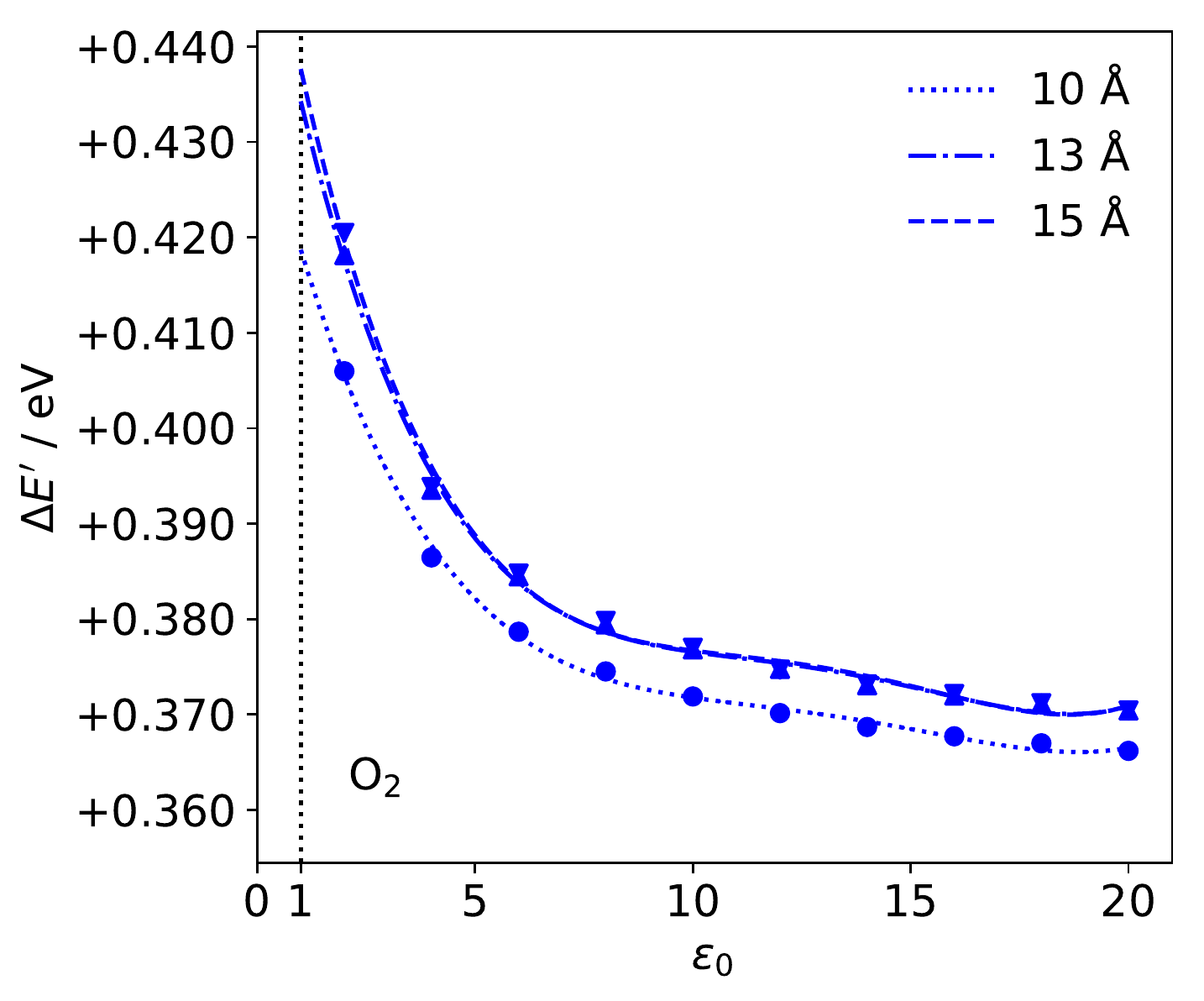}\includegraphics[width=0.26\columnwidth]{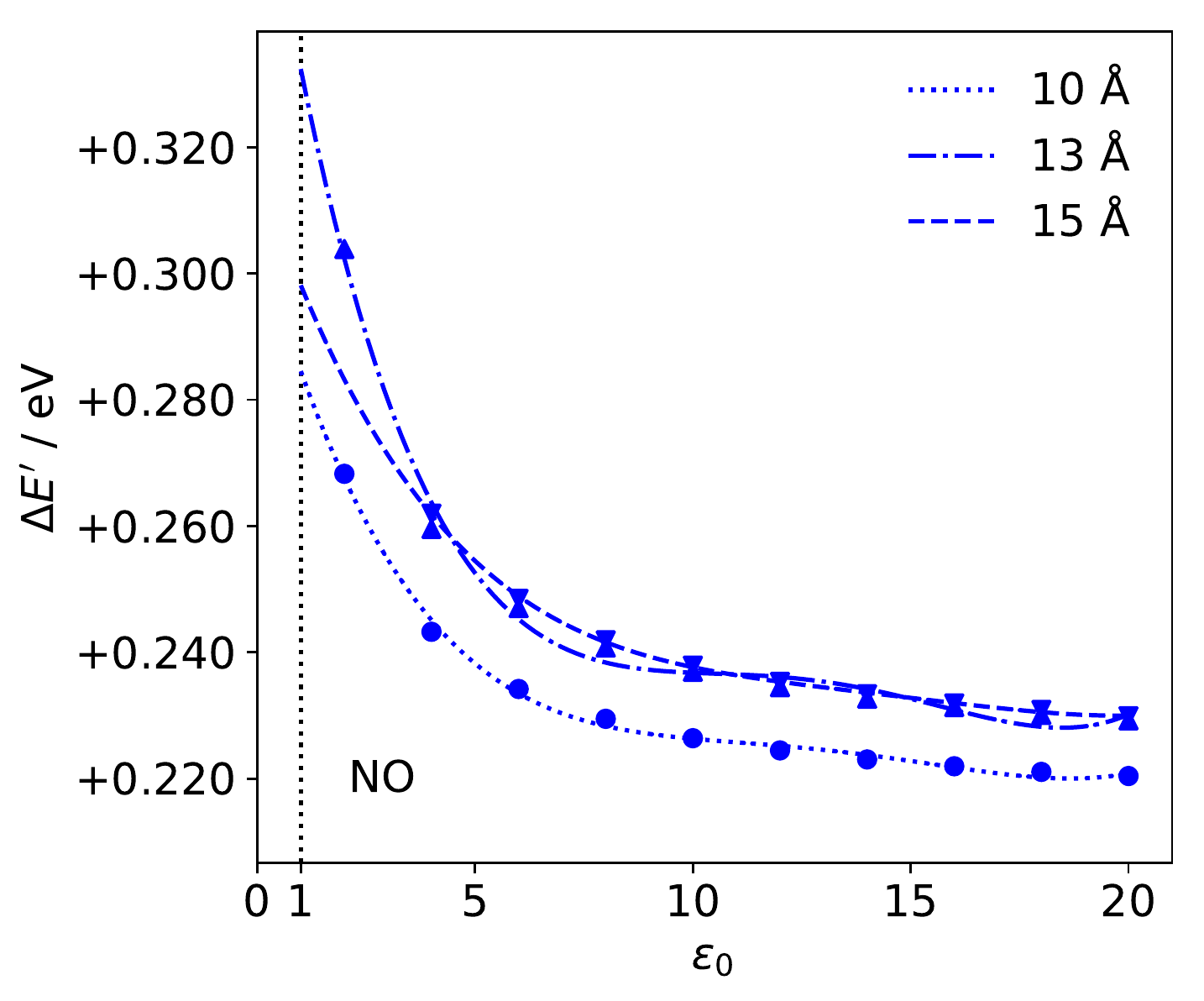}\includegraphics[width=0.26\columnwidth]{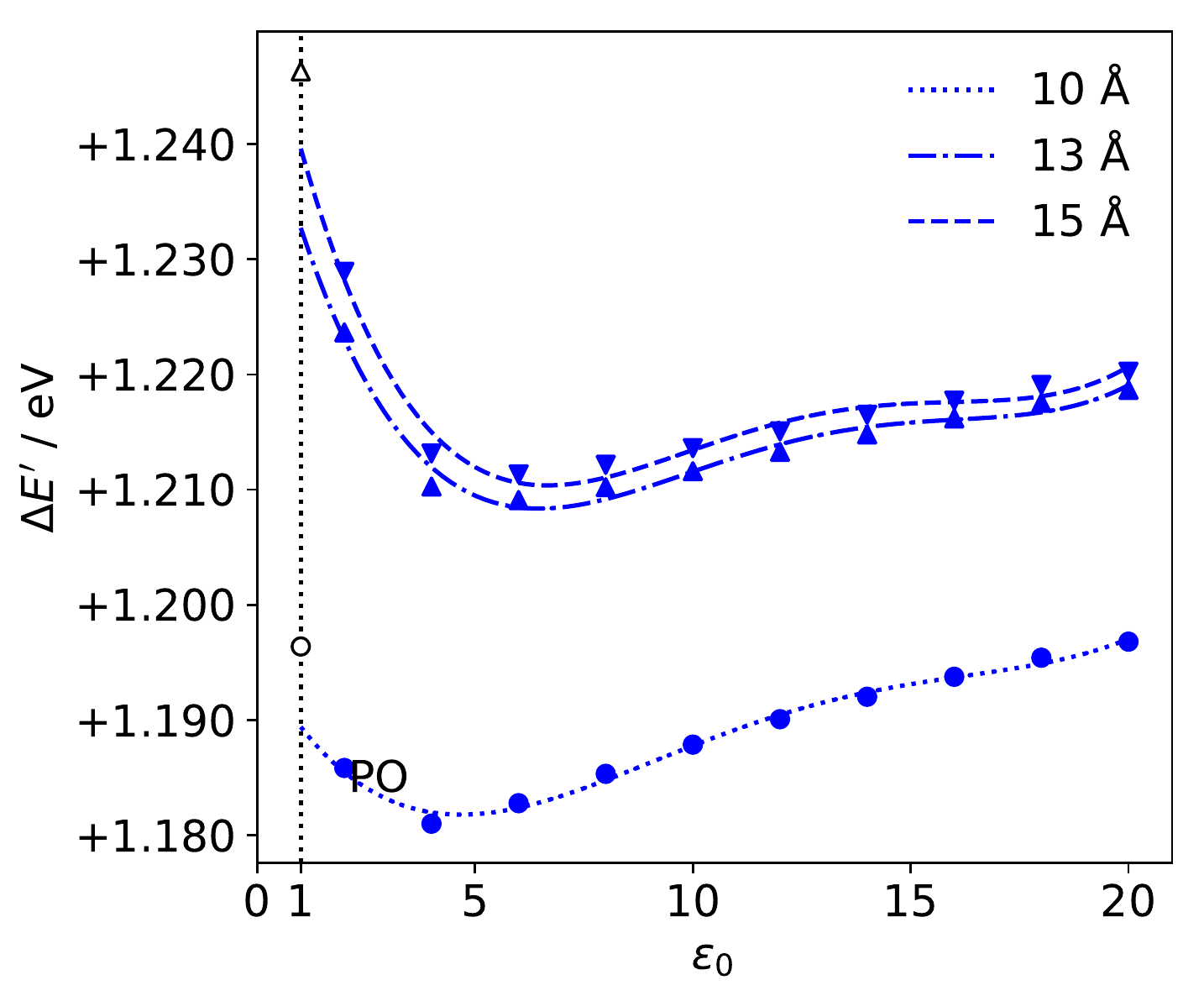} \\
\includegraphics[width=0.26\columnwidth]{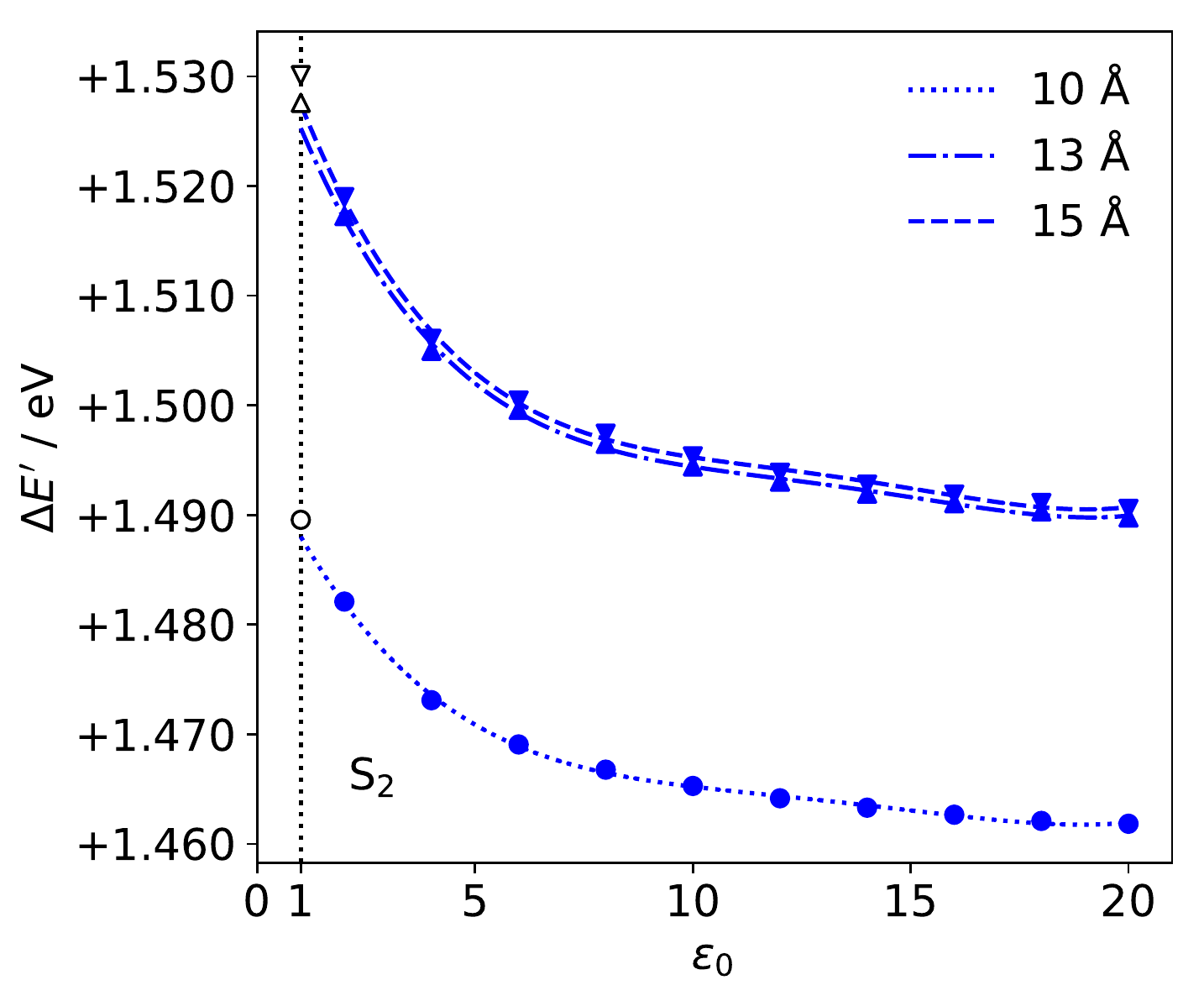}\includegraphics[width=0.26\columnwidth]{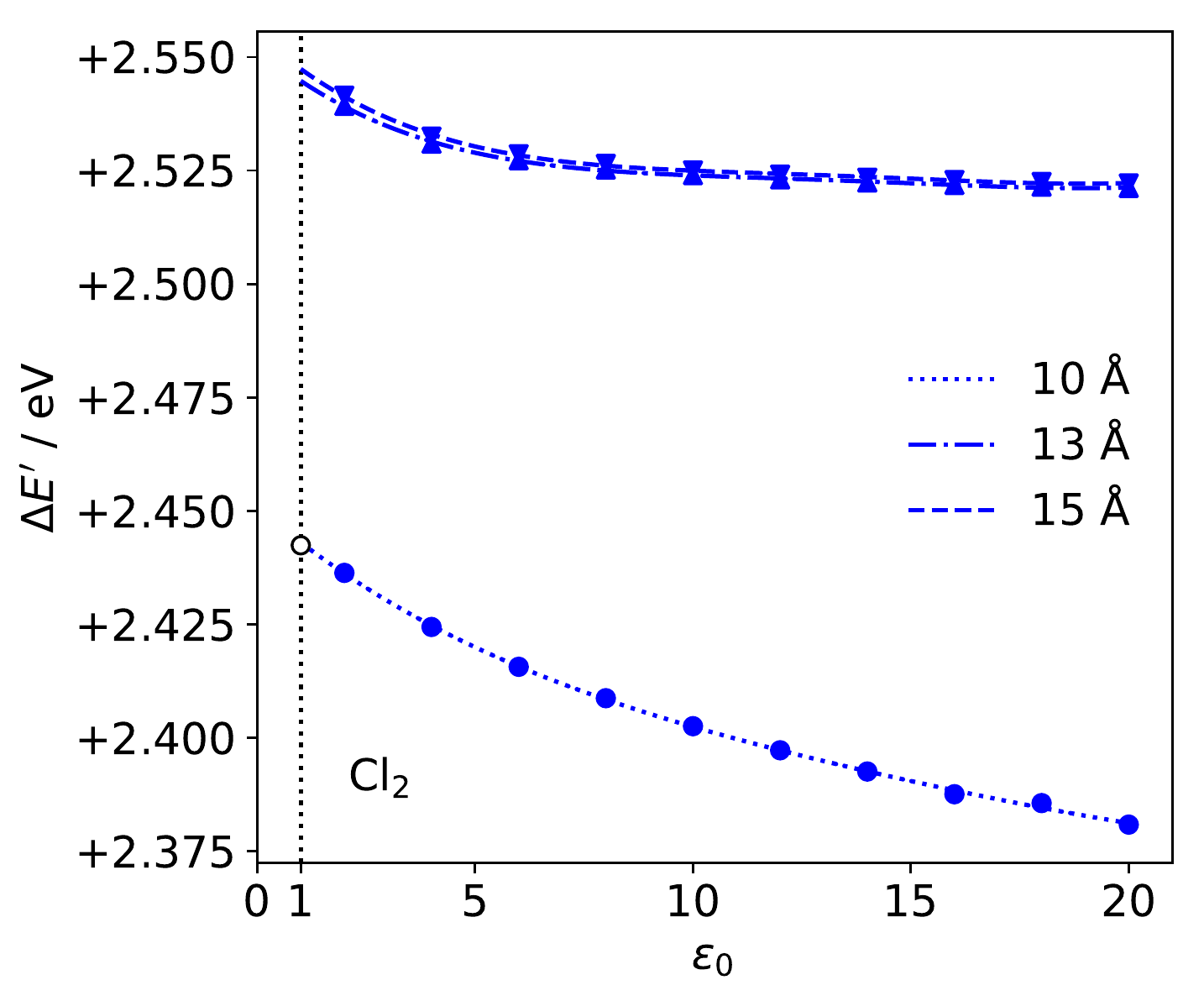}\includegraphics[width=0.26\columnwidth]{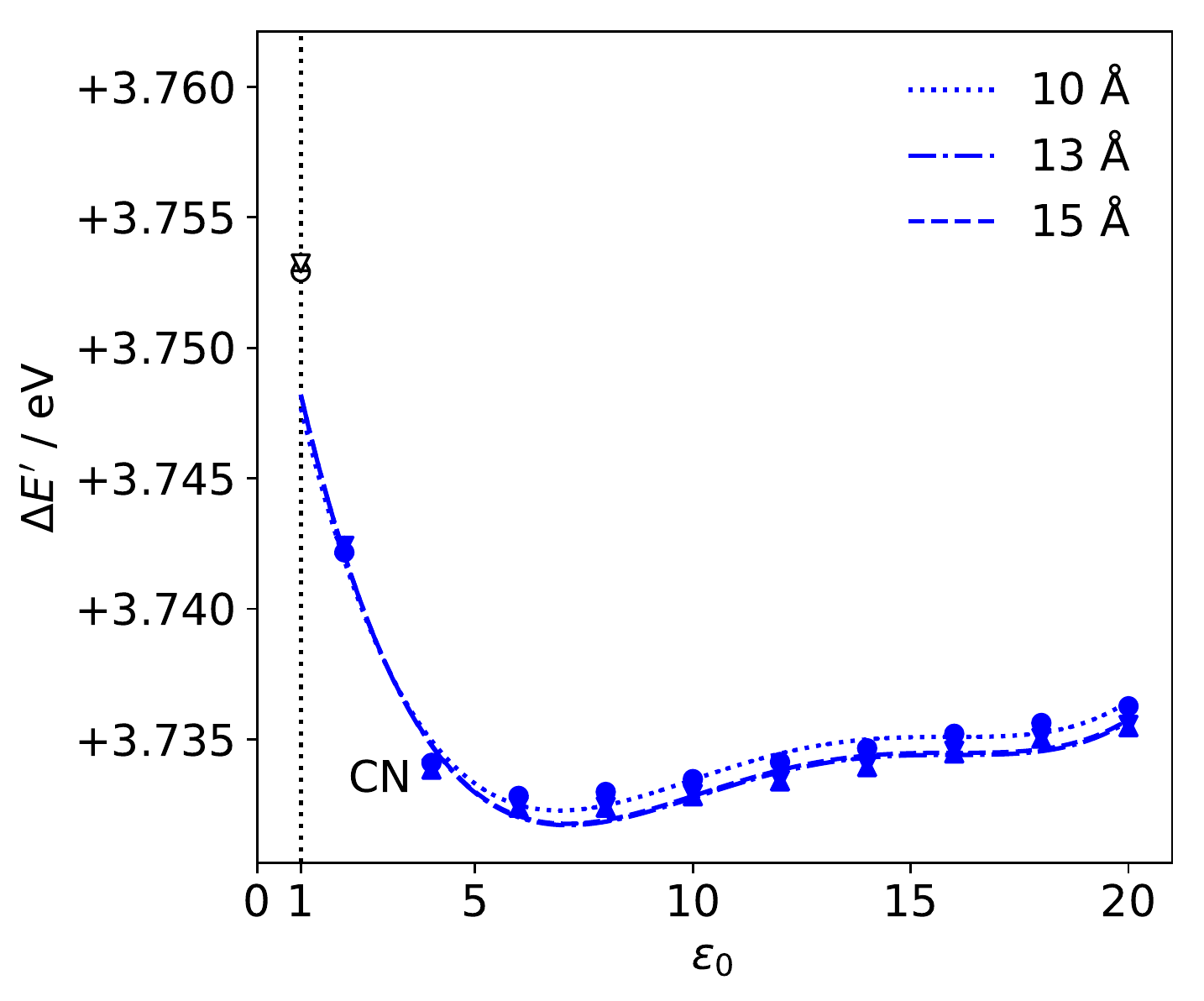} \\
\end{centering}
\caption{$\Delta E'$ extrapolation for the species of the G2-1 dataset, using the dielectric extrapolation method with the SCCS cavity. Lines represent 4th-degree polynomial functions.}
\label{fig:extrapolation-eps-sccs}
\end{figure}

\begin{figure}
\begin{centering}
\includegraphics[width=0.26\columnwidth]{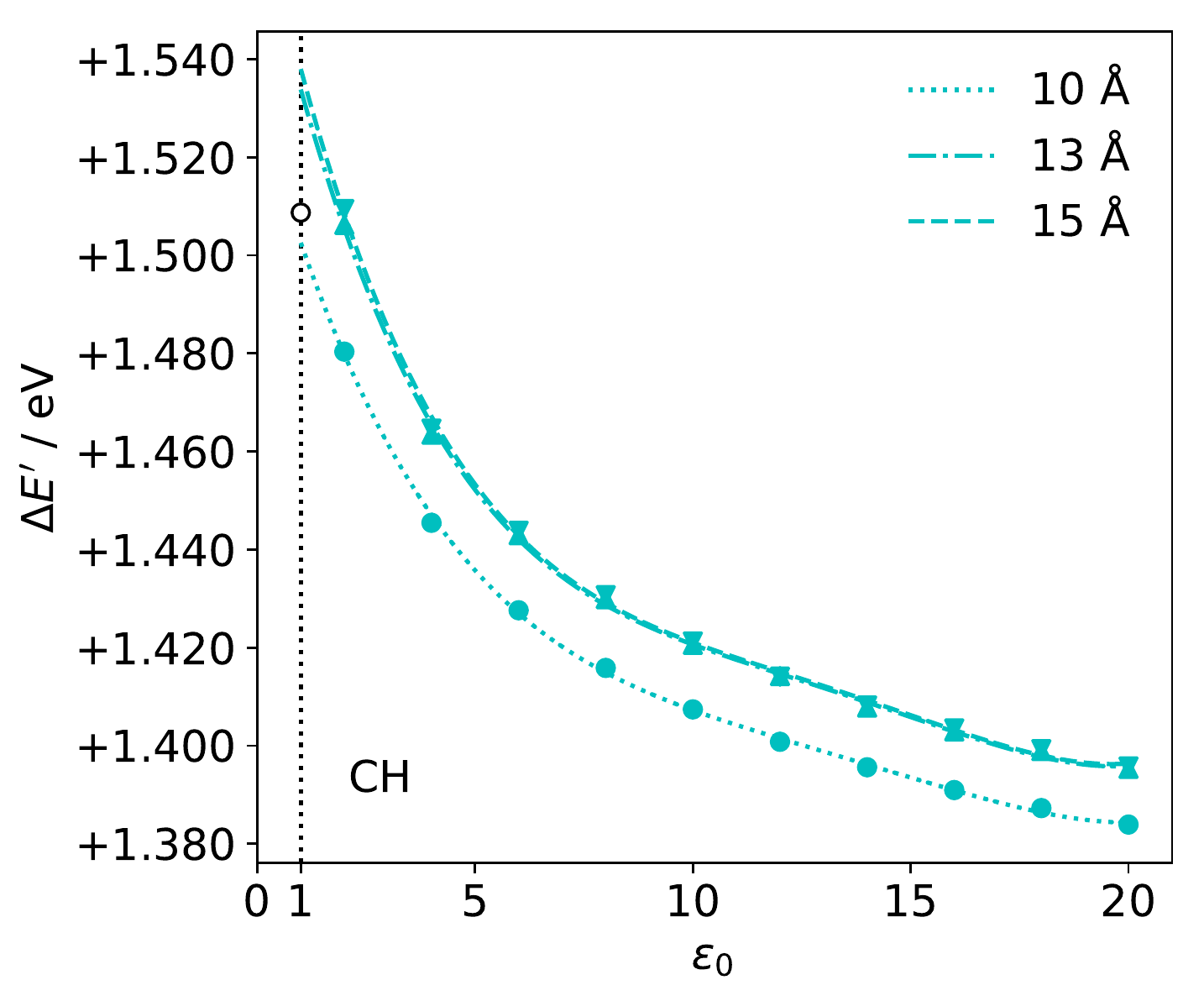}\includegraphics[width=0.26\columnwidth]{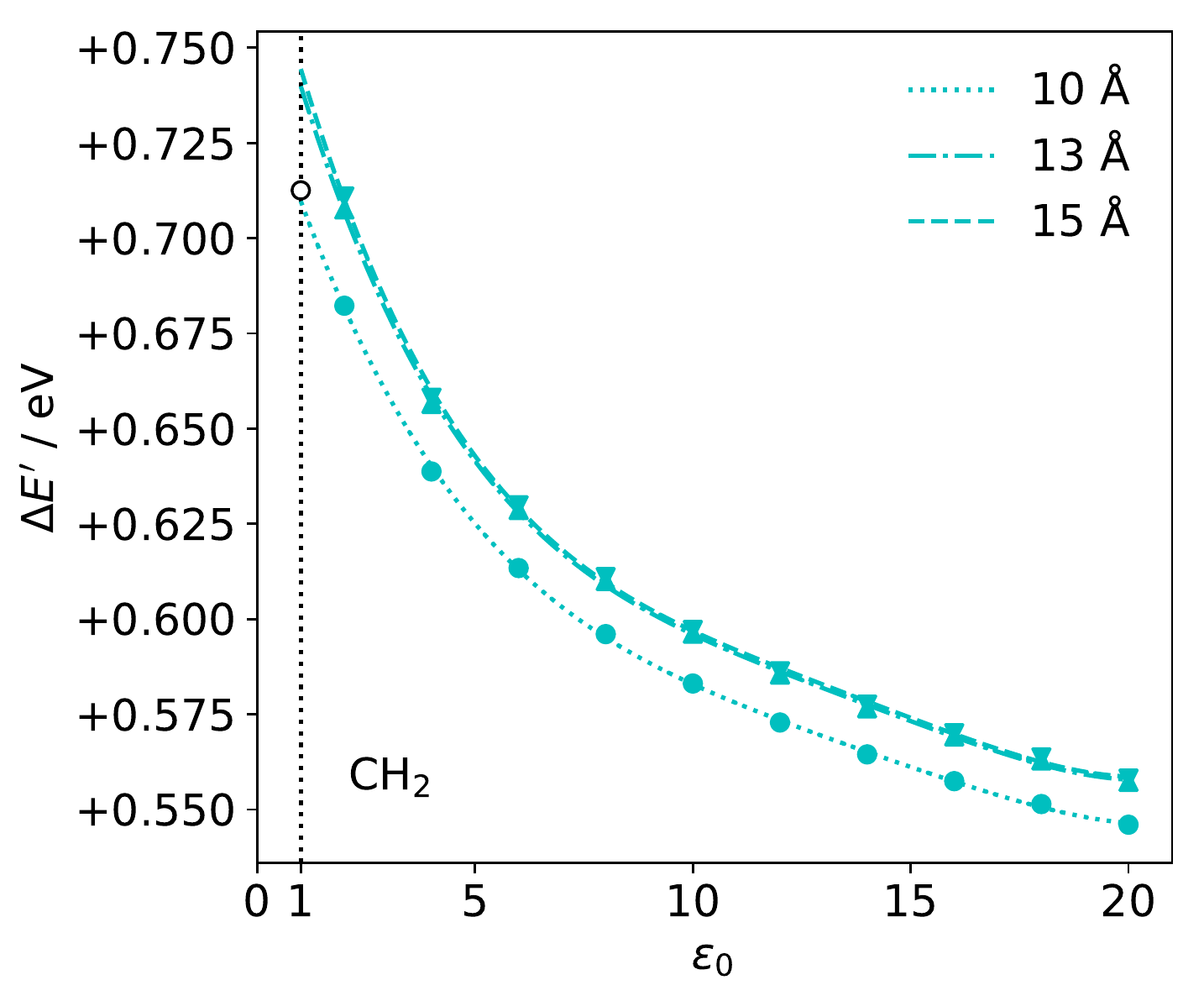}\includegraphics[width=0.26\columnwidth]{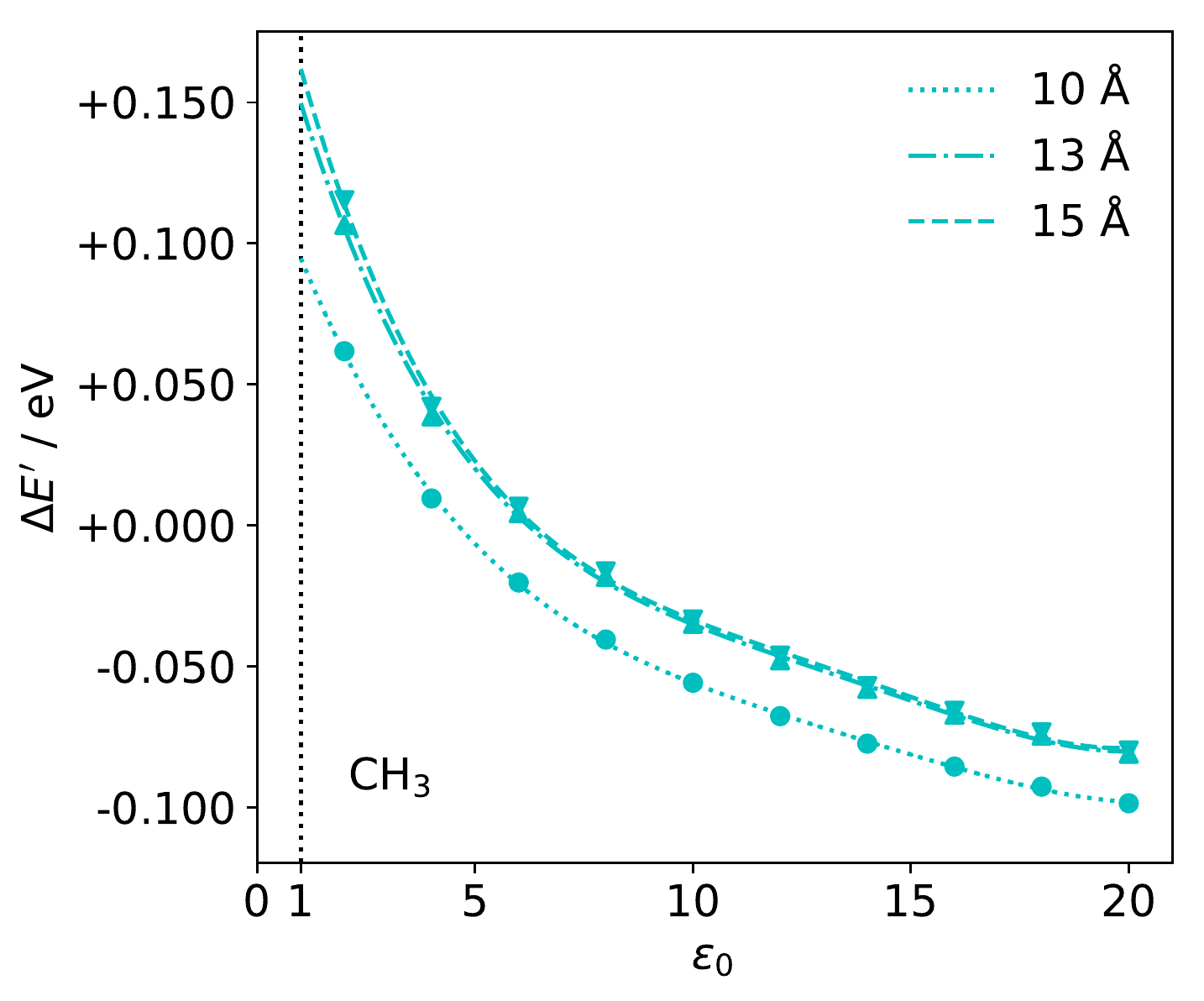} \\
\includegraphics[width=0.26\columnwidth]{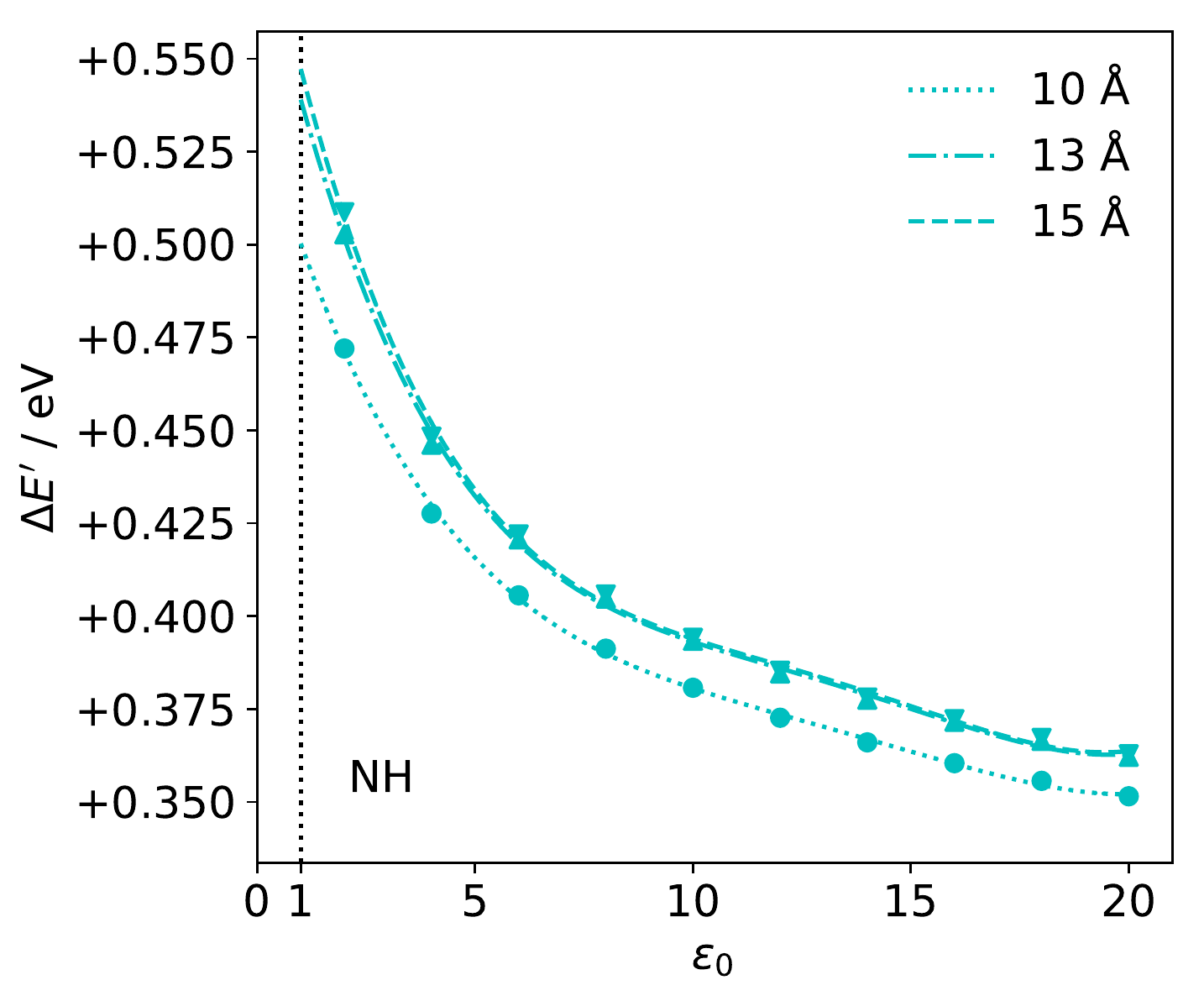}\includegraphics[width=0.26\columnwidth]{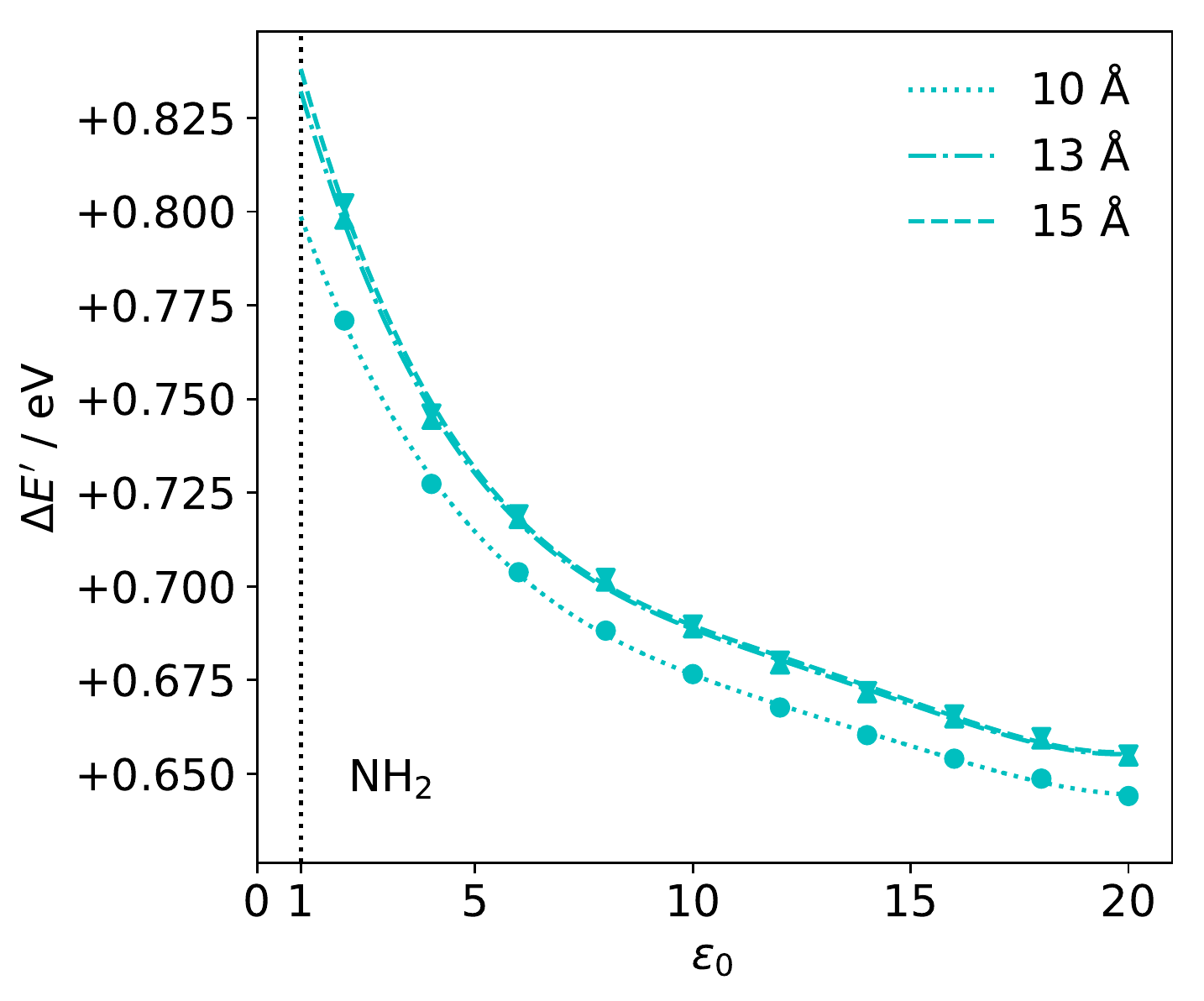}\includegraphics[width=0.26\columnwidth]{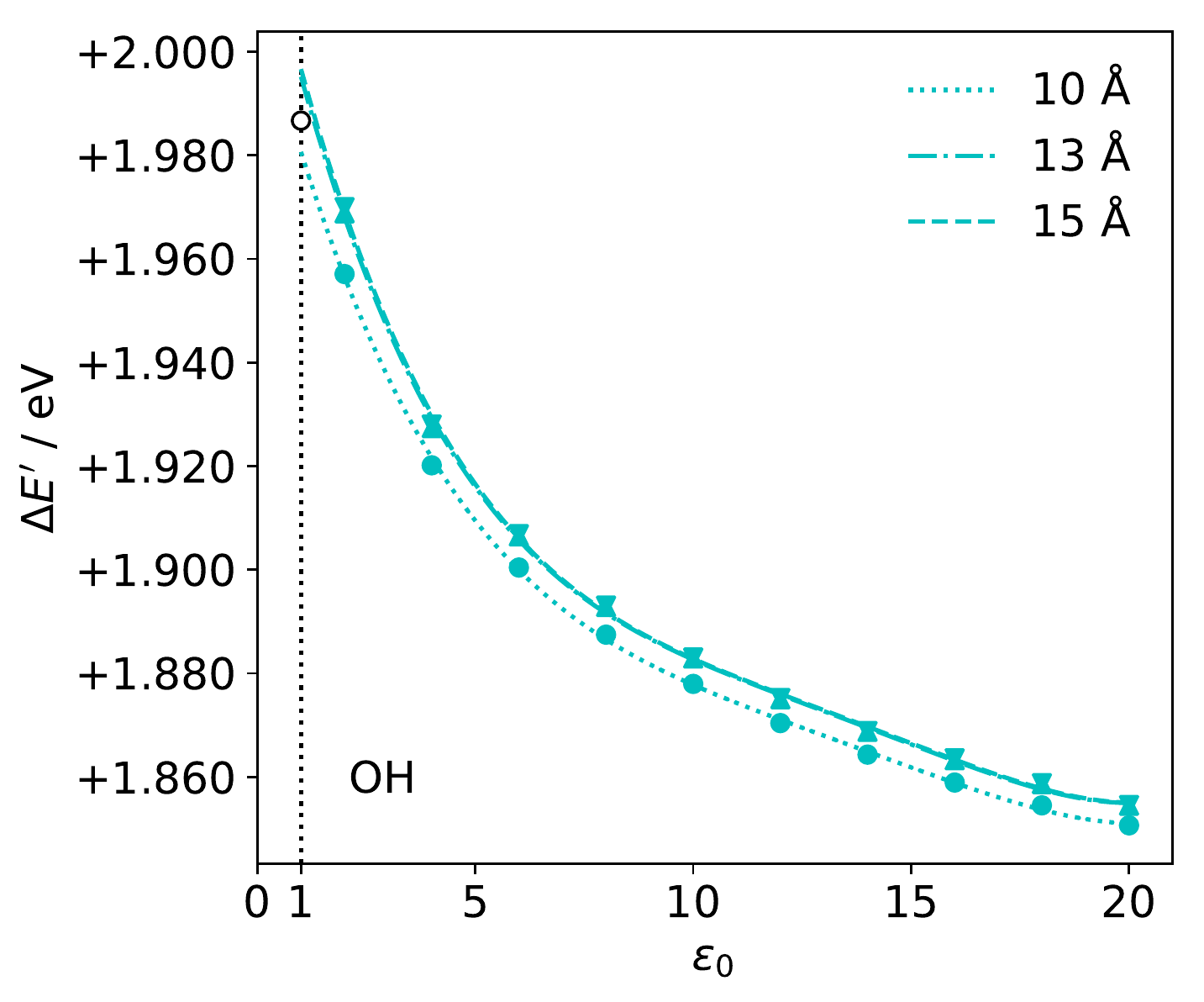} \\
\includegraphics[width=0.26\columnwidth]{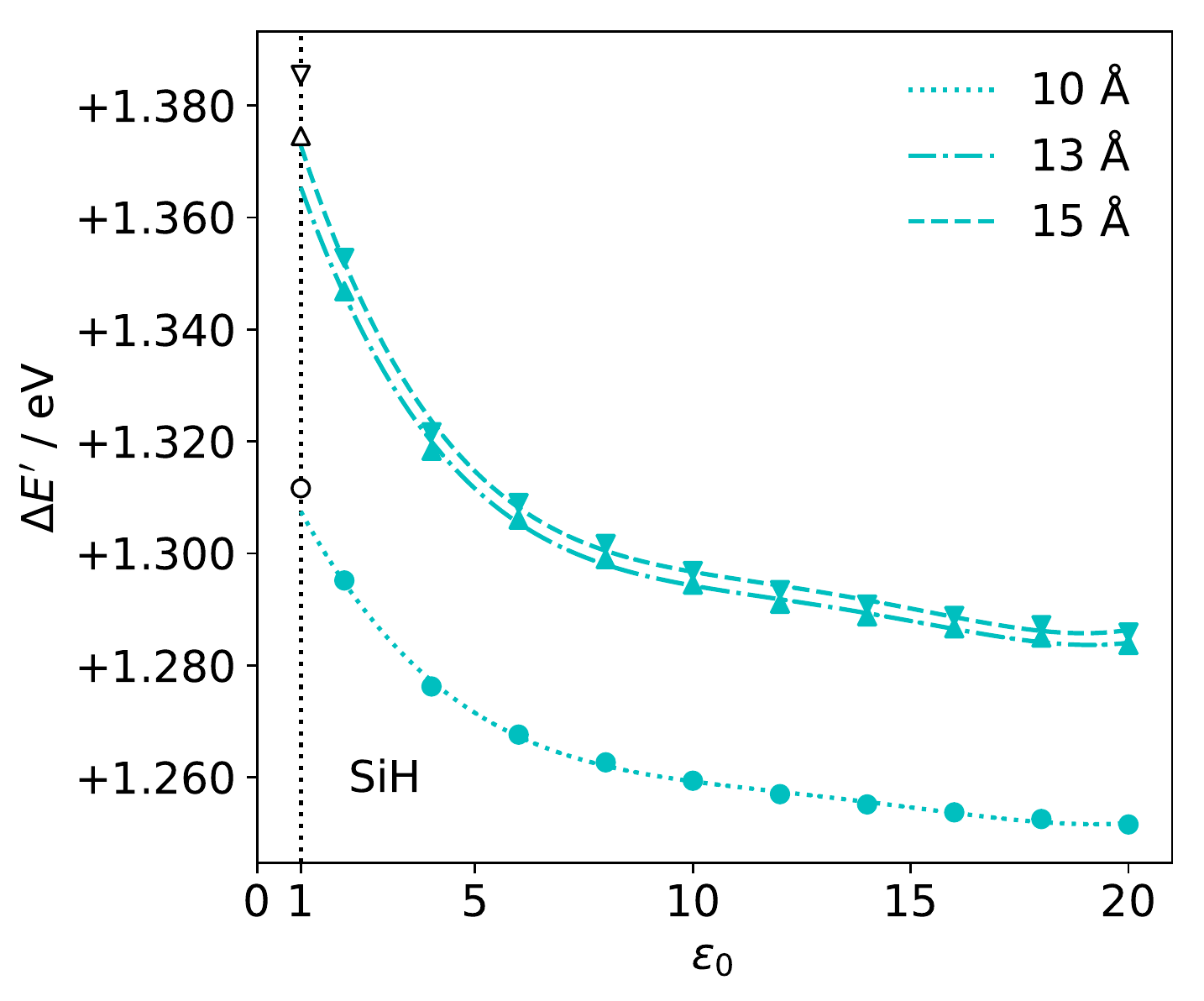}\includegraphics[width=0.26\columnwidth]{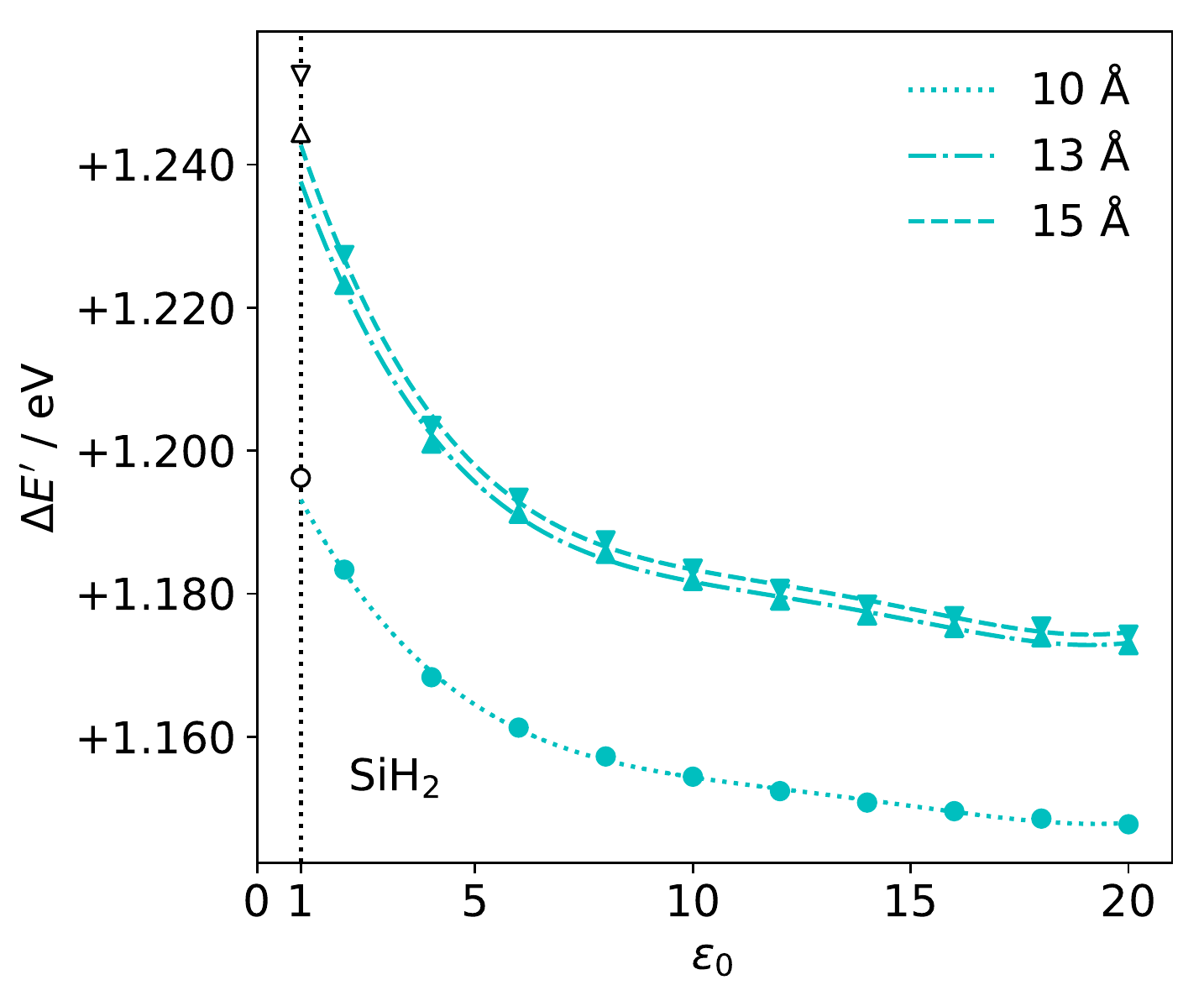}\includegraphics[width=0.26\columnwidth]{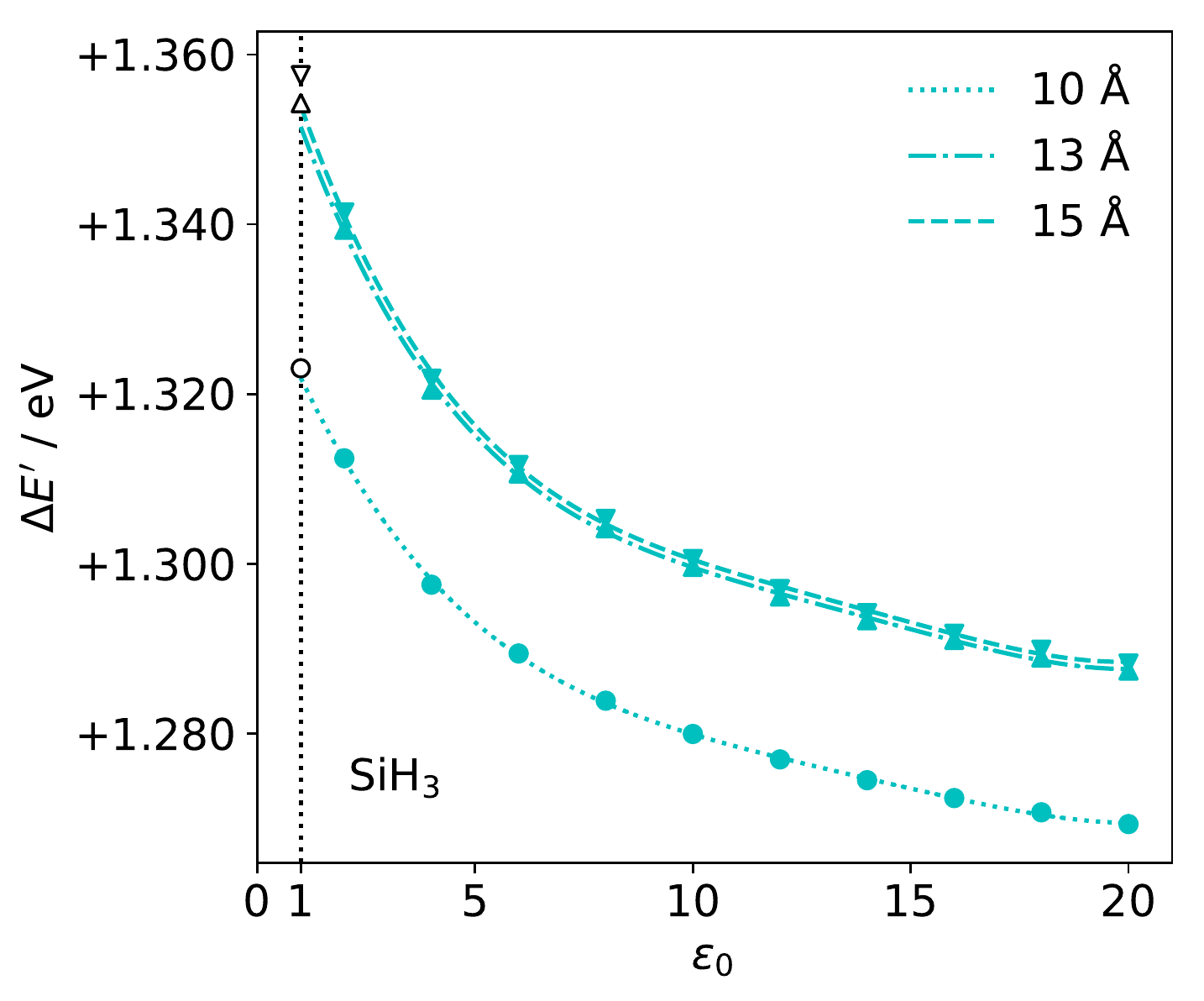} \\
\includegraphics[width=0.26\columnwidth]{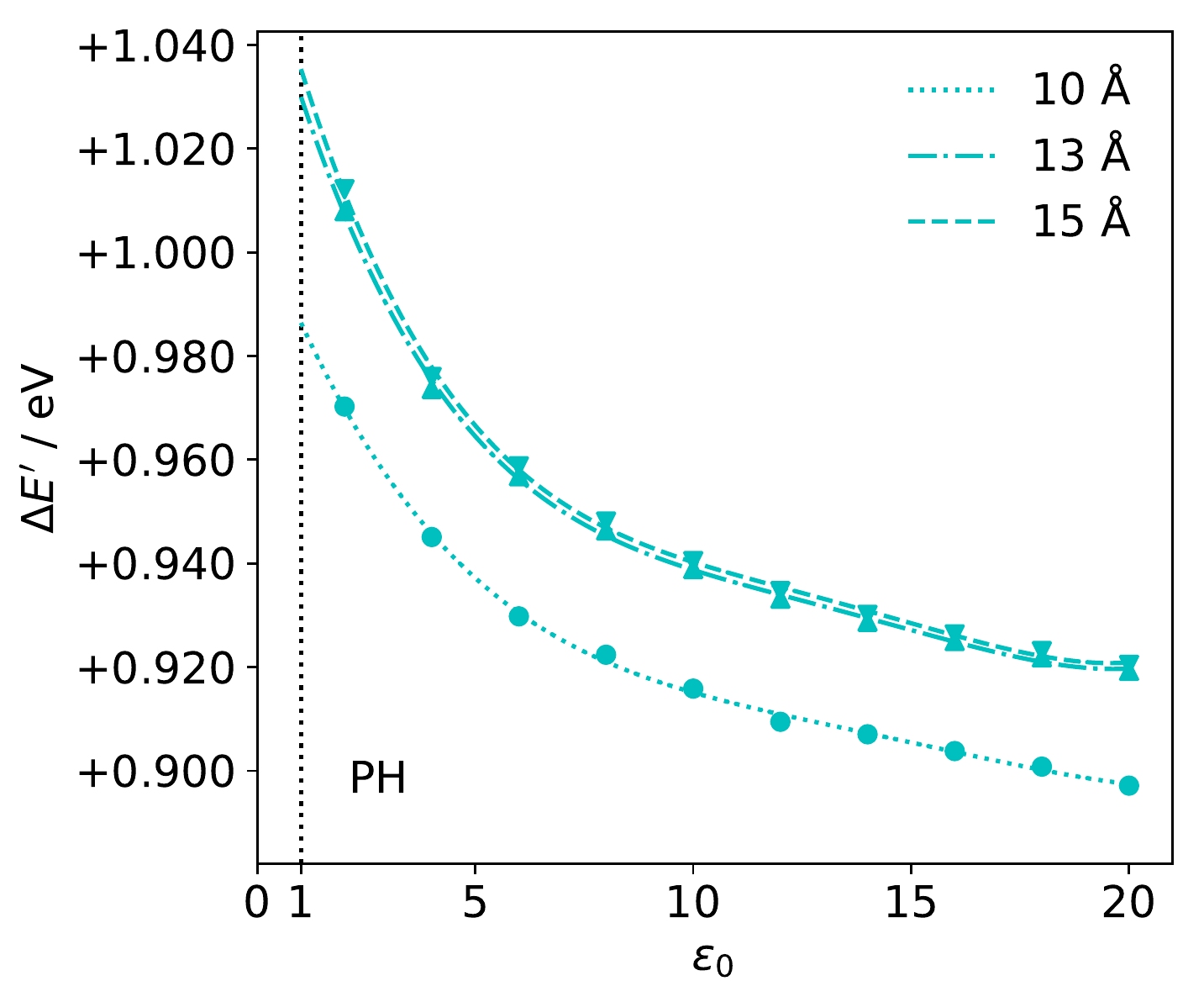}\includegraphics[width=0.26\columnwidth]{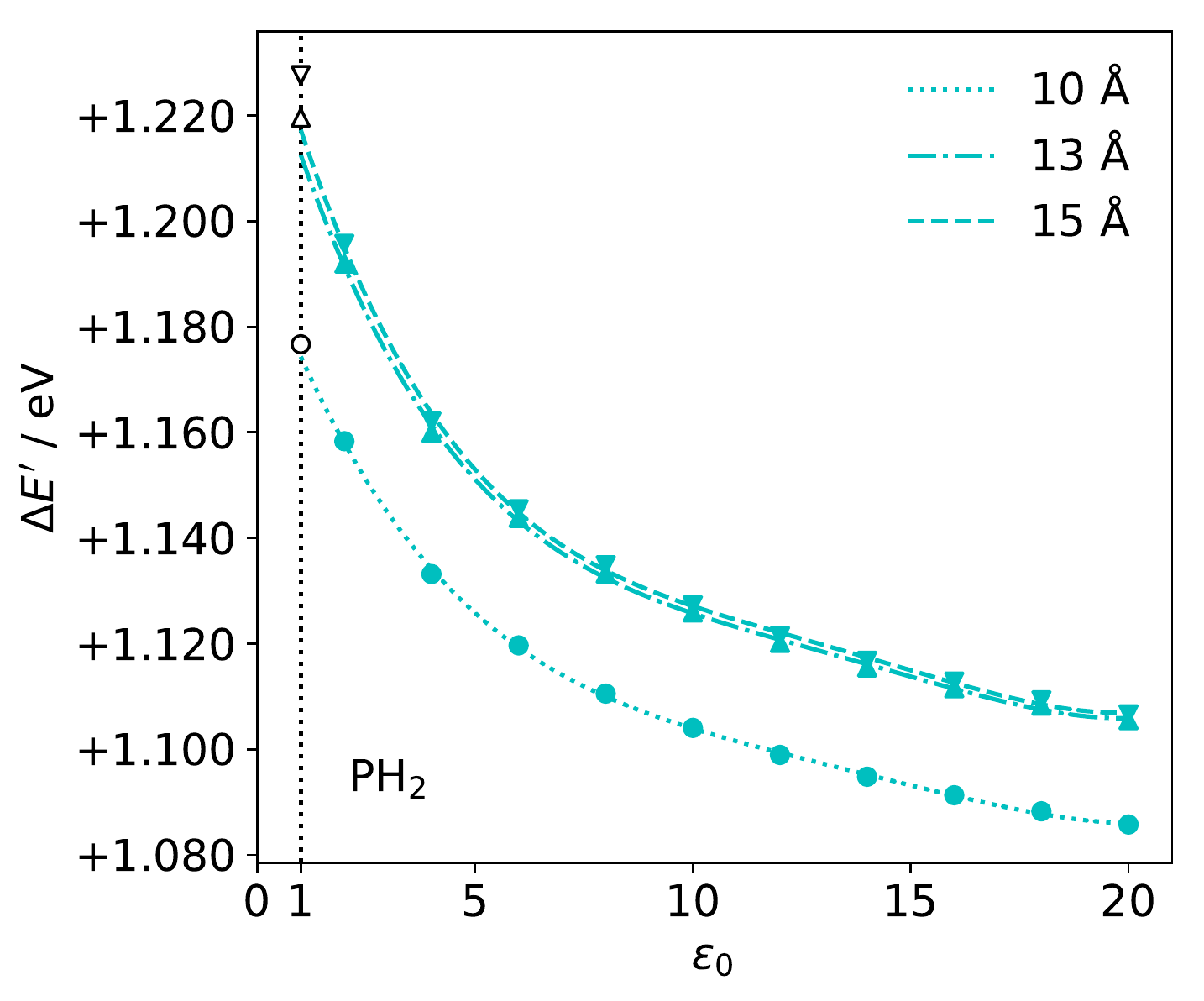}\includegraphics[width=0.26\columnwidth]{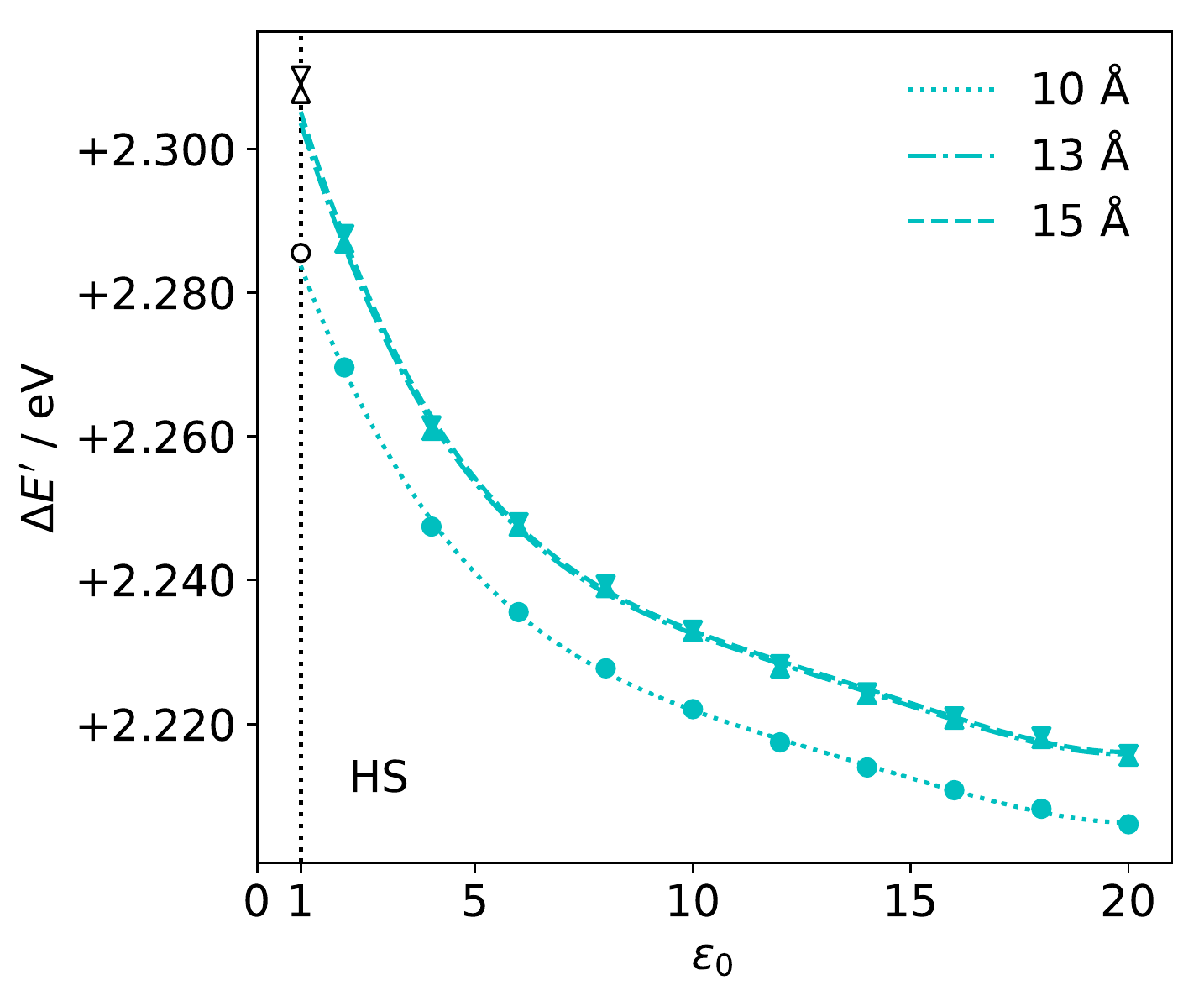} \\
\includegraphics[width=0.26\columnwidth]{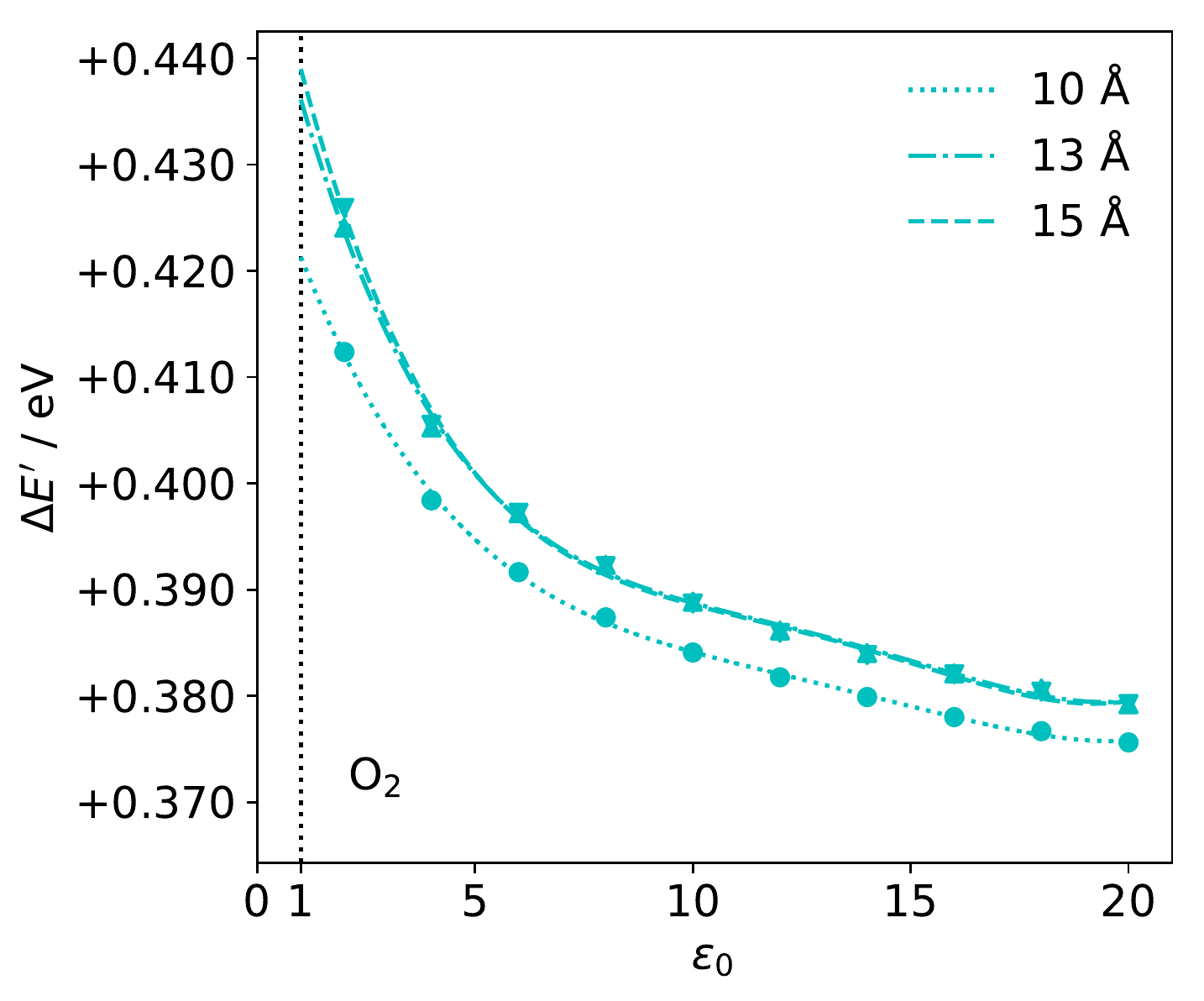}\includegraphics[width=0.26\columnwidth]{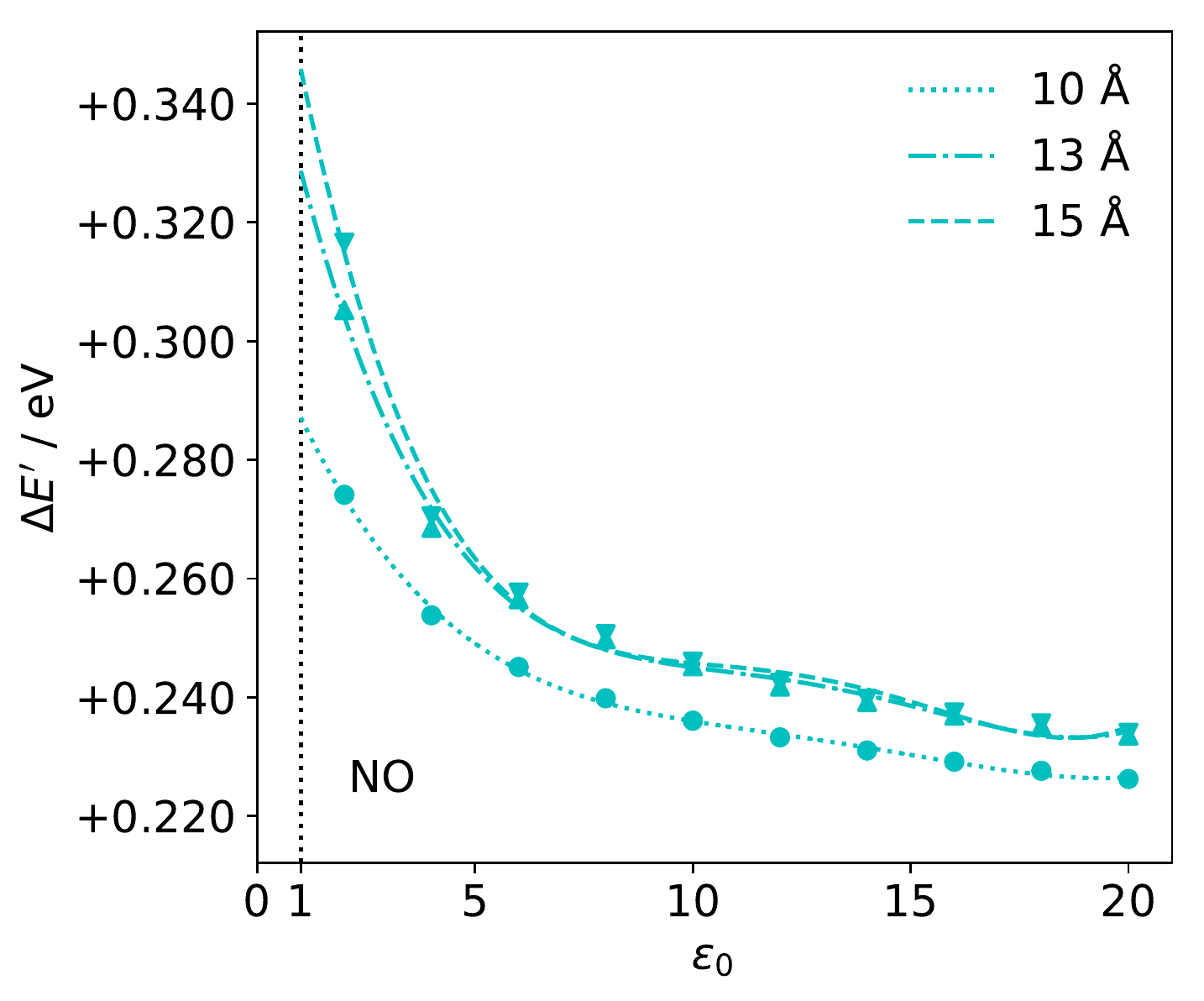}\includegraphics[width=0.26\columnwidth]{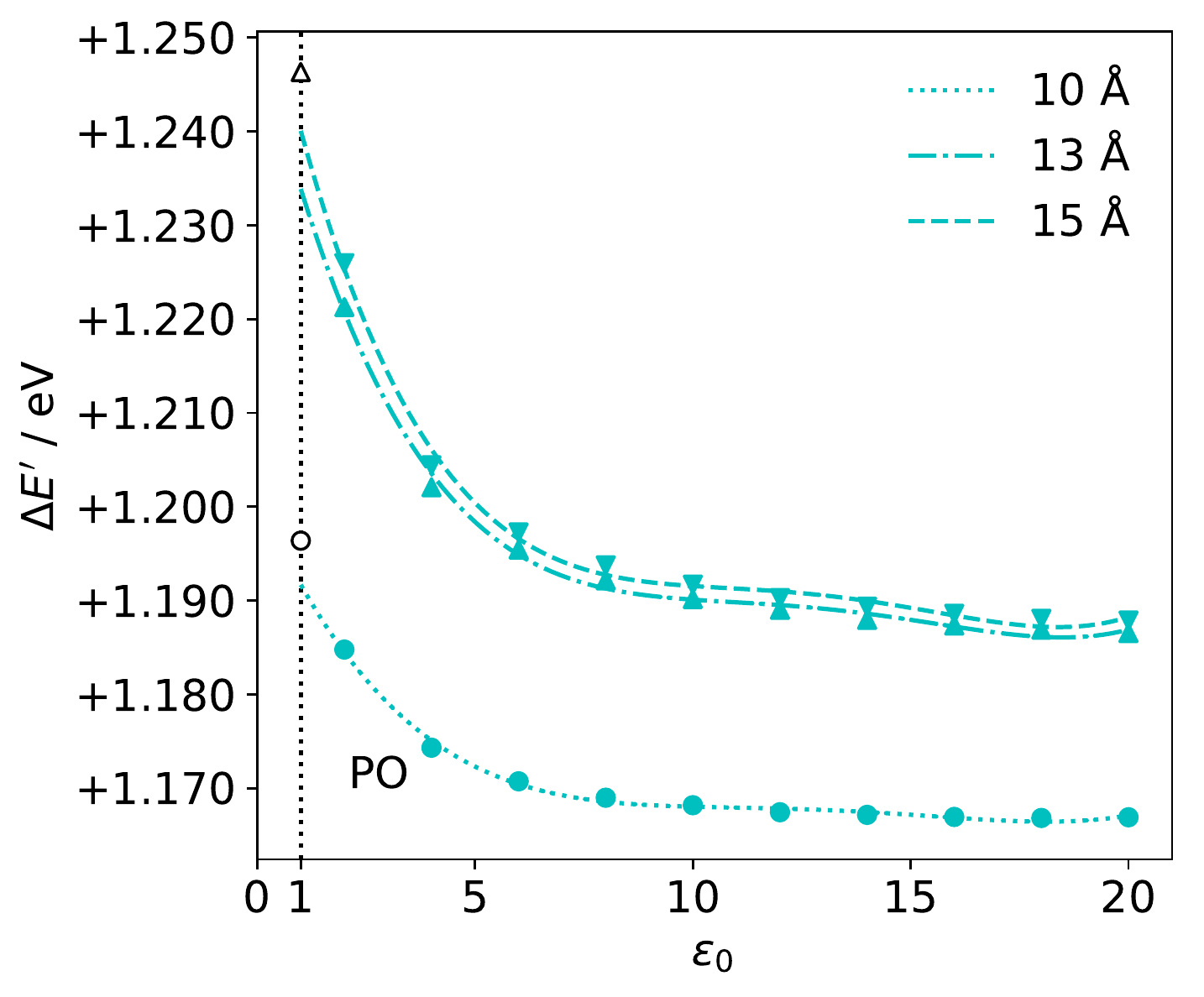} \\
\includegraphics[width=0.26\columnwidth]{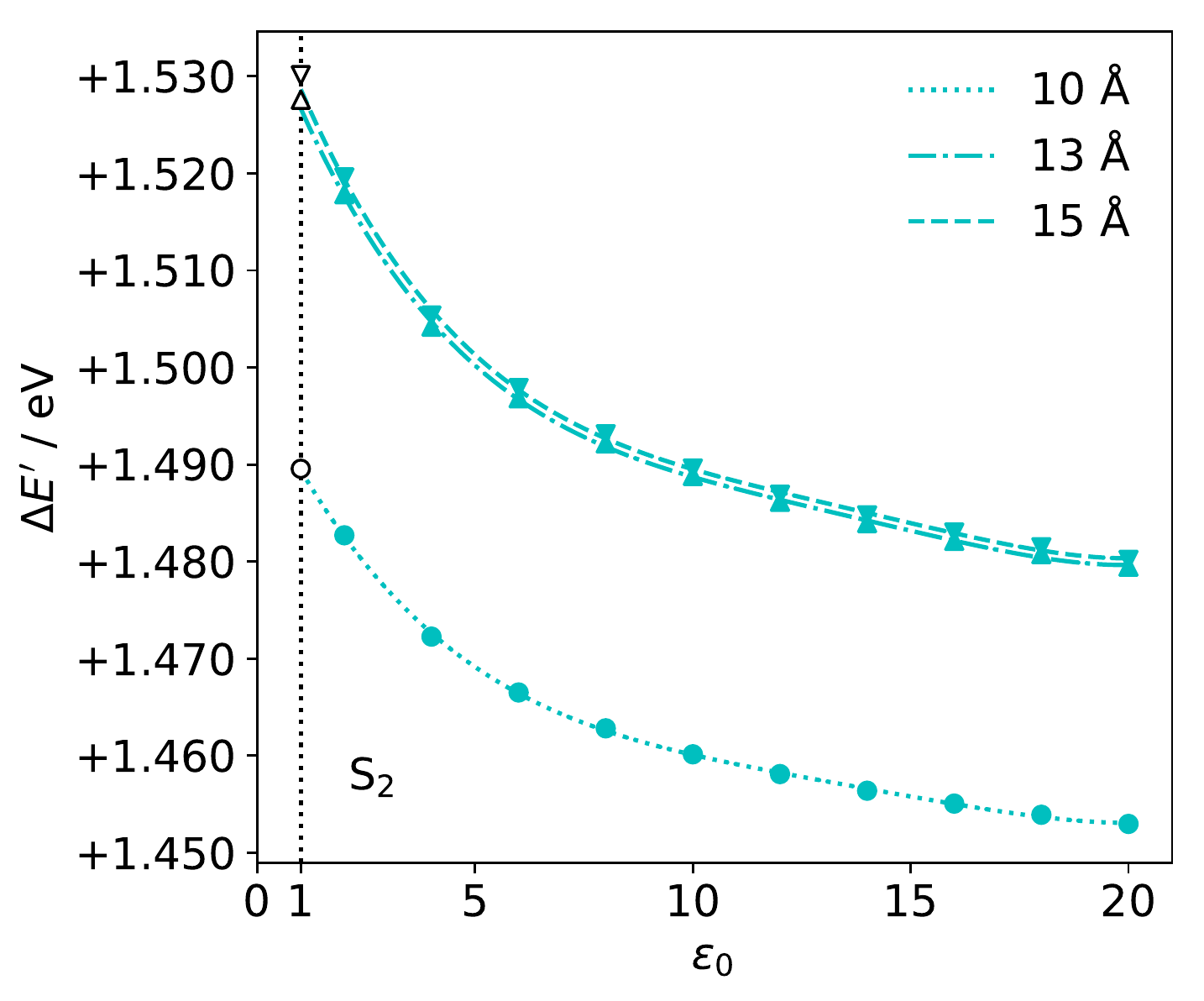}\includegraphics[width=0.26\columnwidth]{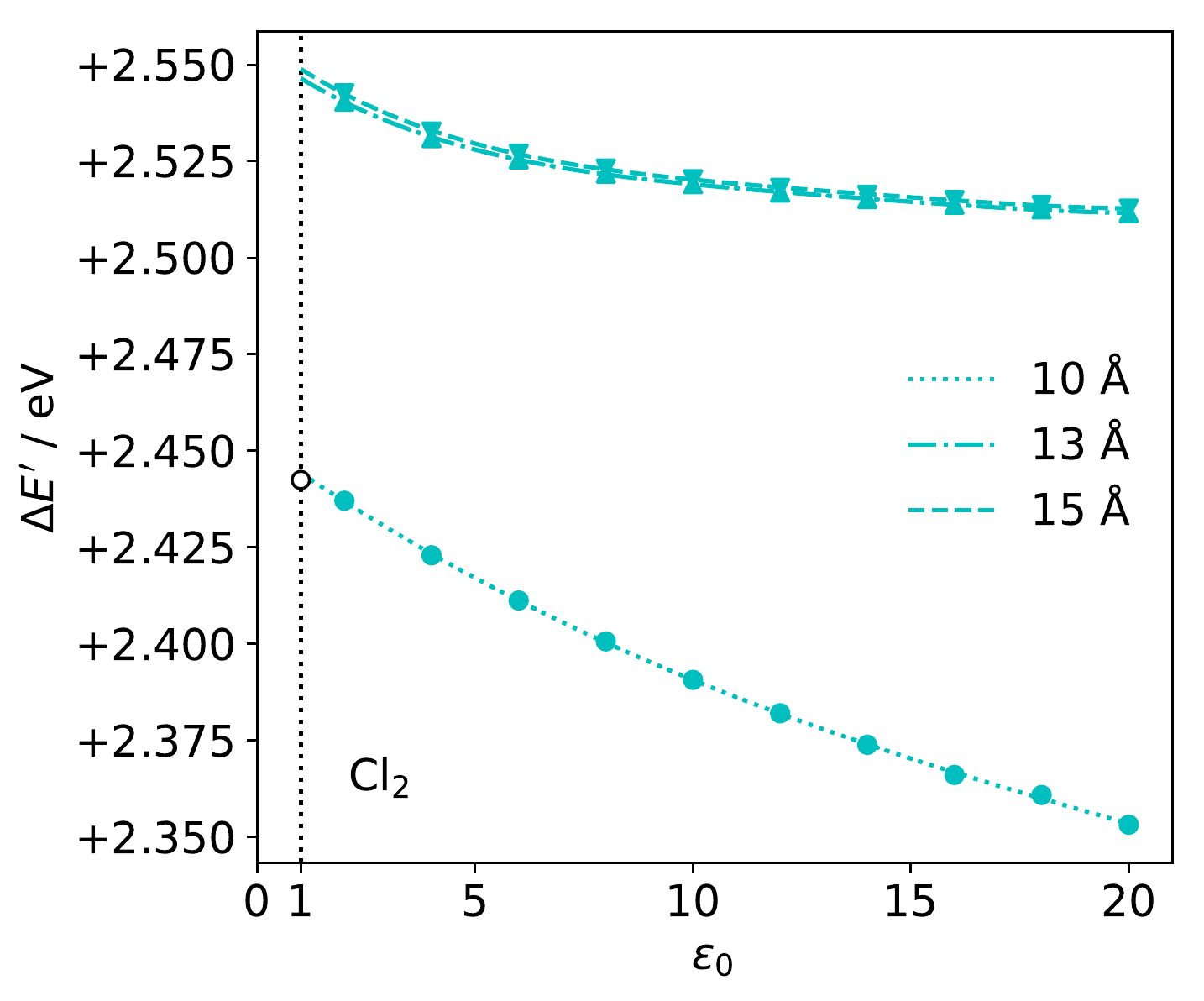}\includegraphics[width=0.26\columnwidth]{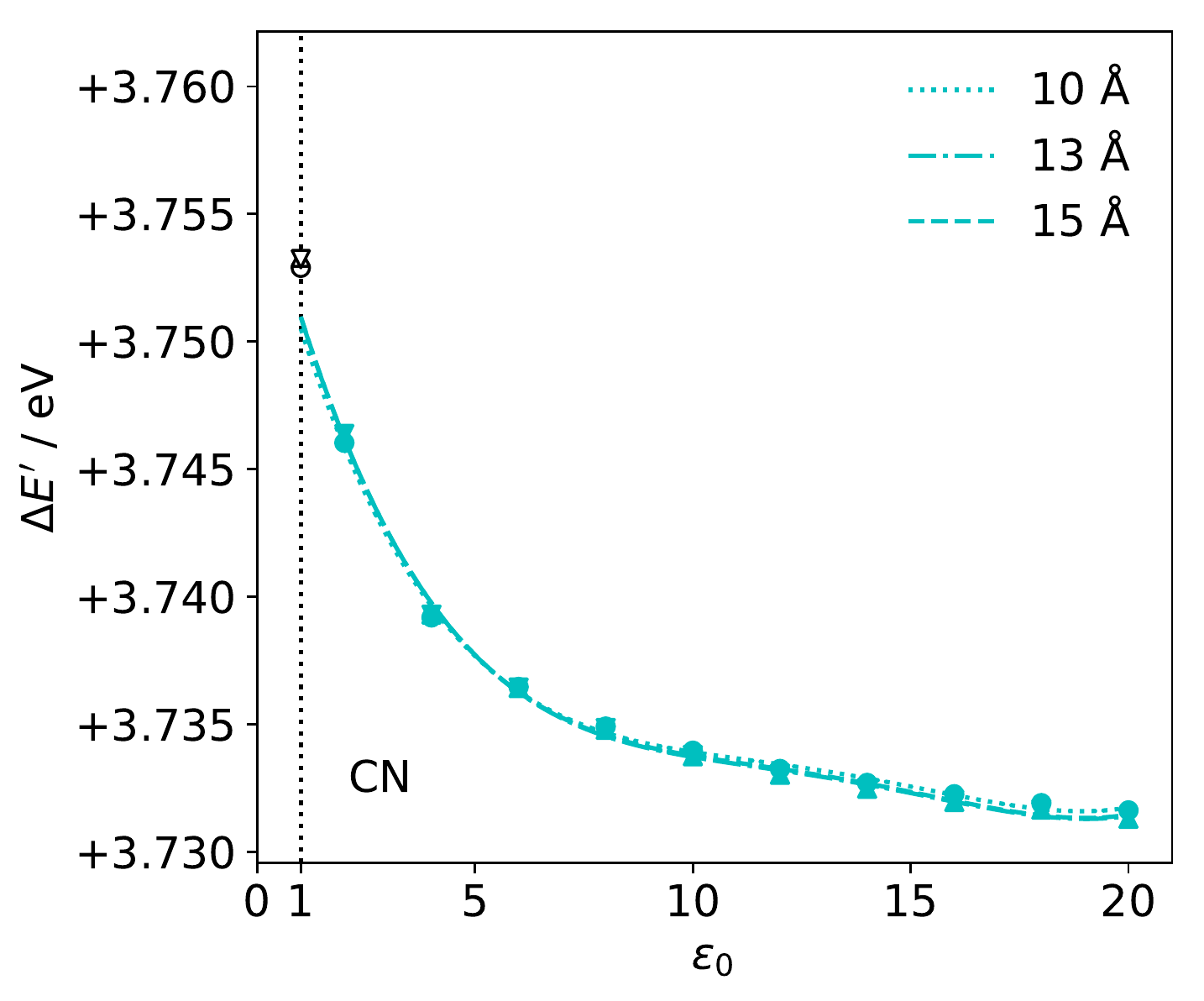} \\
\end{centering}
\caption{$\Delta E'$ extrapolation for the species of the G2-1 dataset, using the dielectric extrapolation method with the SSCS cavity. Lines represent 4th-degree polynomial functions.}
\label{fig:extrapolation-eps-sscs}
\end{figure}

\begin{figure}
\begin{centering}
\includegraphics[width=0.26\columnwidth]{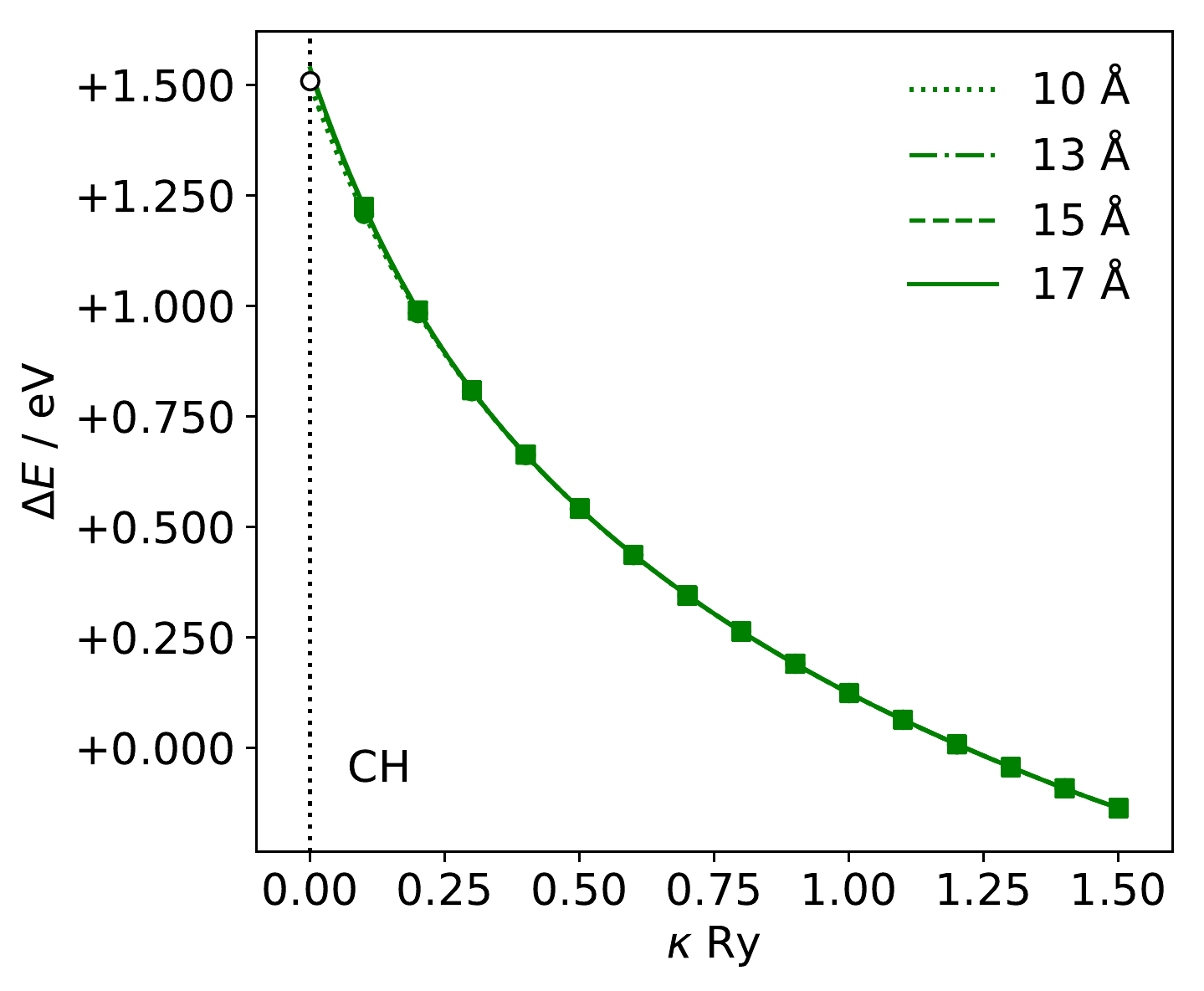}\includegraphics[width=0.26\columnwidth]{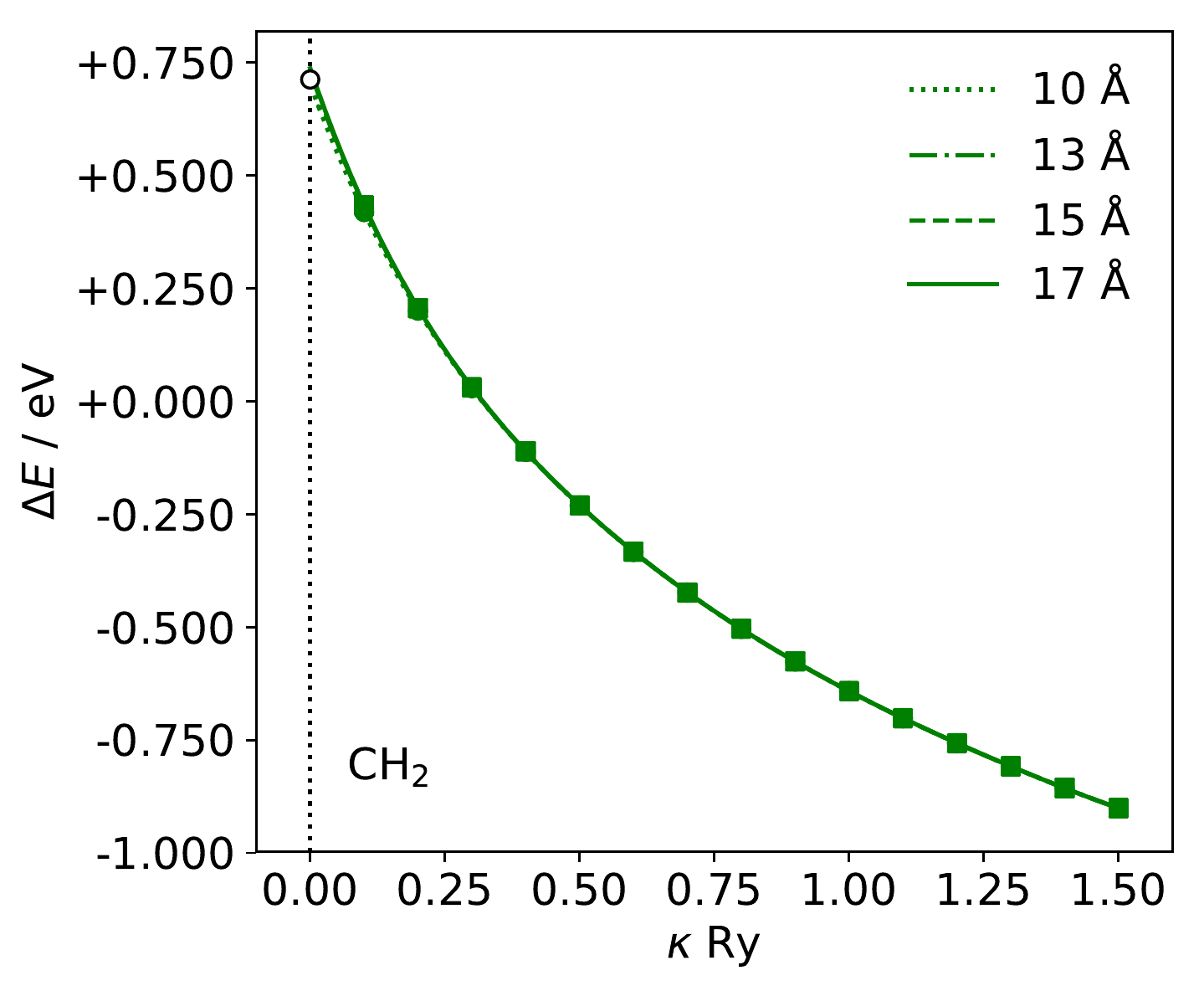}\includegraphics[width=0.26\columnwidth]{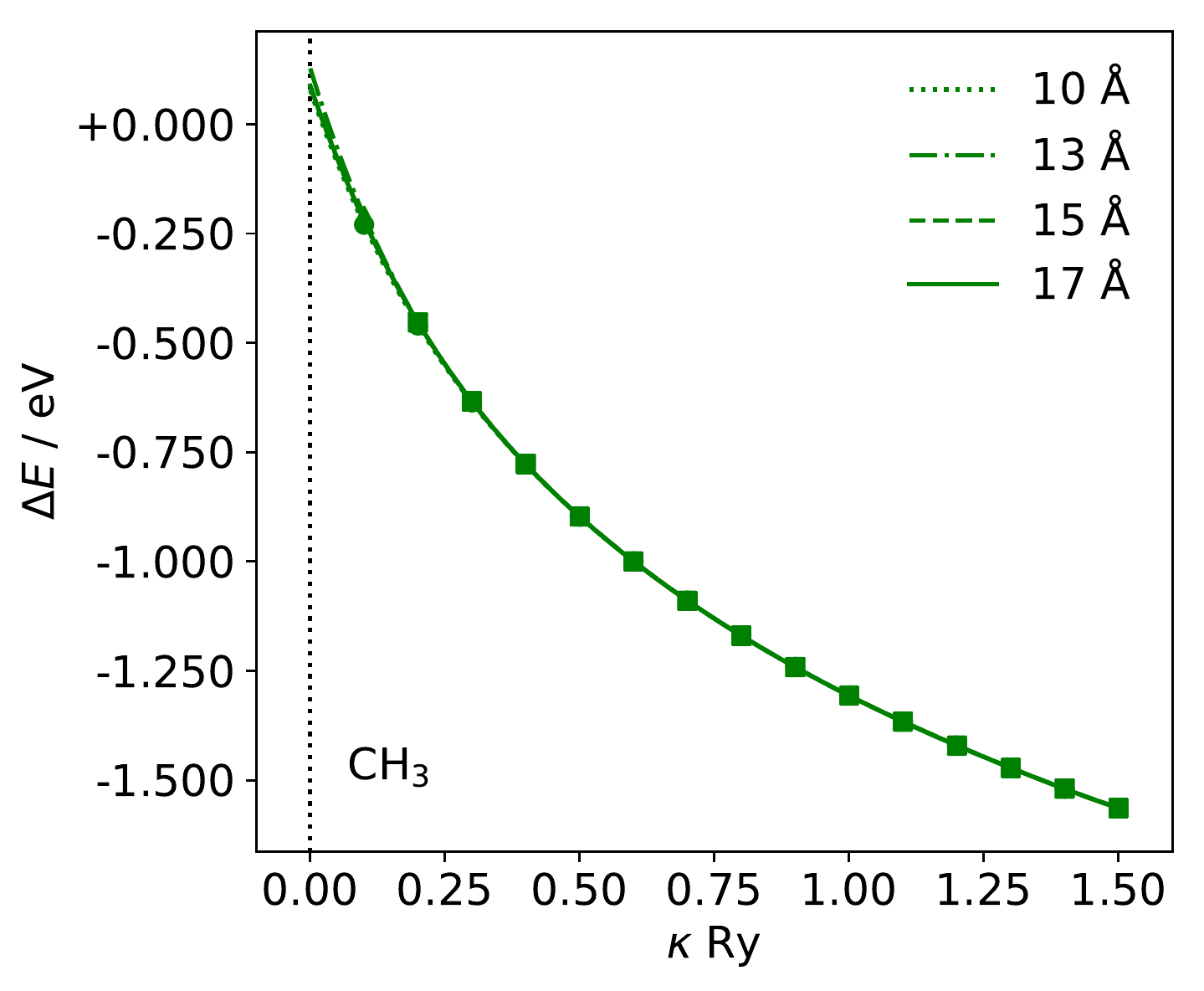} \\
\includegraphics[width=0.26\columnwidth]{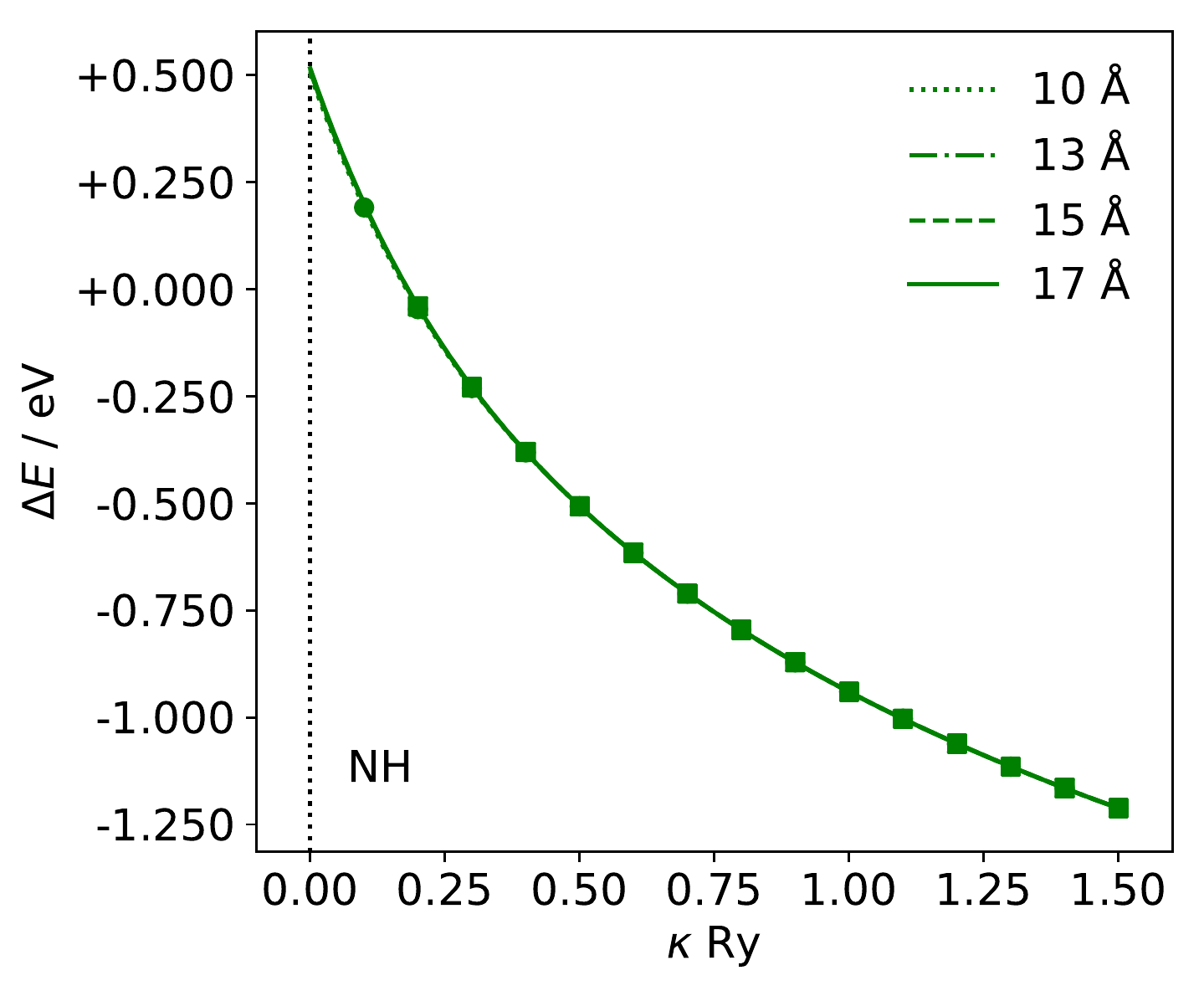}\includegraphics[width=0.26\columnwidth]{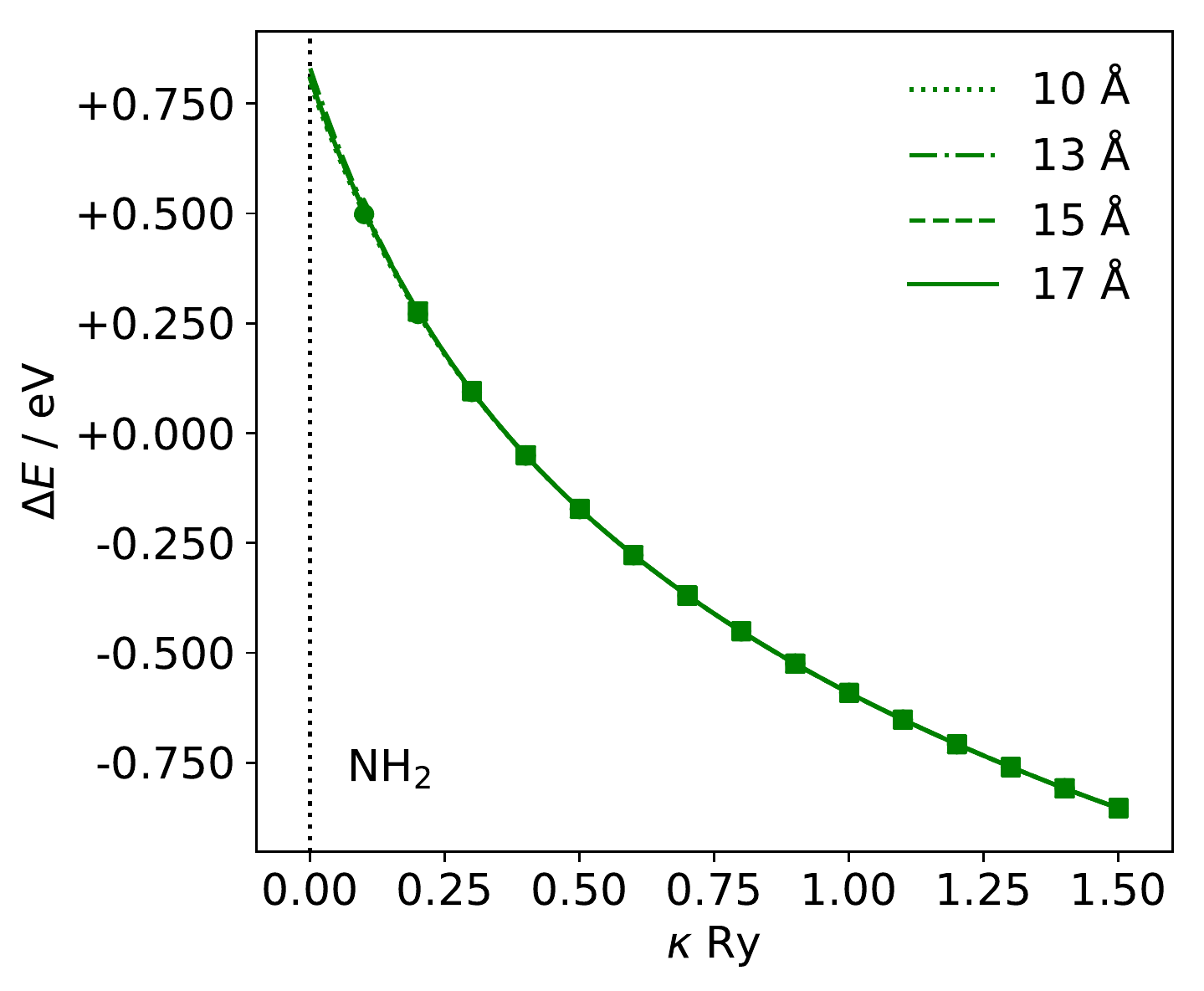}\includegraphics[width=0.26\columnwidth]{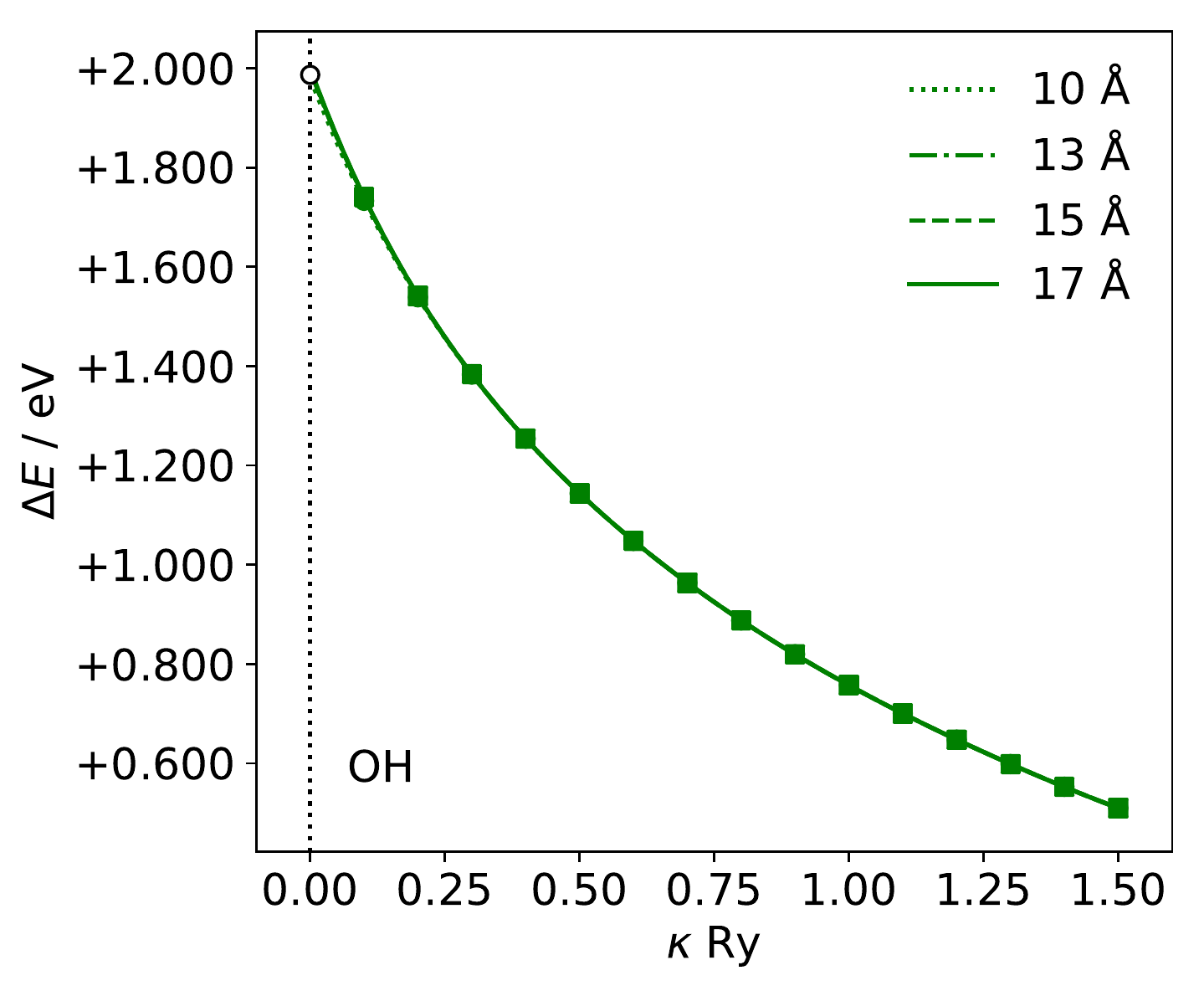} \\
\includegraphics[width=0.26\columnwidth]{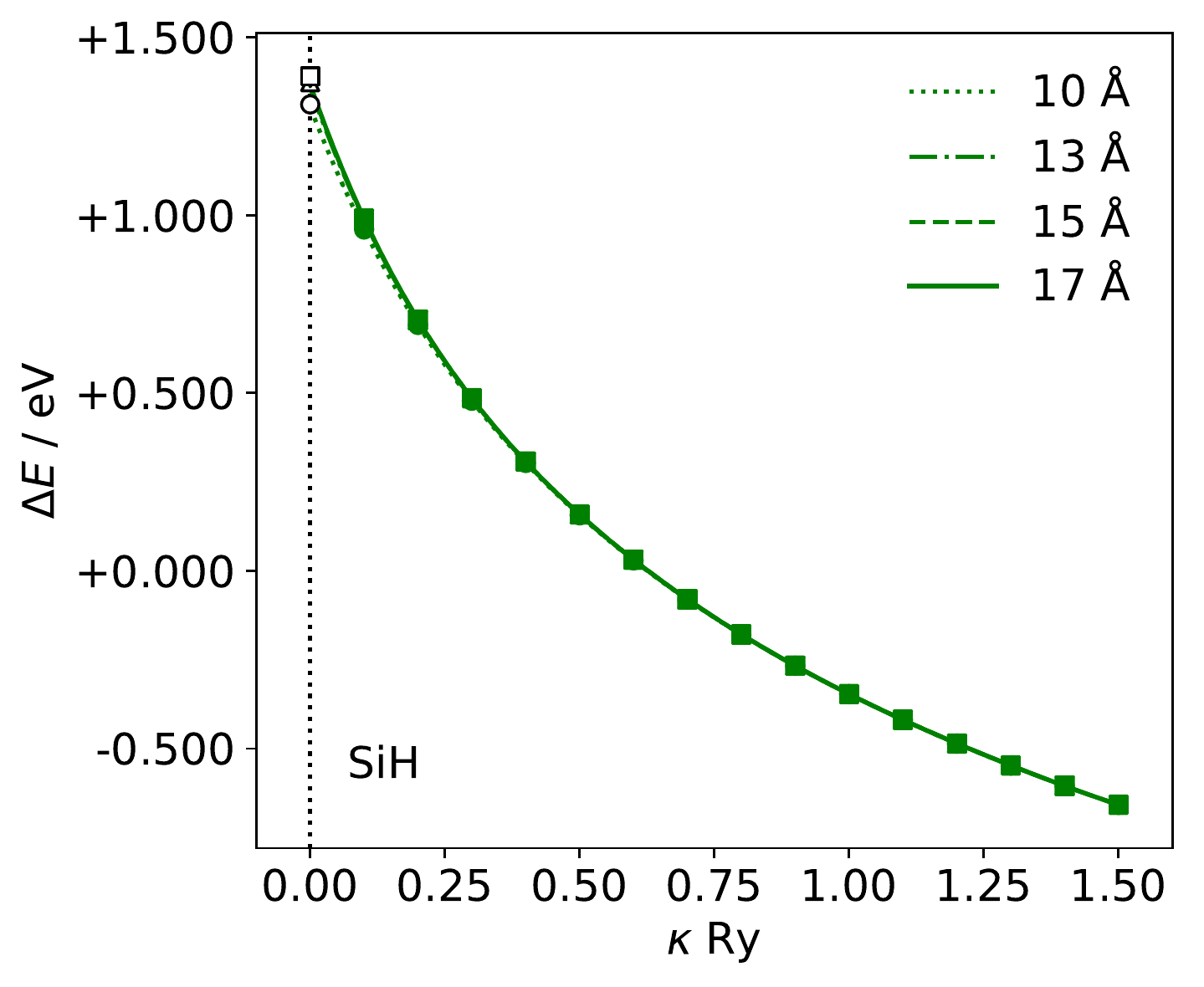}\includegraphics[width=0.26\columnwidth]{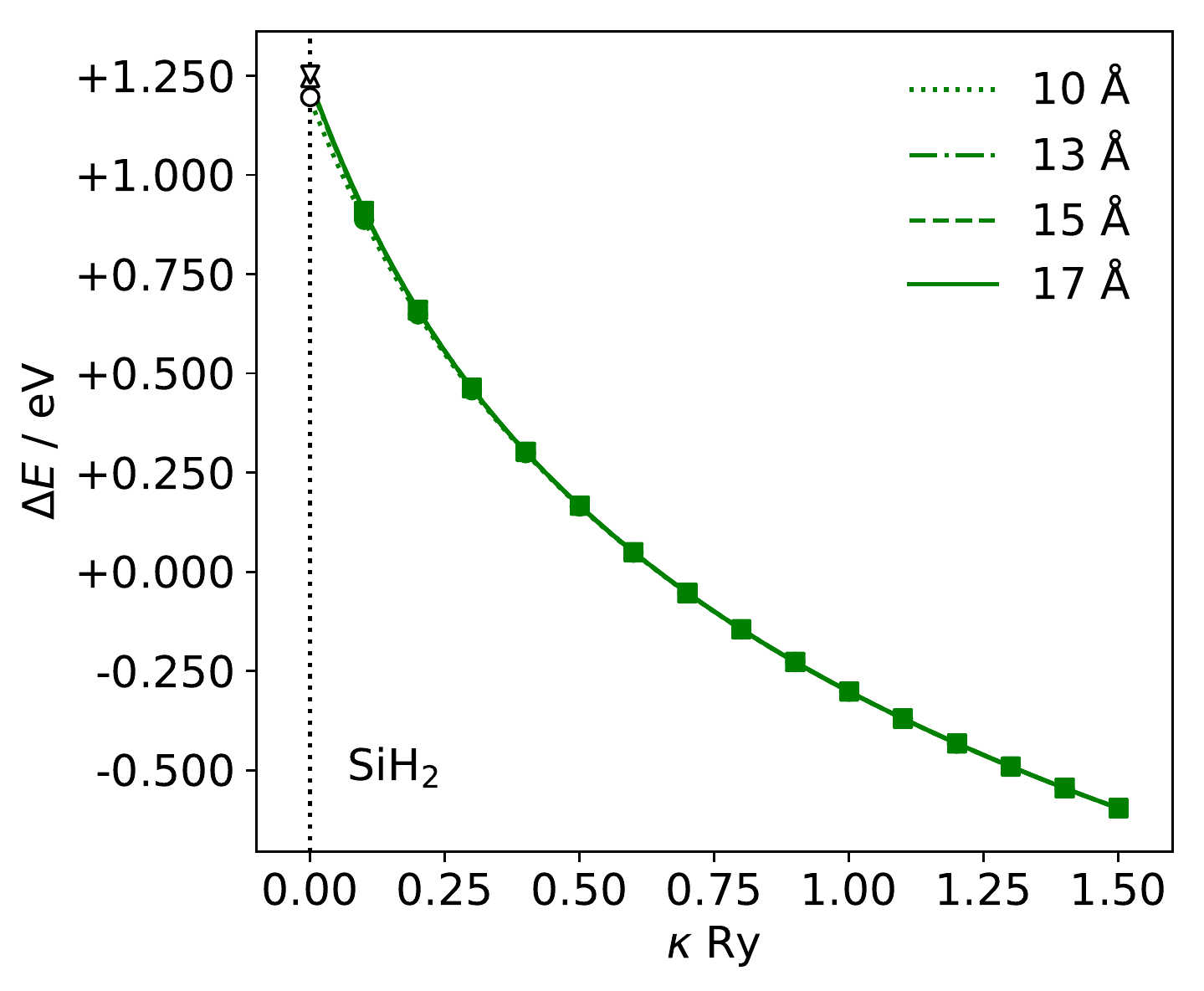}\includegraphics[width=0.26\columnwidth]{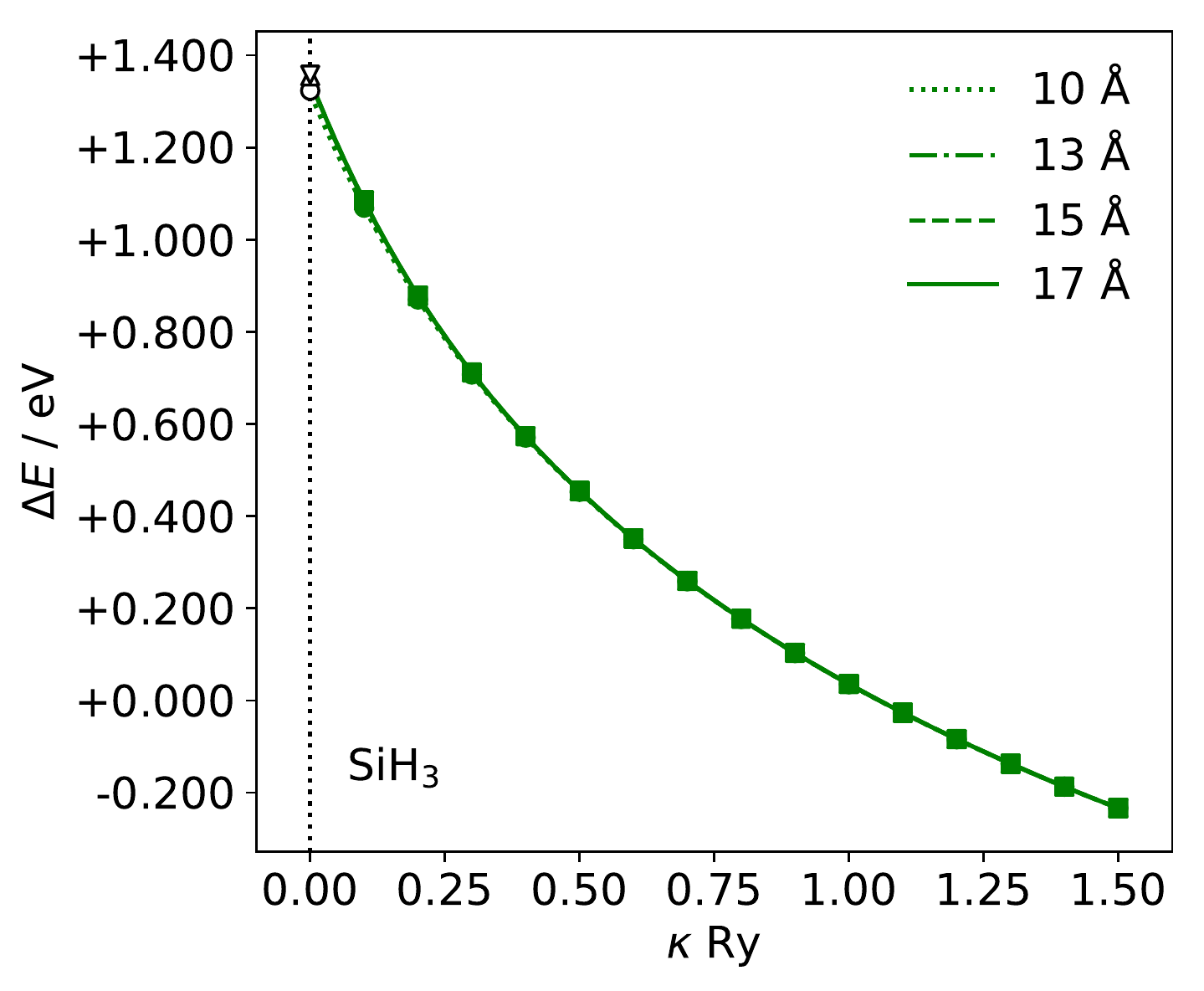} \\
\includegraphics[width=0.26\columnwidth]{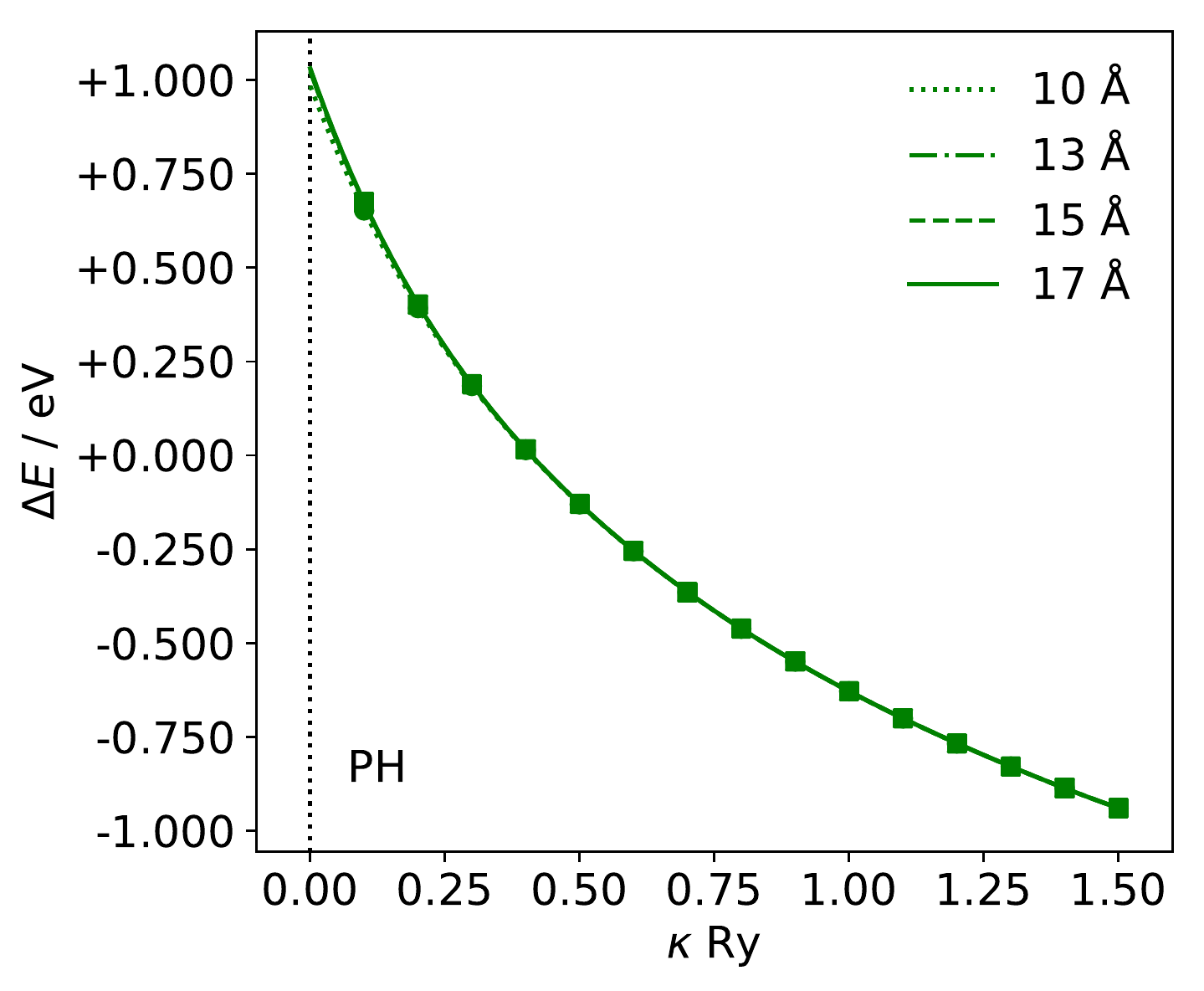}\includegraphics[width=0.26\columnwidth]{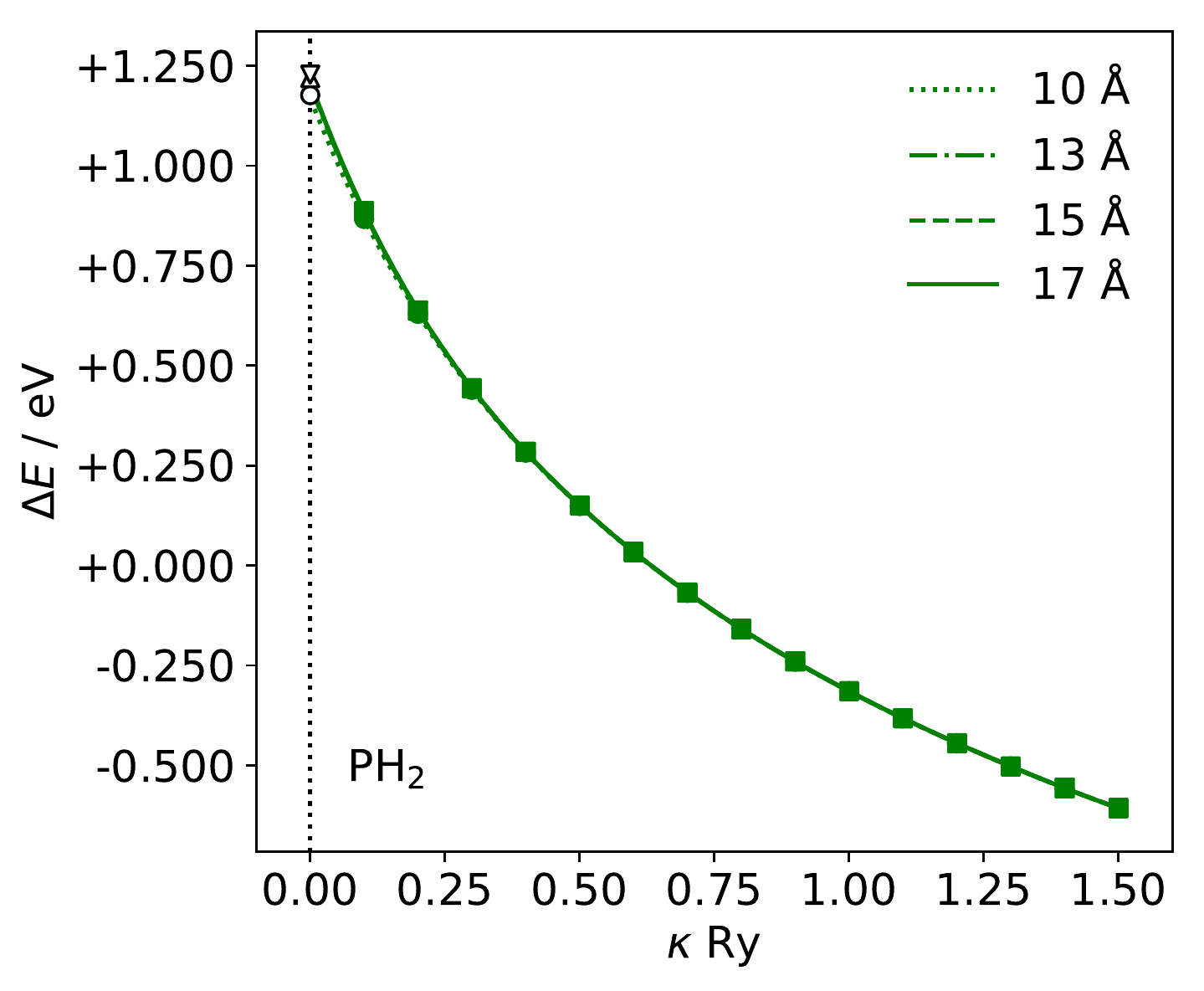}\includegraphics[width=0.26\columnwidth]{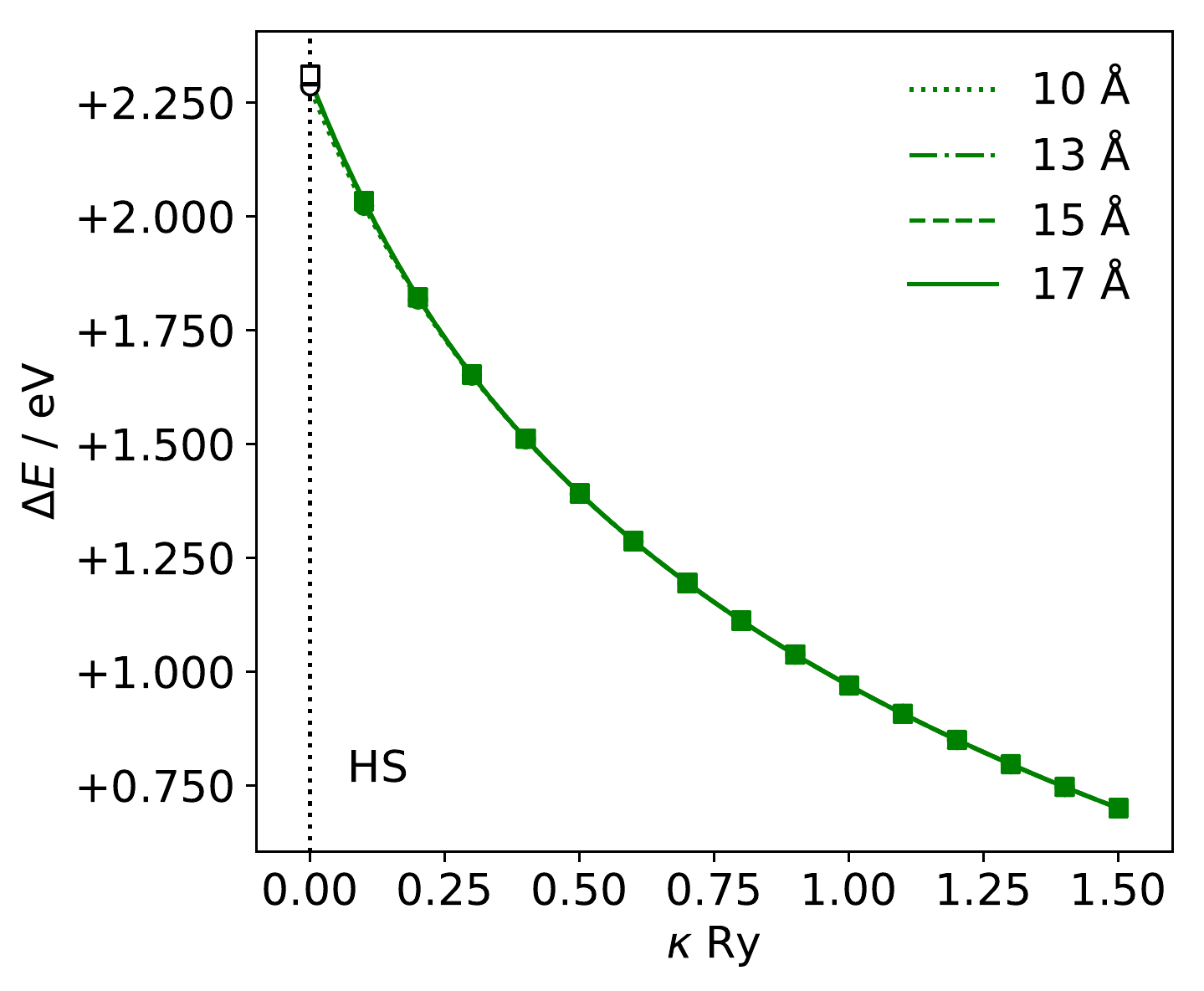} \\
\includegraphics[width=0.26\columnwidth]{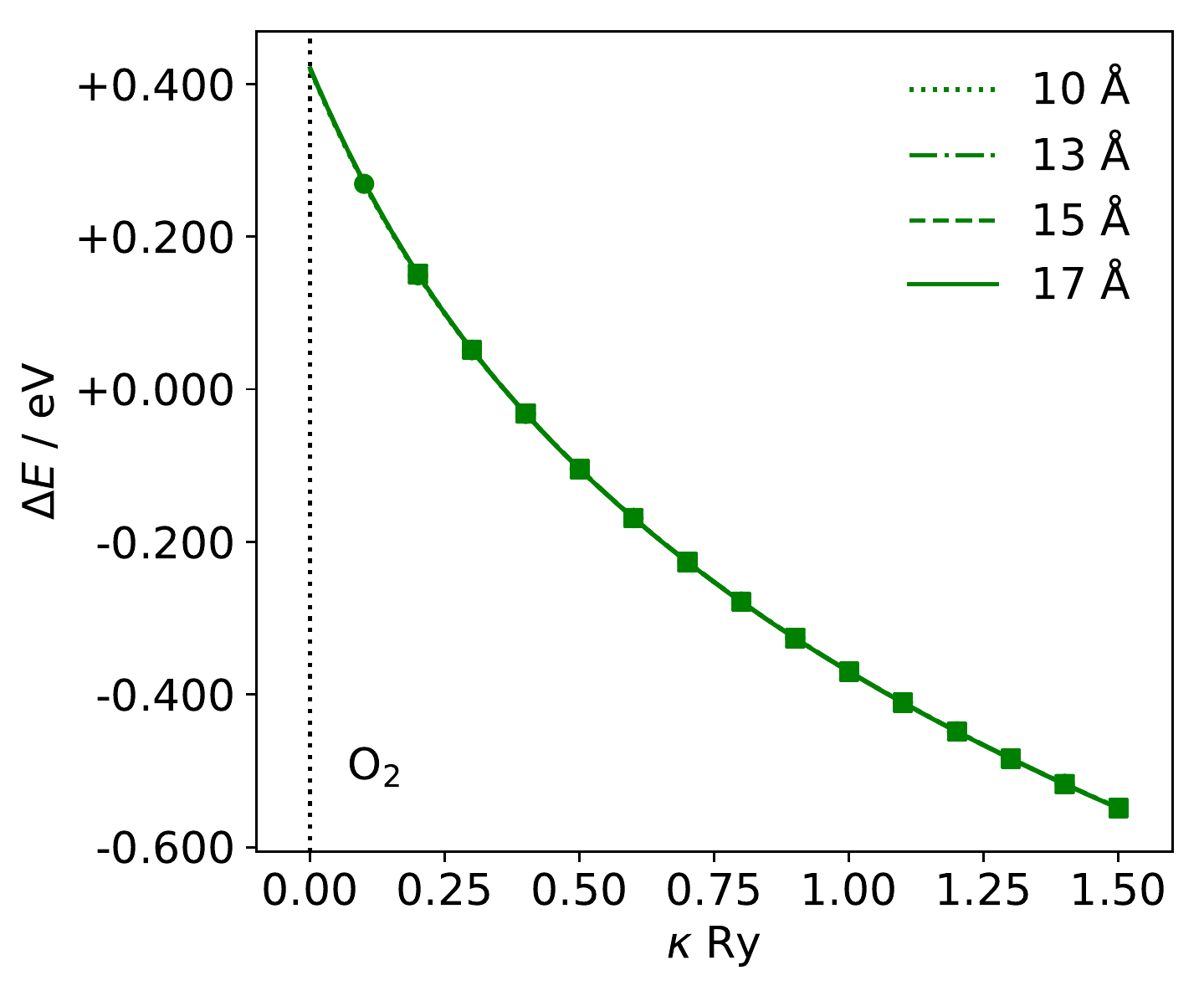}\includegraphics[width=0.26\columnwidth]{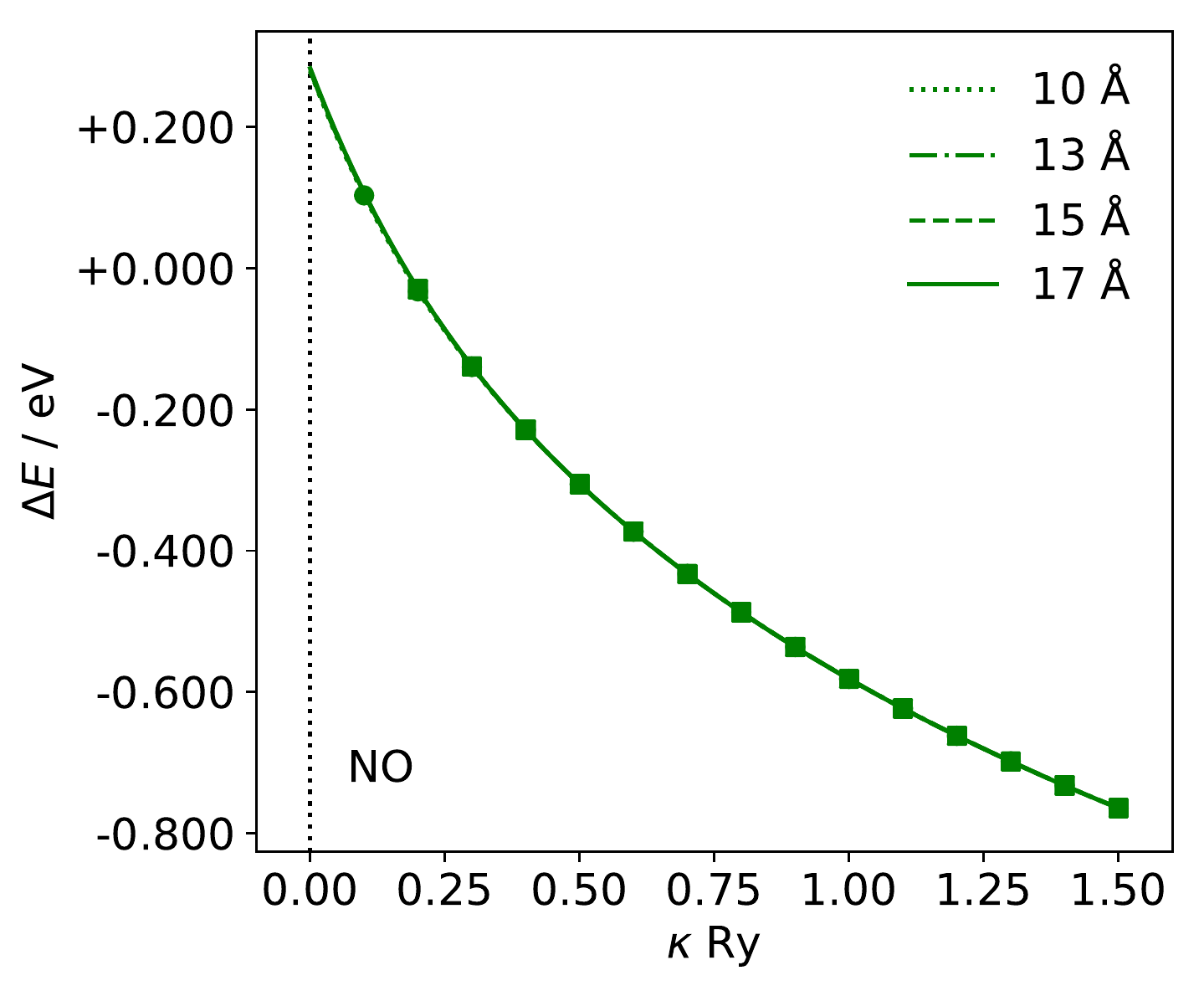}\includegraphics[width=0.26\columnwidth]{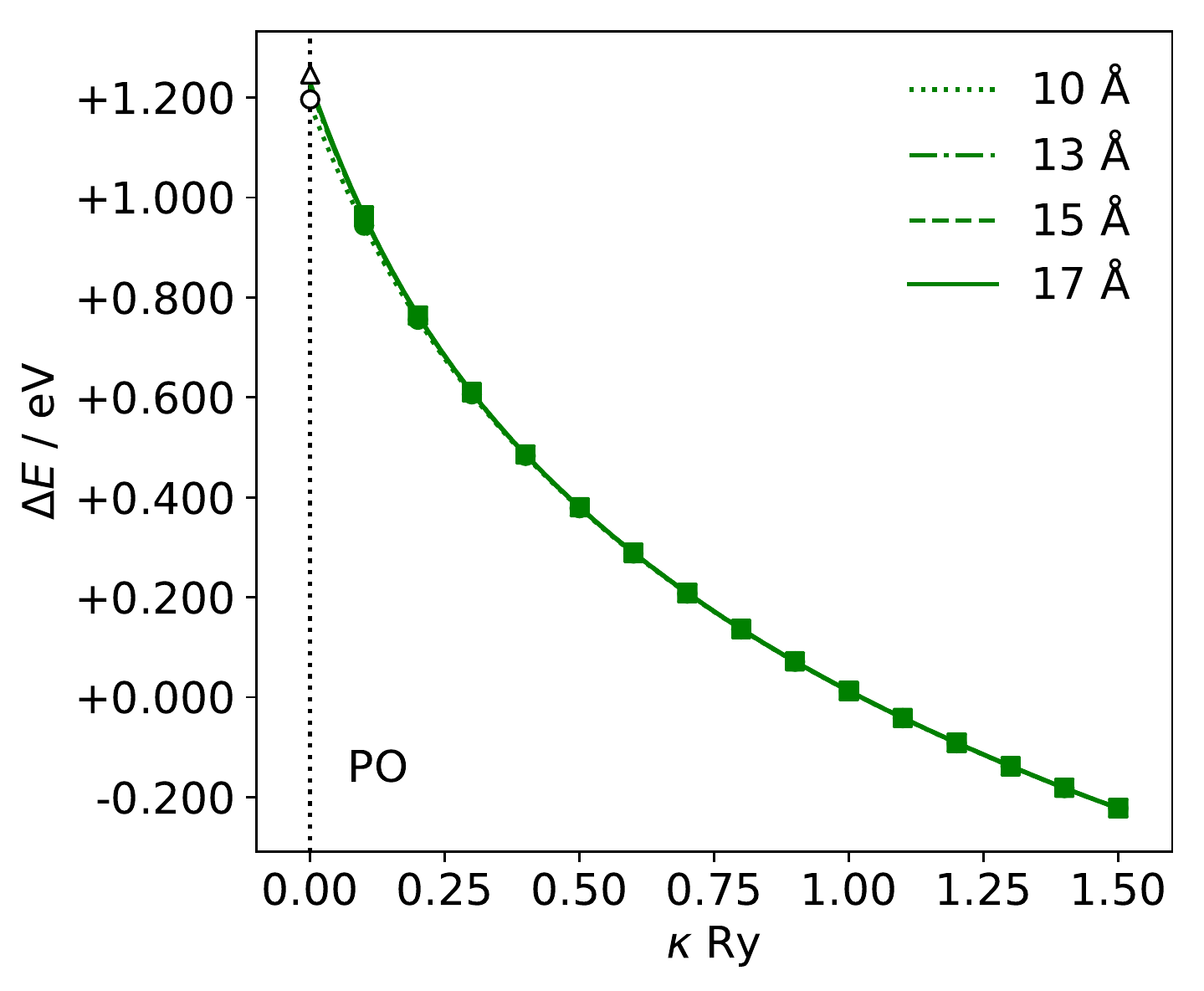} \\
\includegraphics[width=0.26\columnwidth]{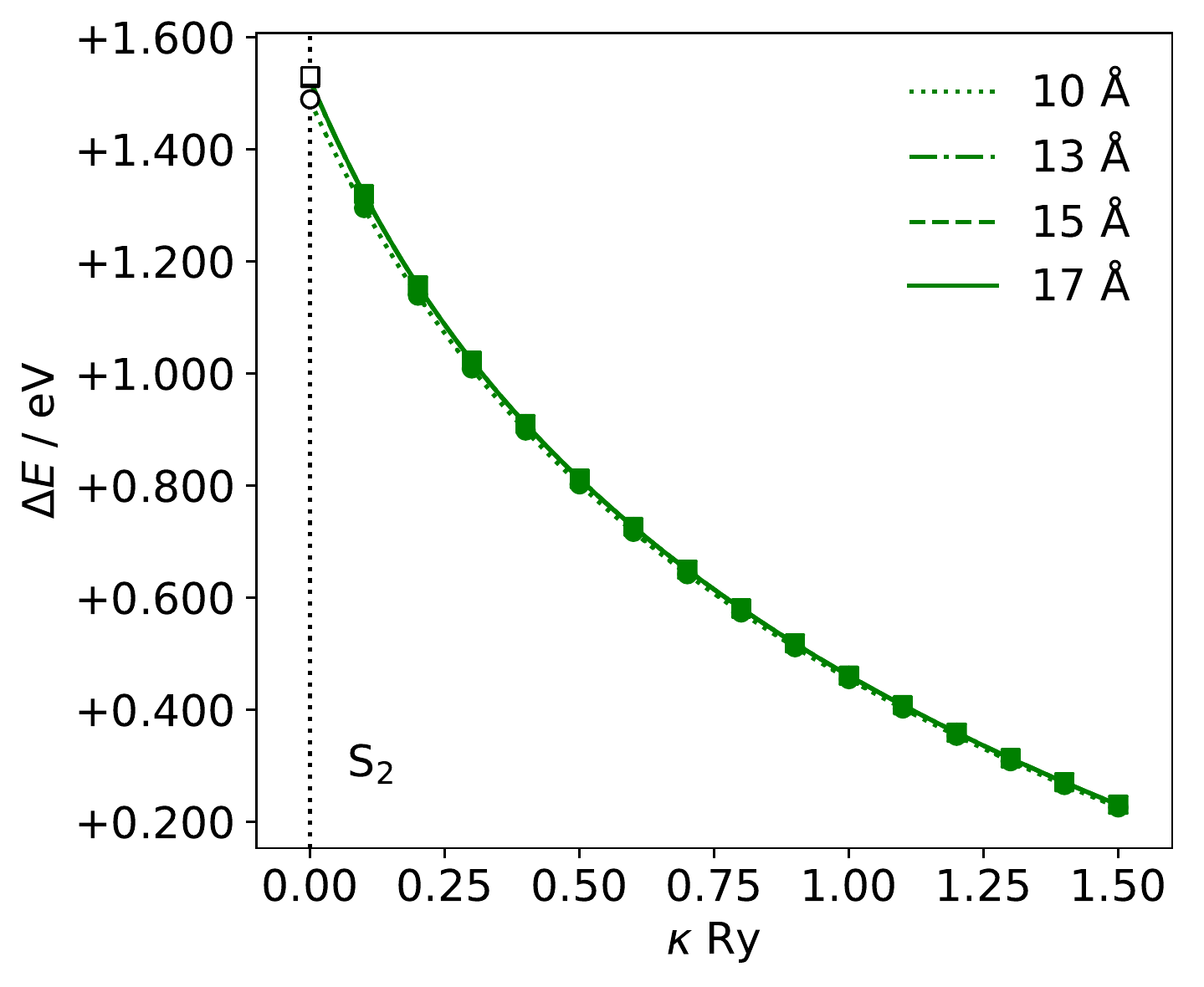}\includegraphics[width=0.26\columnwidth]{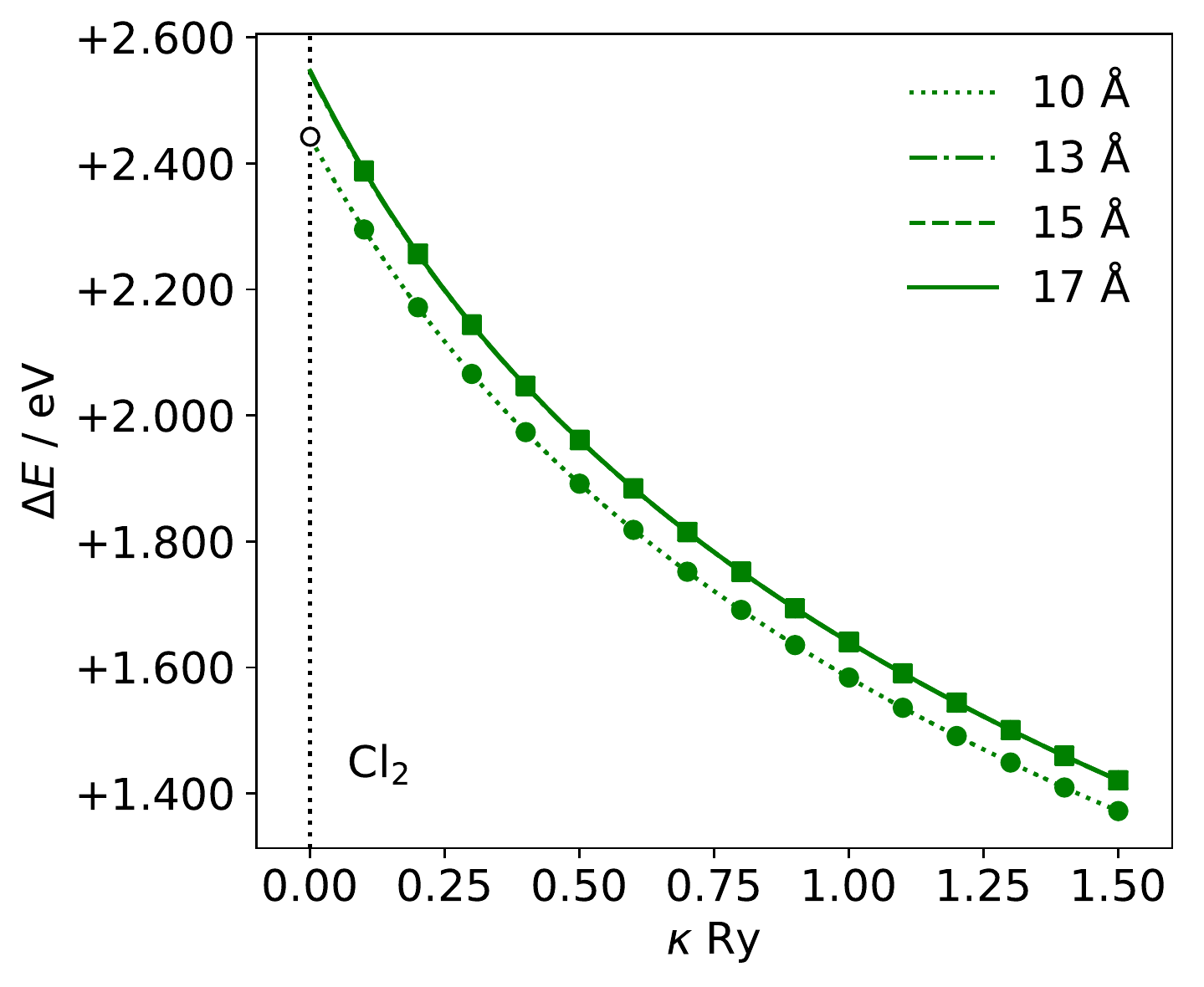}\includegraphics[width=0.26\columnwidth]{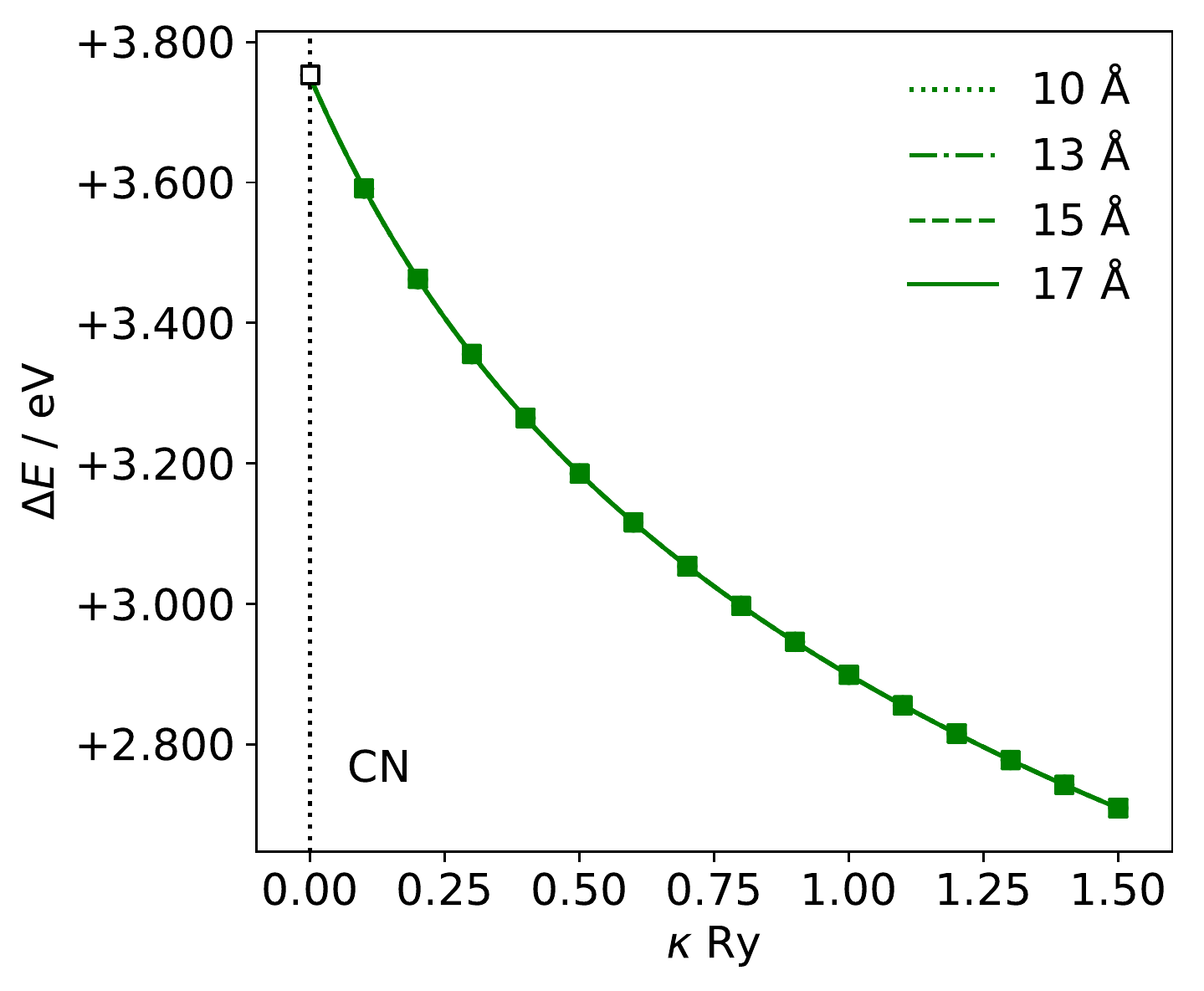} \\
\end{centering}
\caption{$\Delta E$ extrapolation for the species of the G2-1 dataset, using the confining-potential extrapolation towards vanishing potentials. Lines represent 6th-degree polynomial functions.}
\label{fig:extrapolation-confining-energy}
\end{figure}

\begin{figure}
\begin{centering}
\includegraphics[width=0.26\columnwidth]{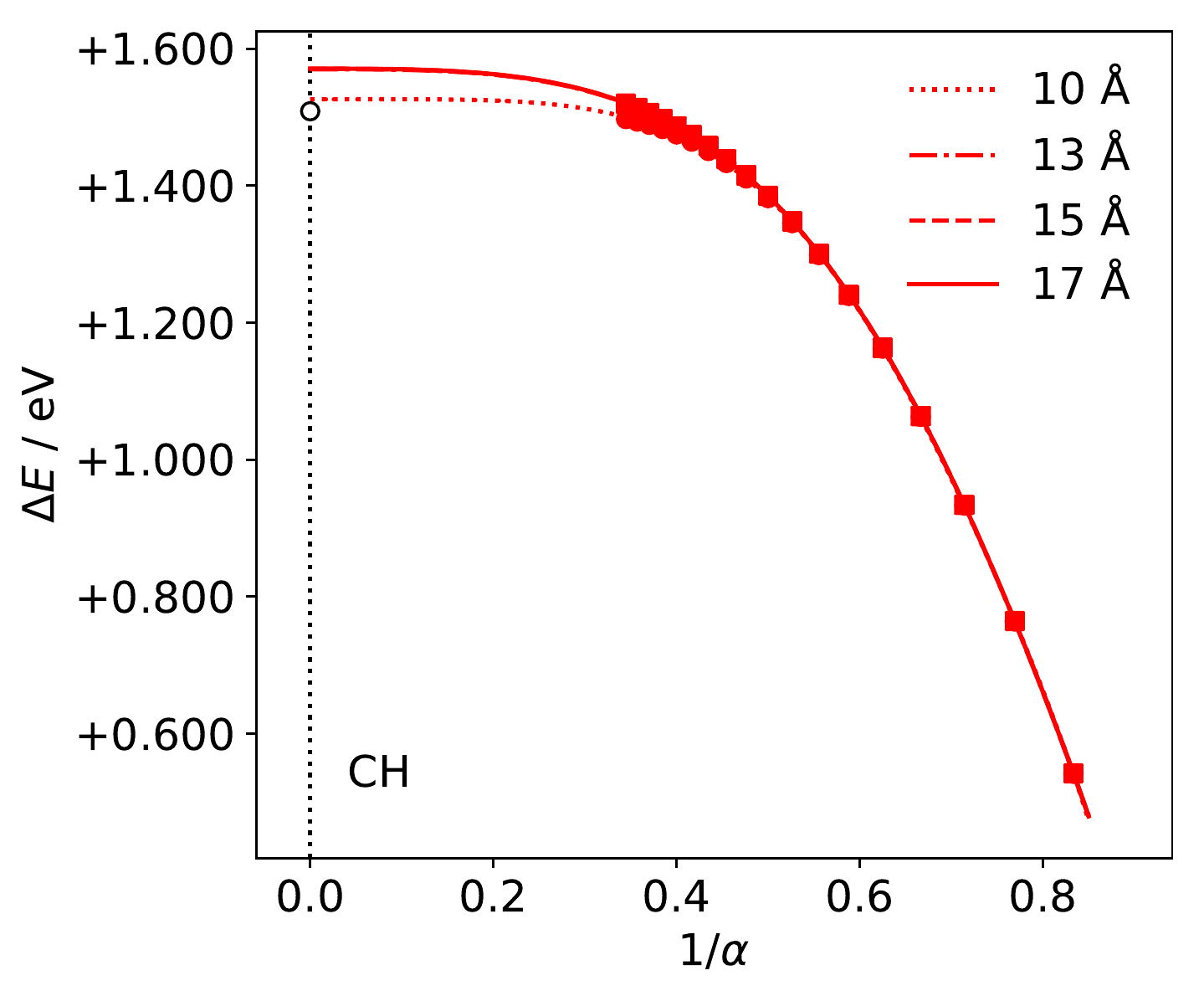}\includegraphics[width=0.26\columnwidth]{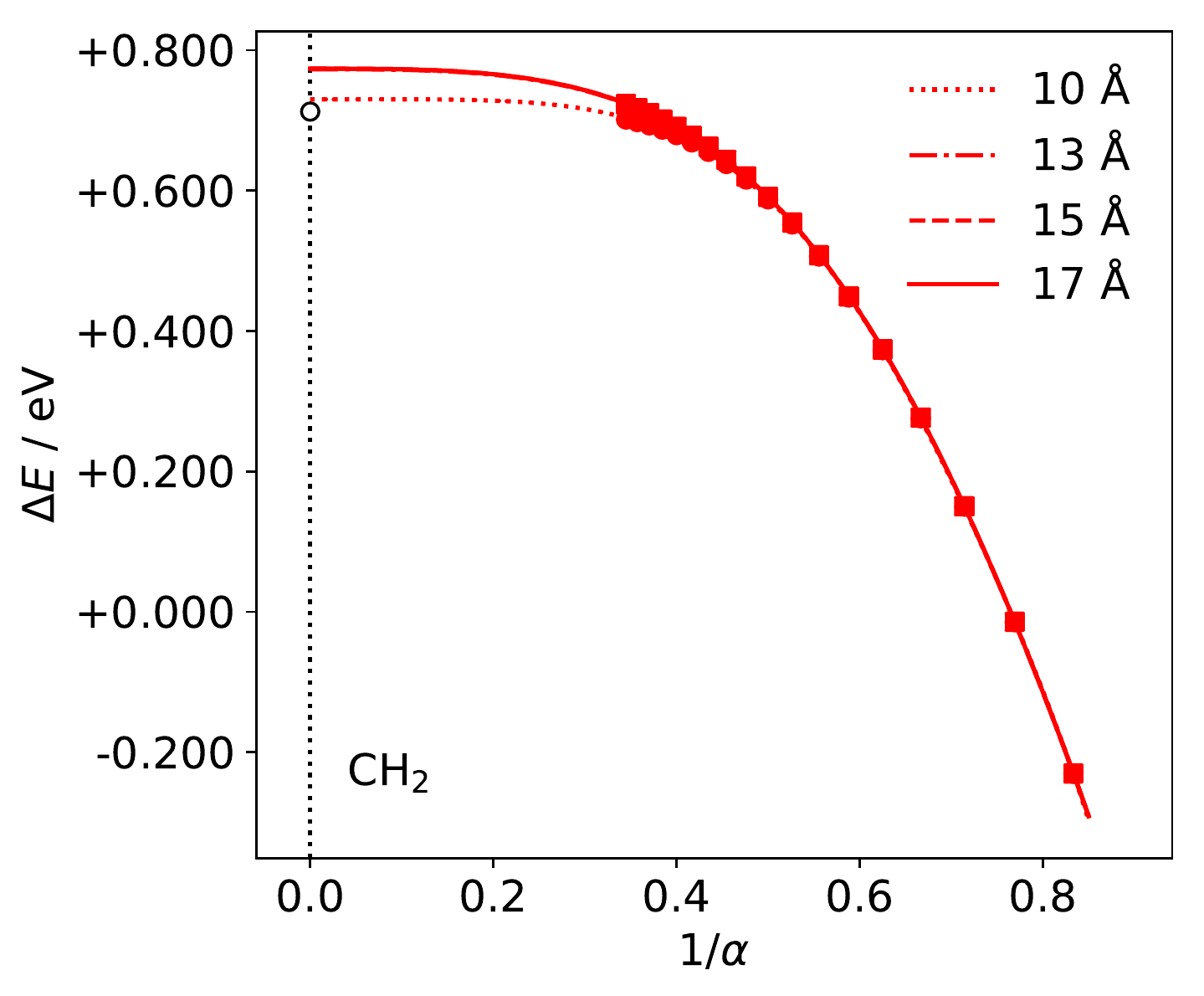}\includegraphics[width=0.26\columnwidth]{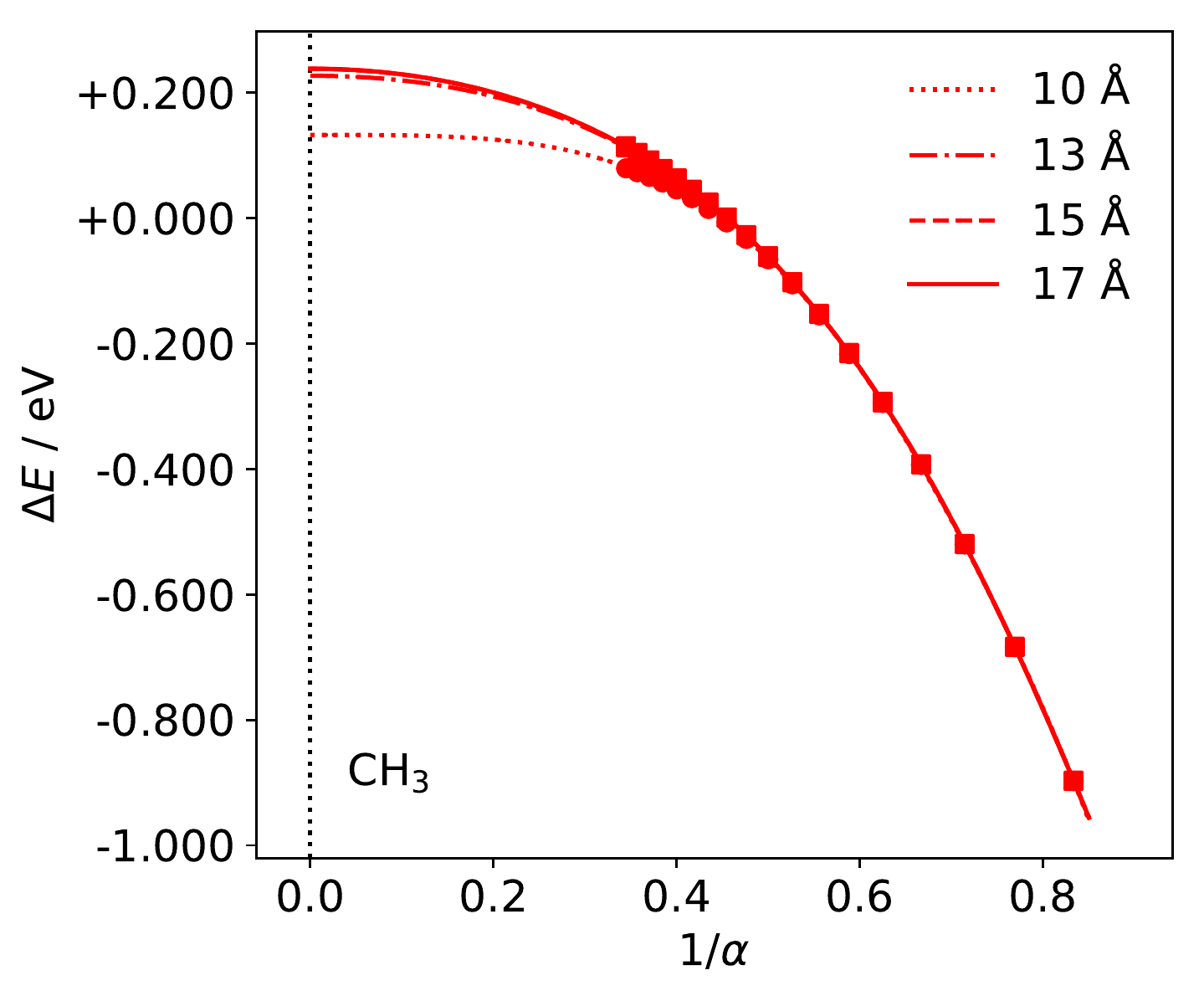} \\
\includegraphics[width=0.26\columnwidth]{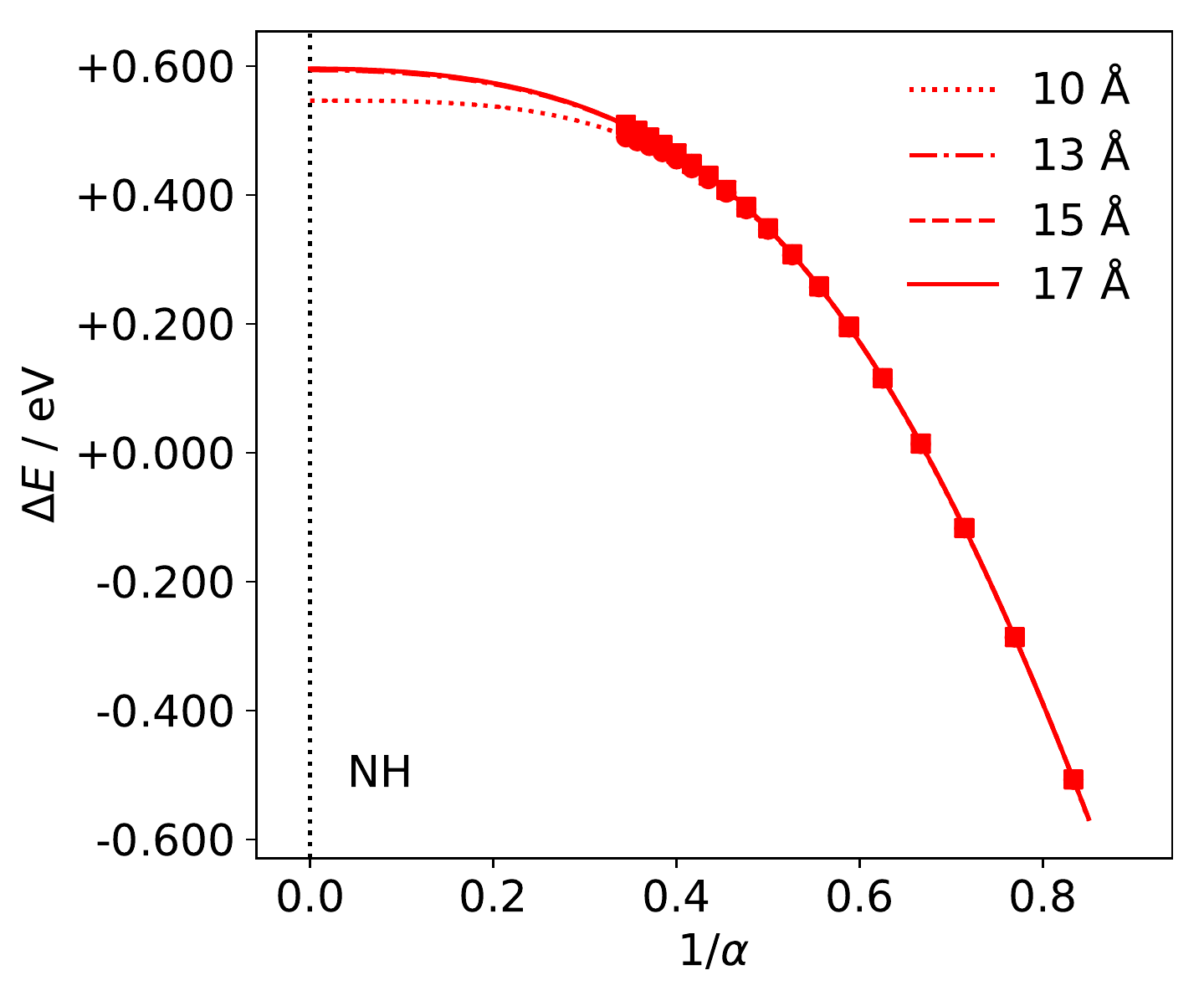}\includegraphics[width=0.26\columnwidth]{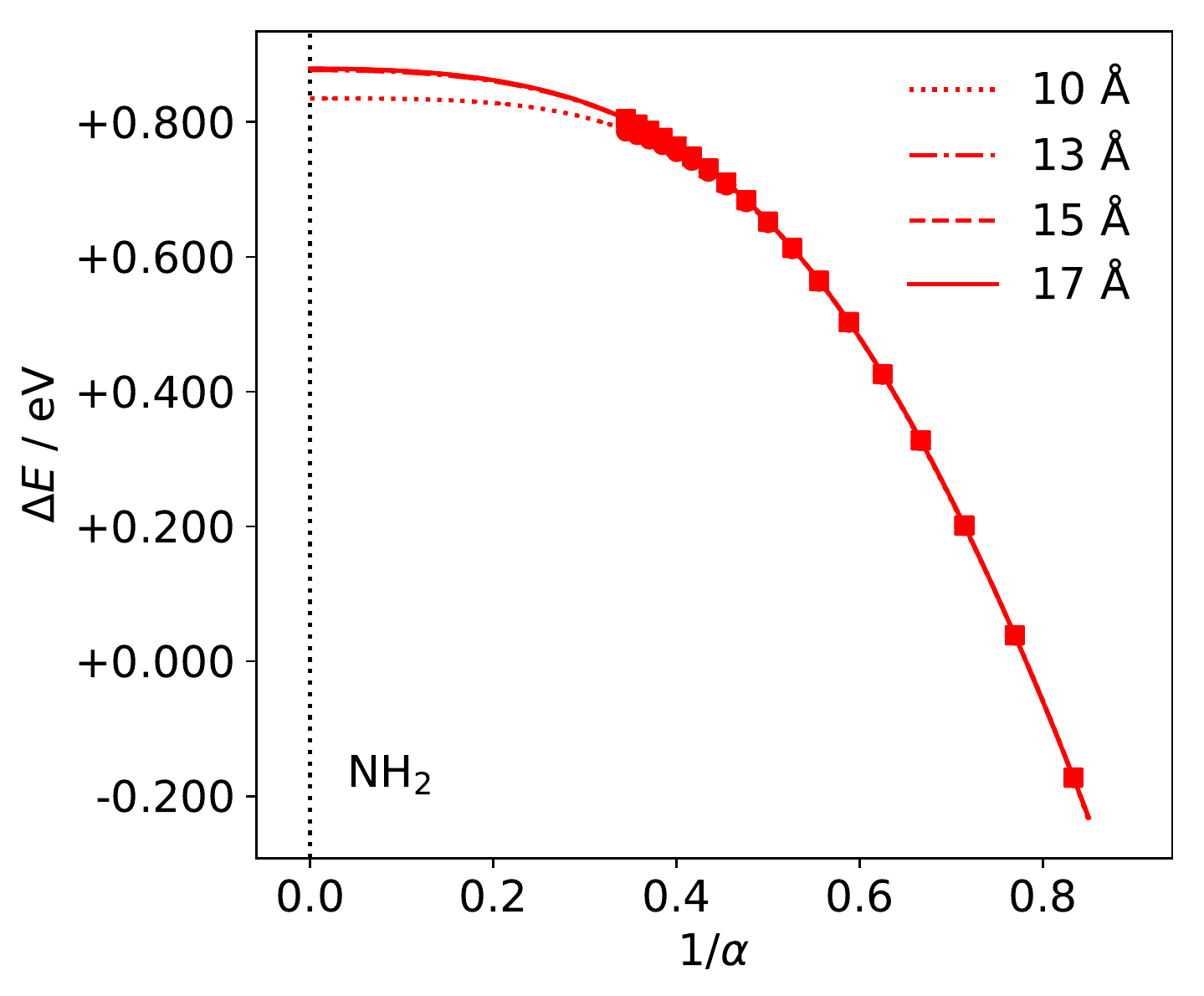}\includegraphics[width=0.26\columnwidth]{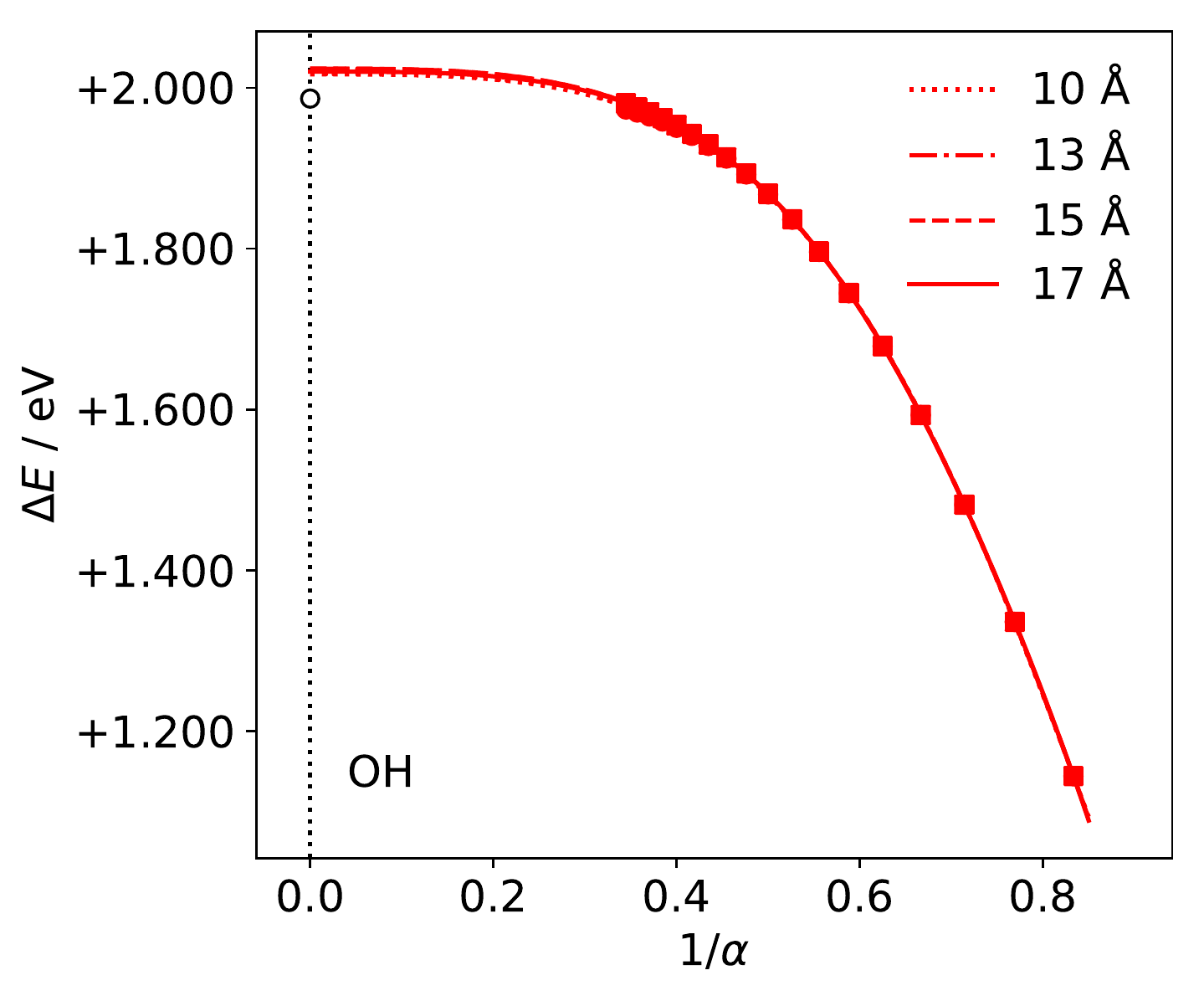} \\
\includegraphics[width=0.26\columnwidth]{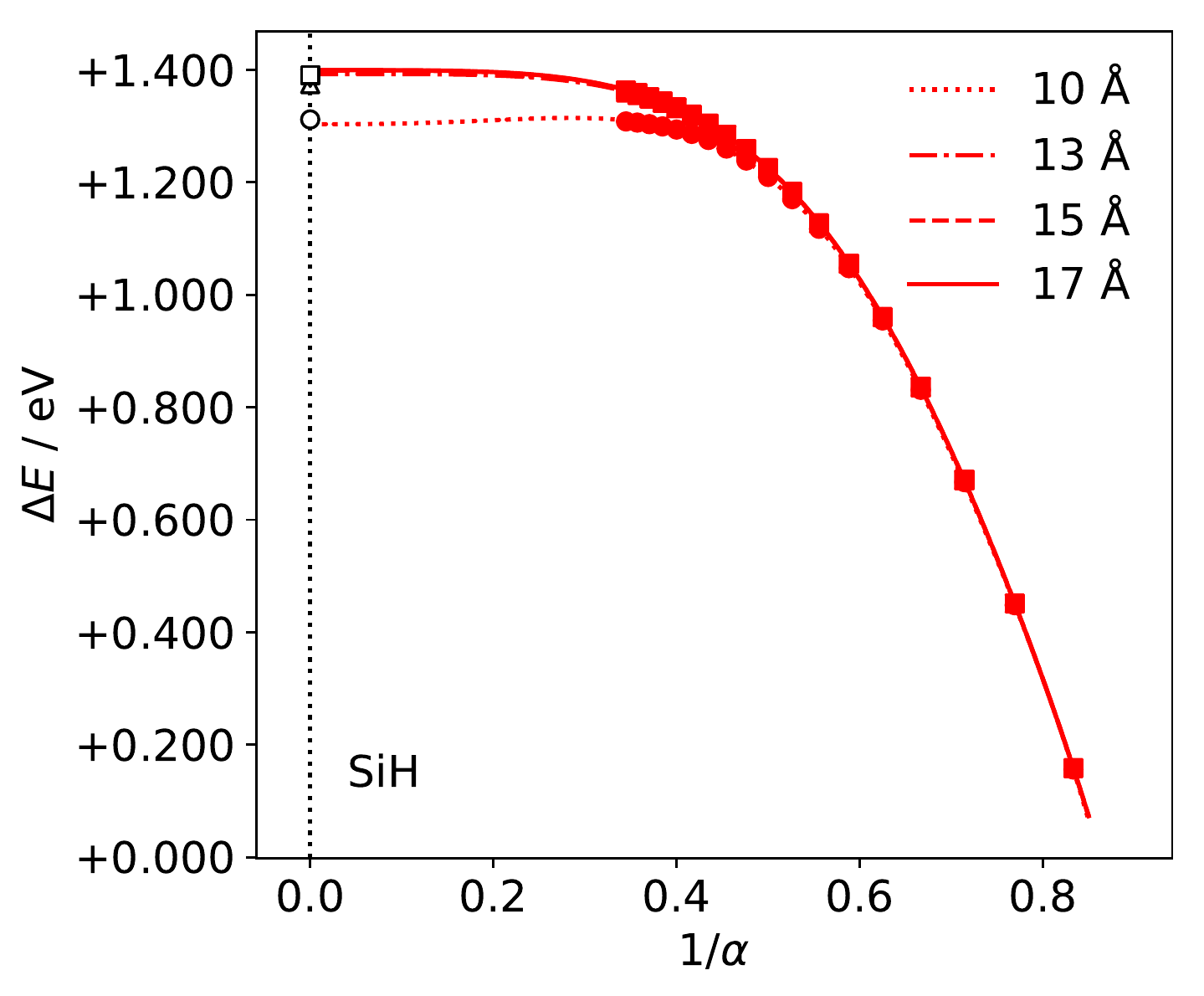}\includegraphics[width=0.26\columnwidth]{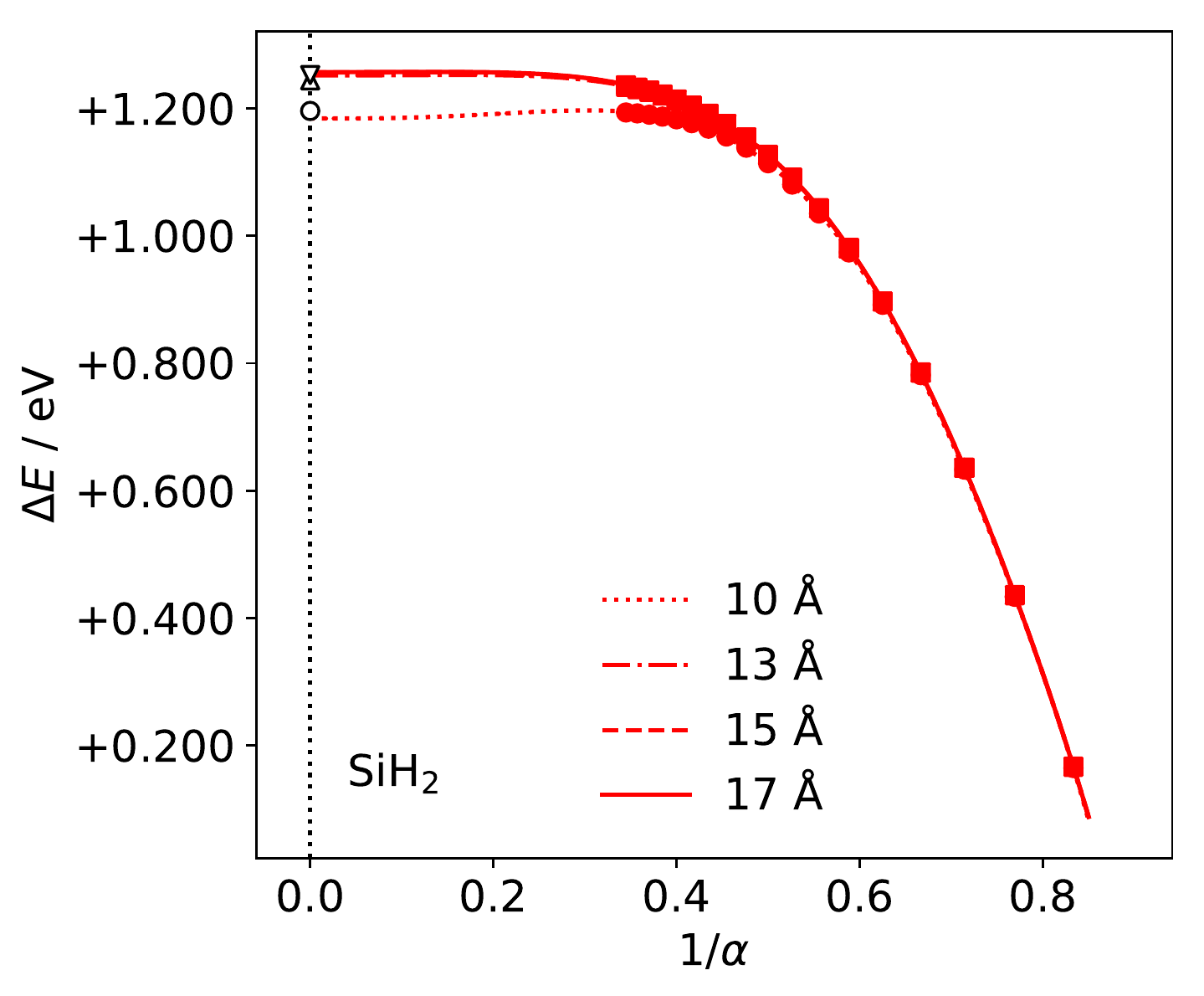}\includegraphics[width=0.26\columnwidth]{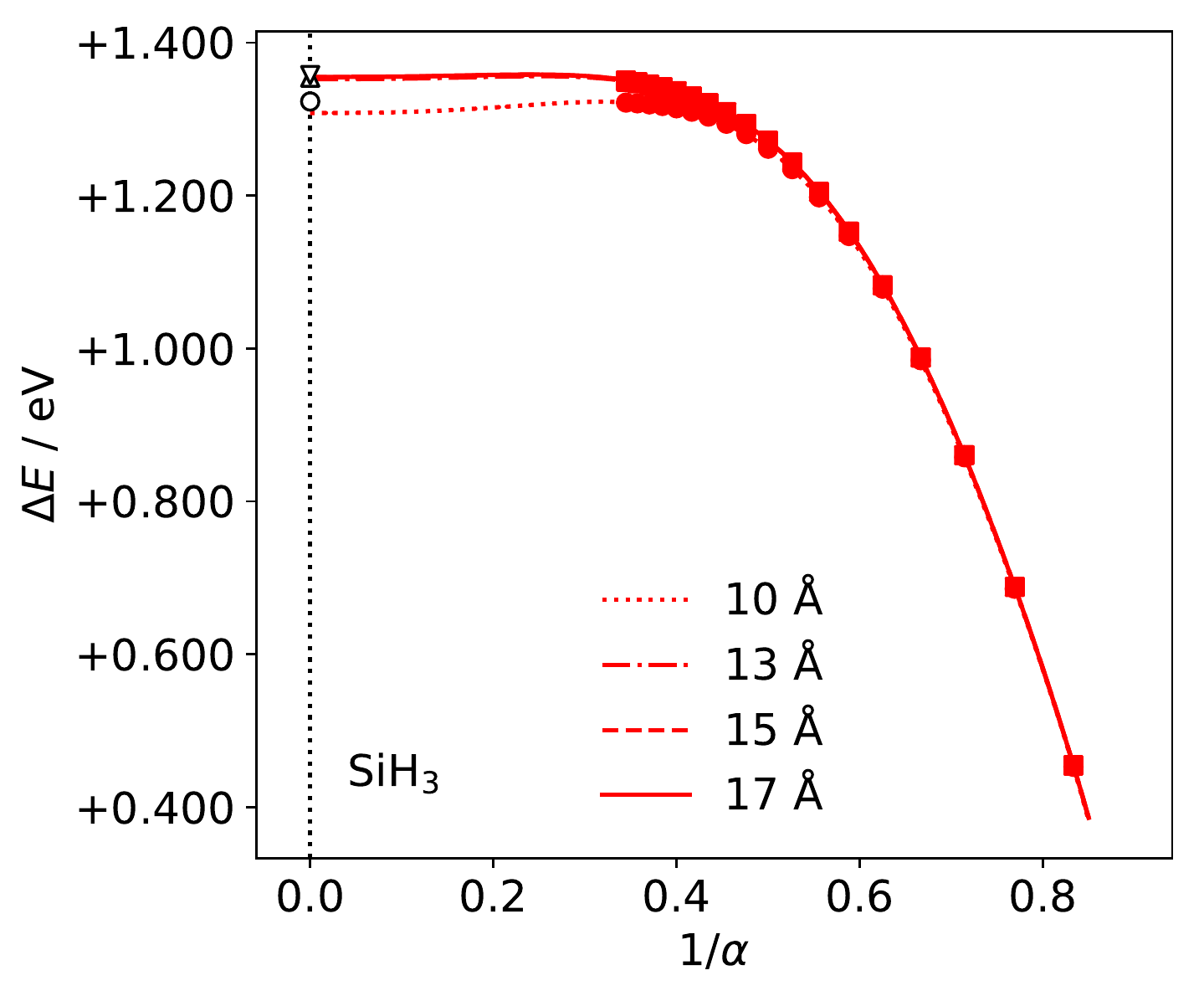} \\
\includegraphics[width=0.26\columnwidth]{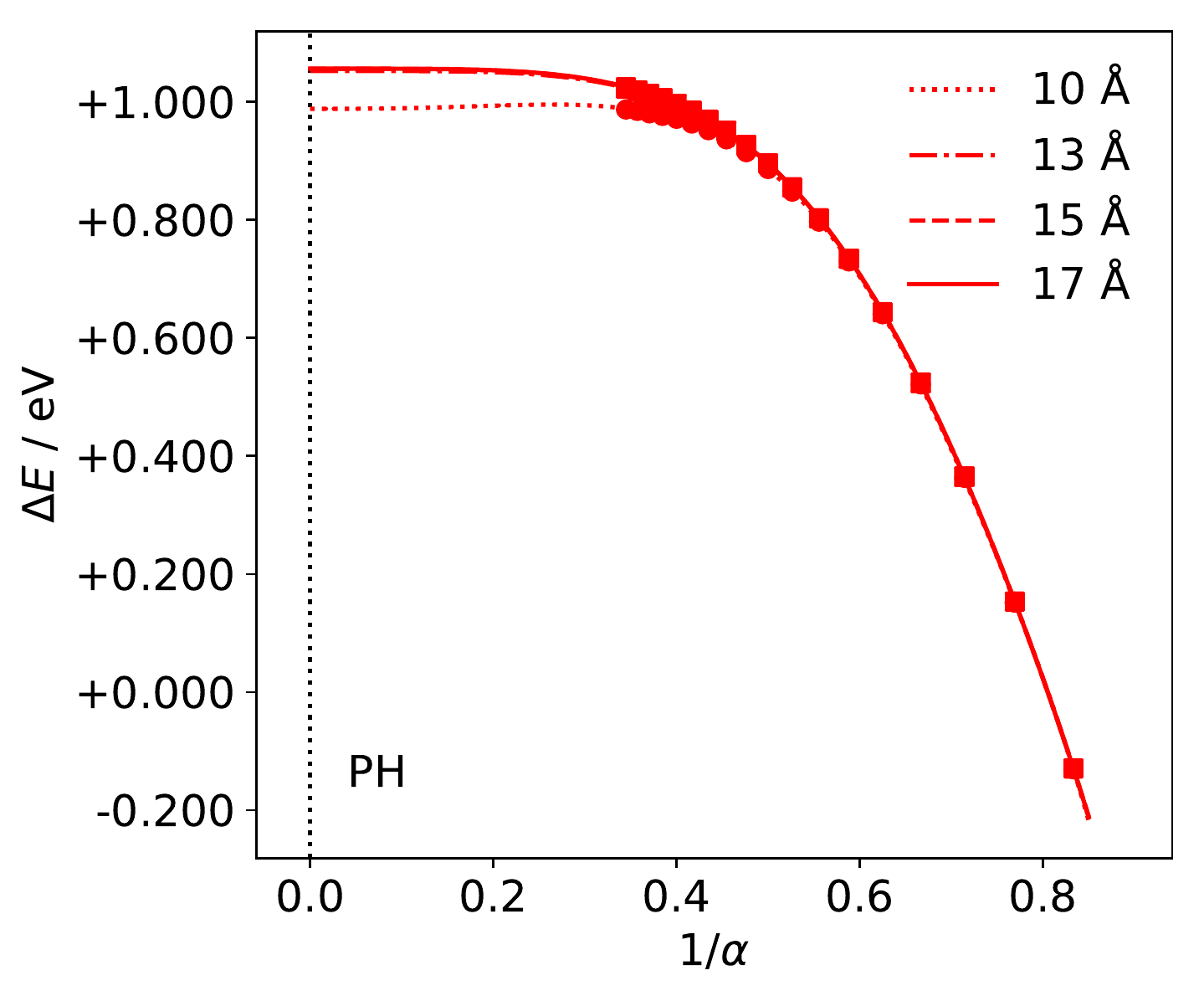}\includegraphics[width=0.26\columnwidth]{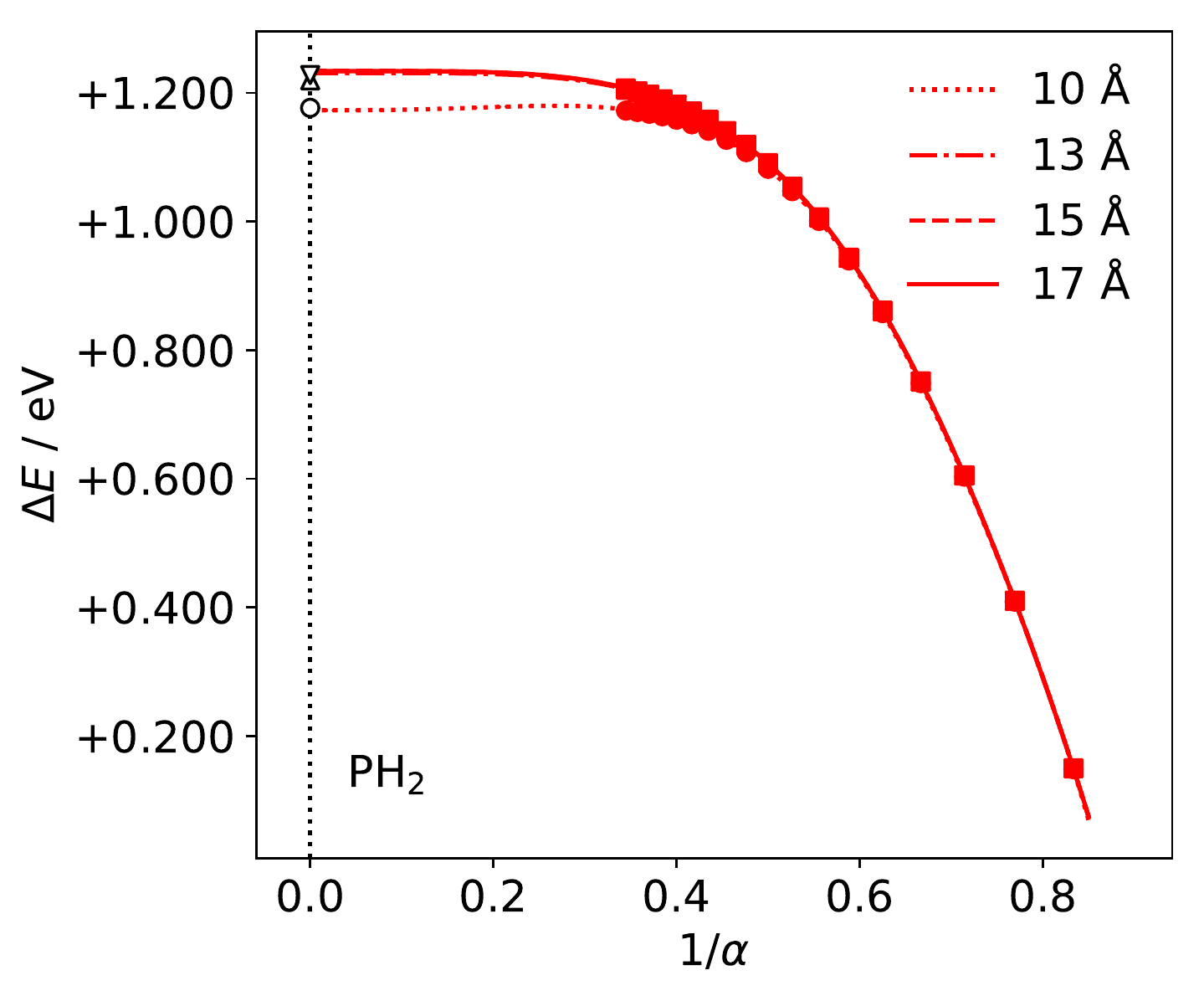}\includegraphics[width=0.26\columnwidth]{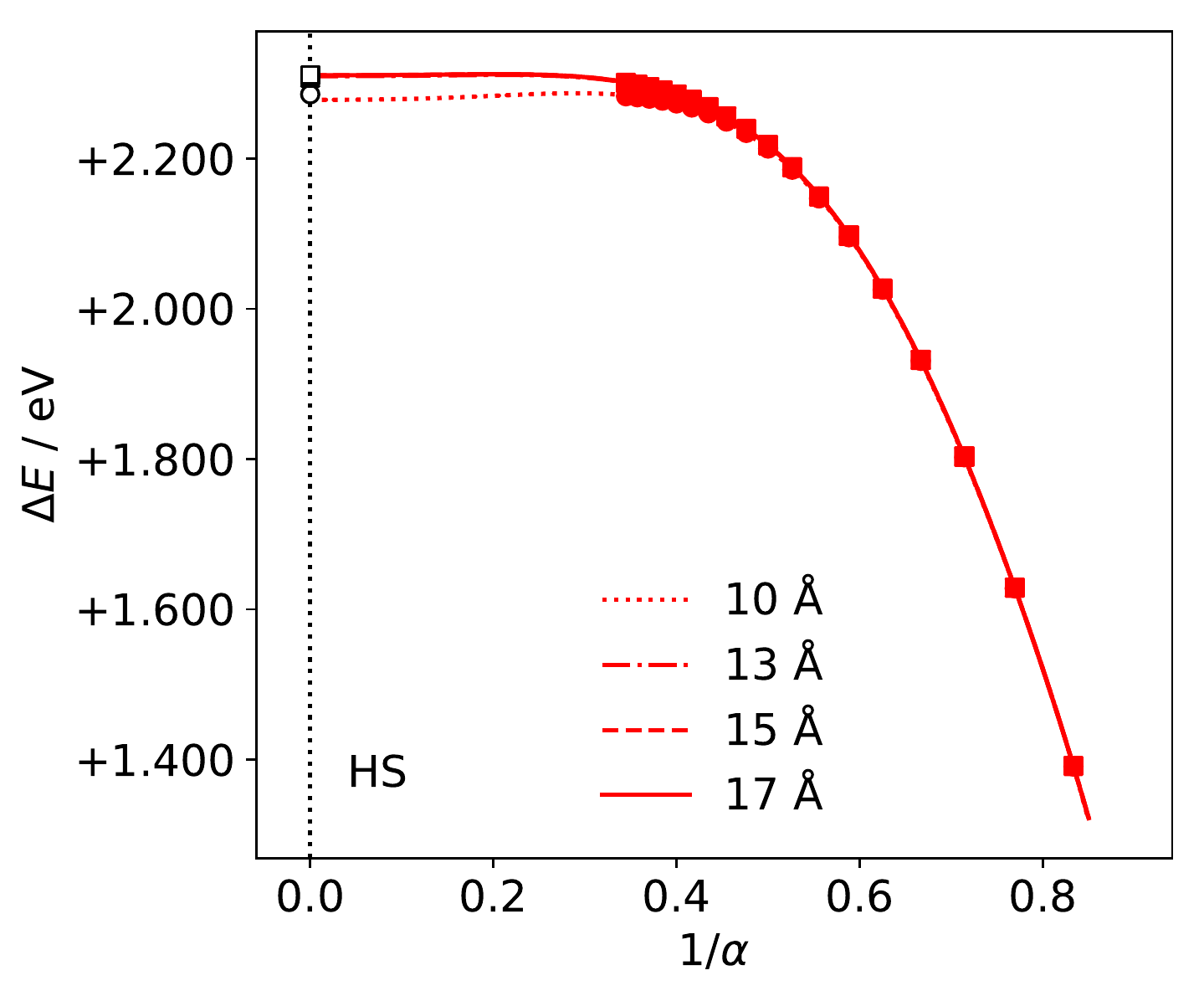} \\
\includegraphics[width=0.26\columnwidth]{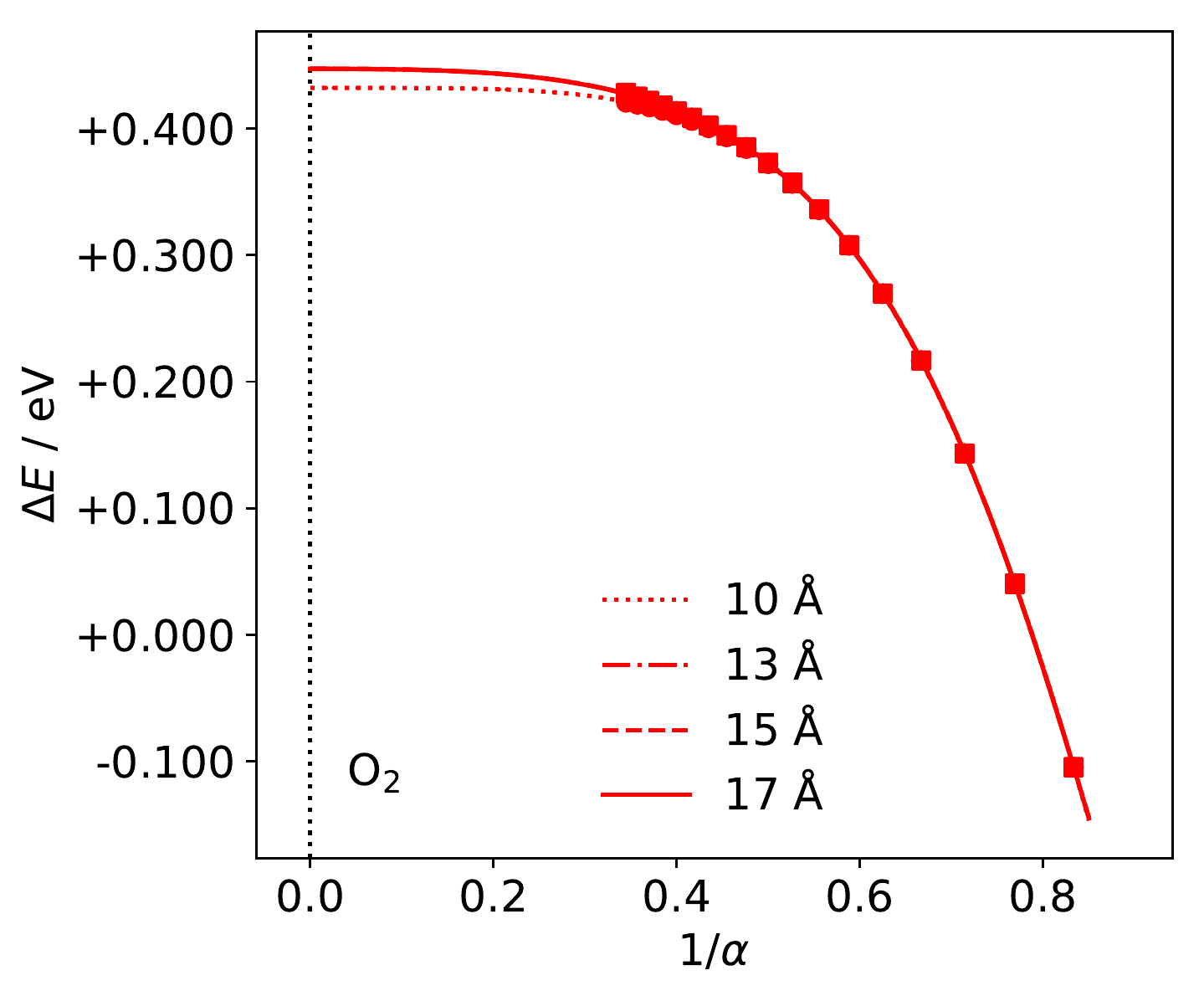}\includegraphics[width=0.26\columnwidth]{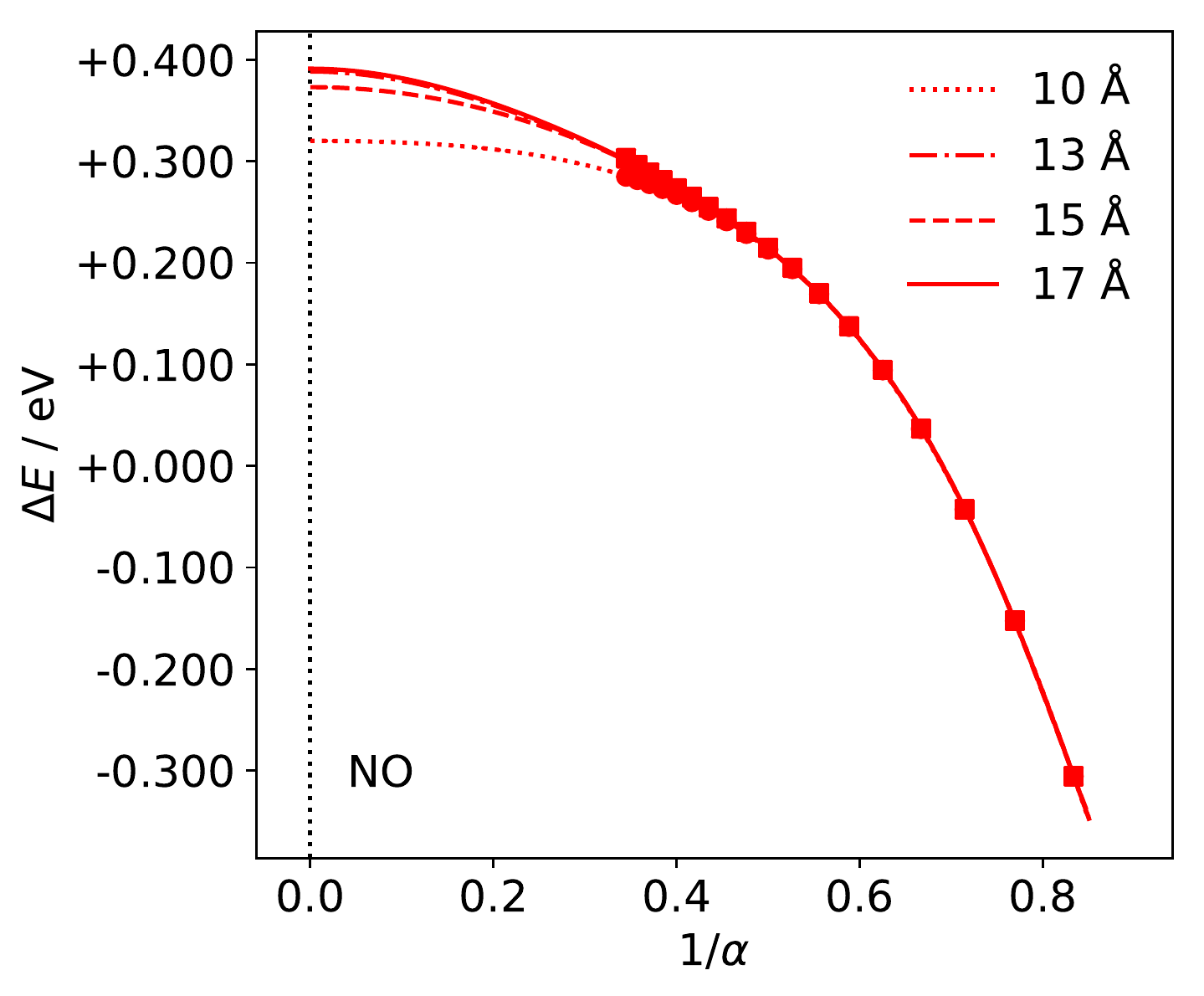}\includegraphics[width=0.26\columnwidth]{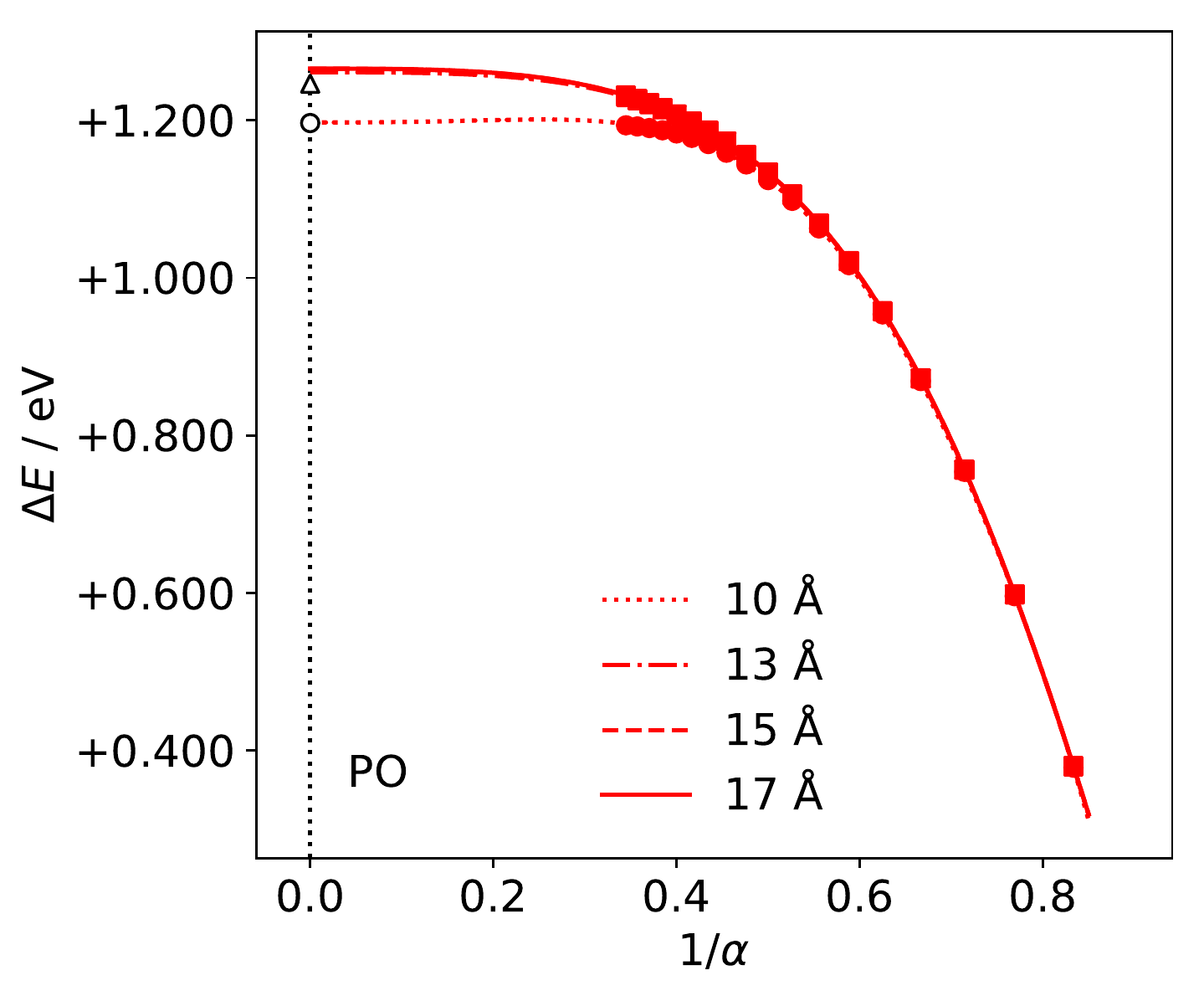} \\
\includegraphics[width=0.26\columnwidth]{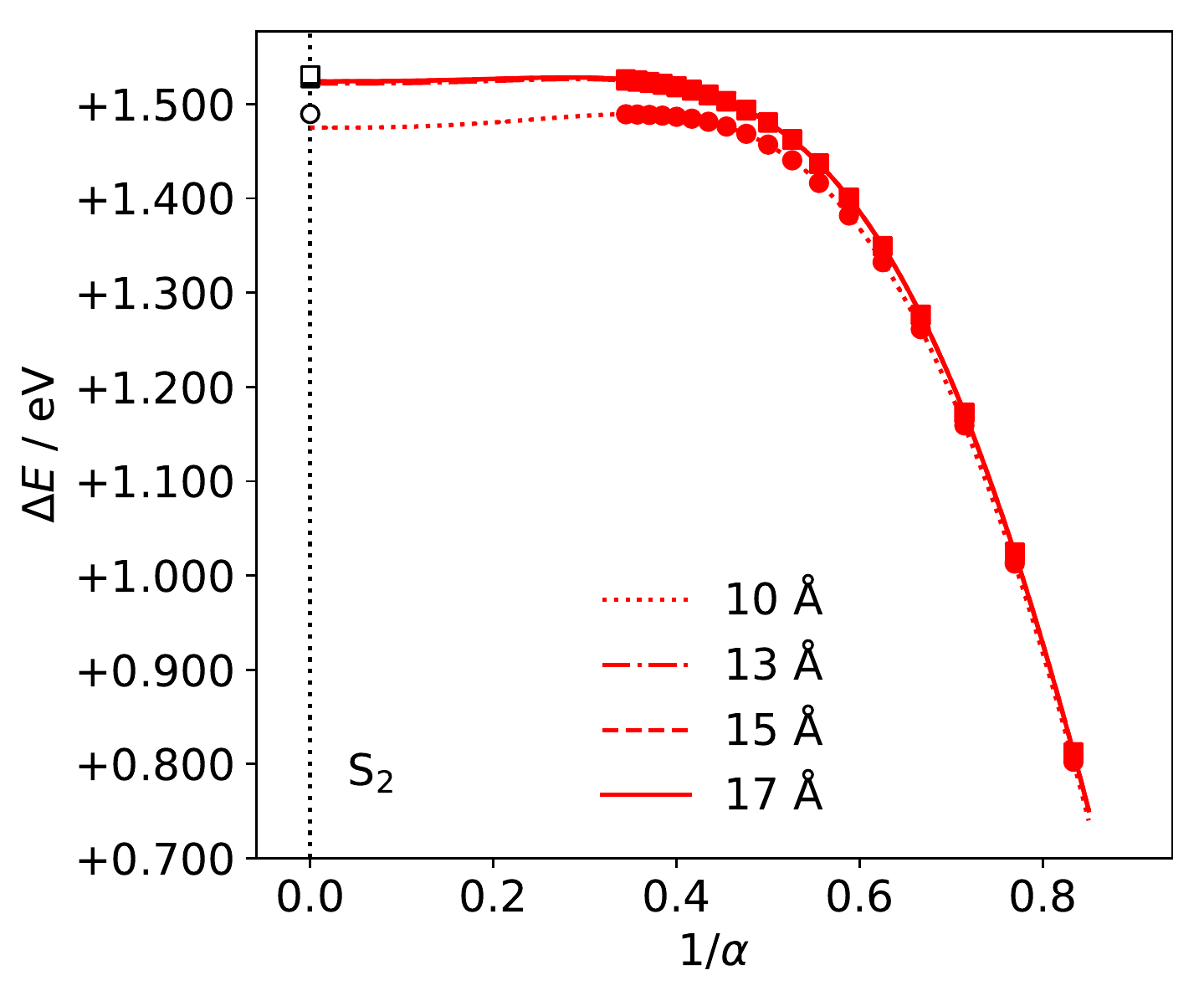}\includegraphics[width=0.26\columnwidth]{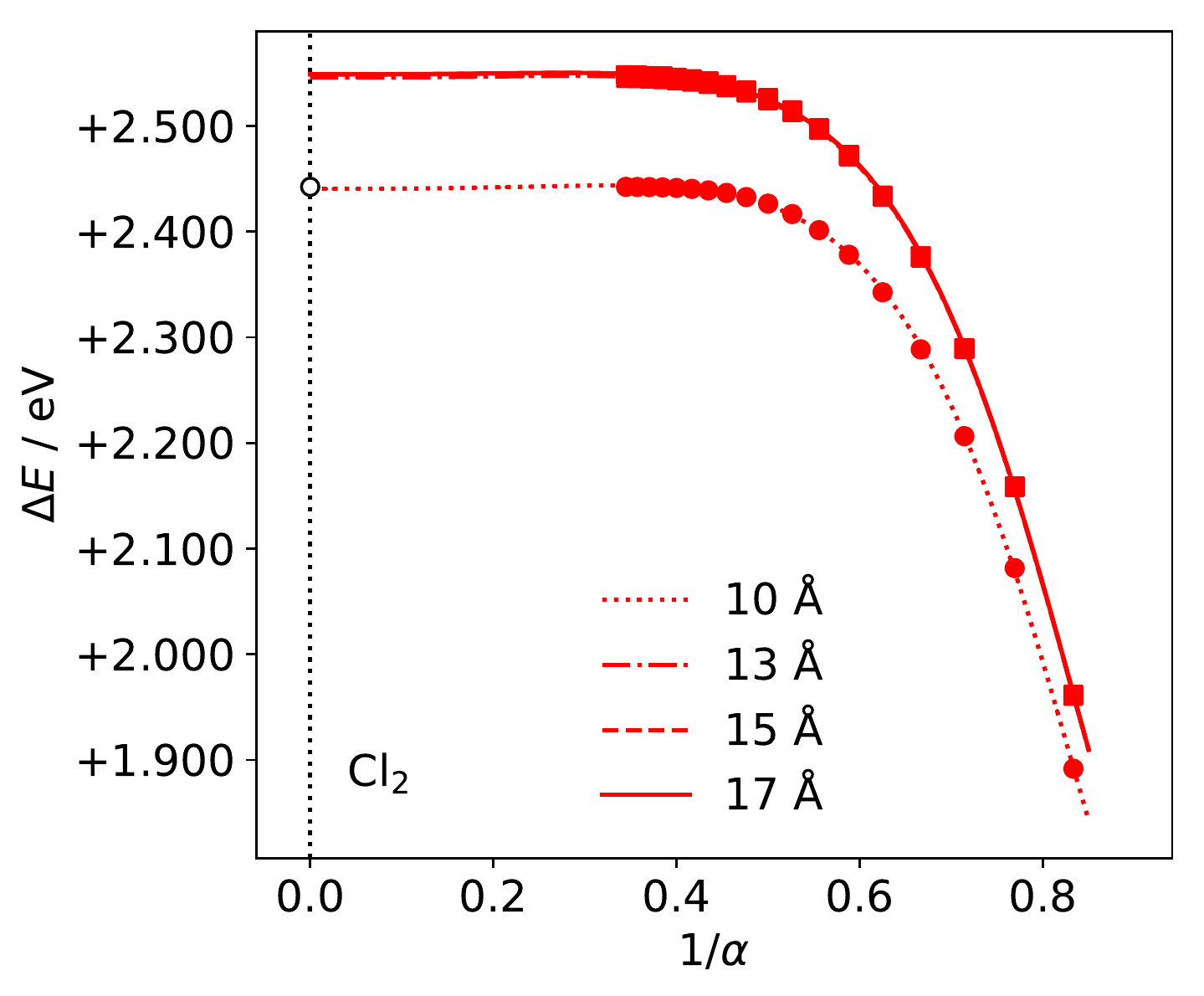}\includegraphics[width=0.26\columnwidth]{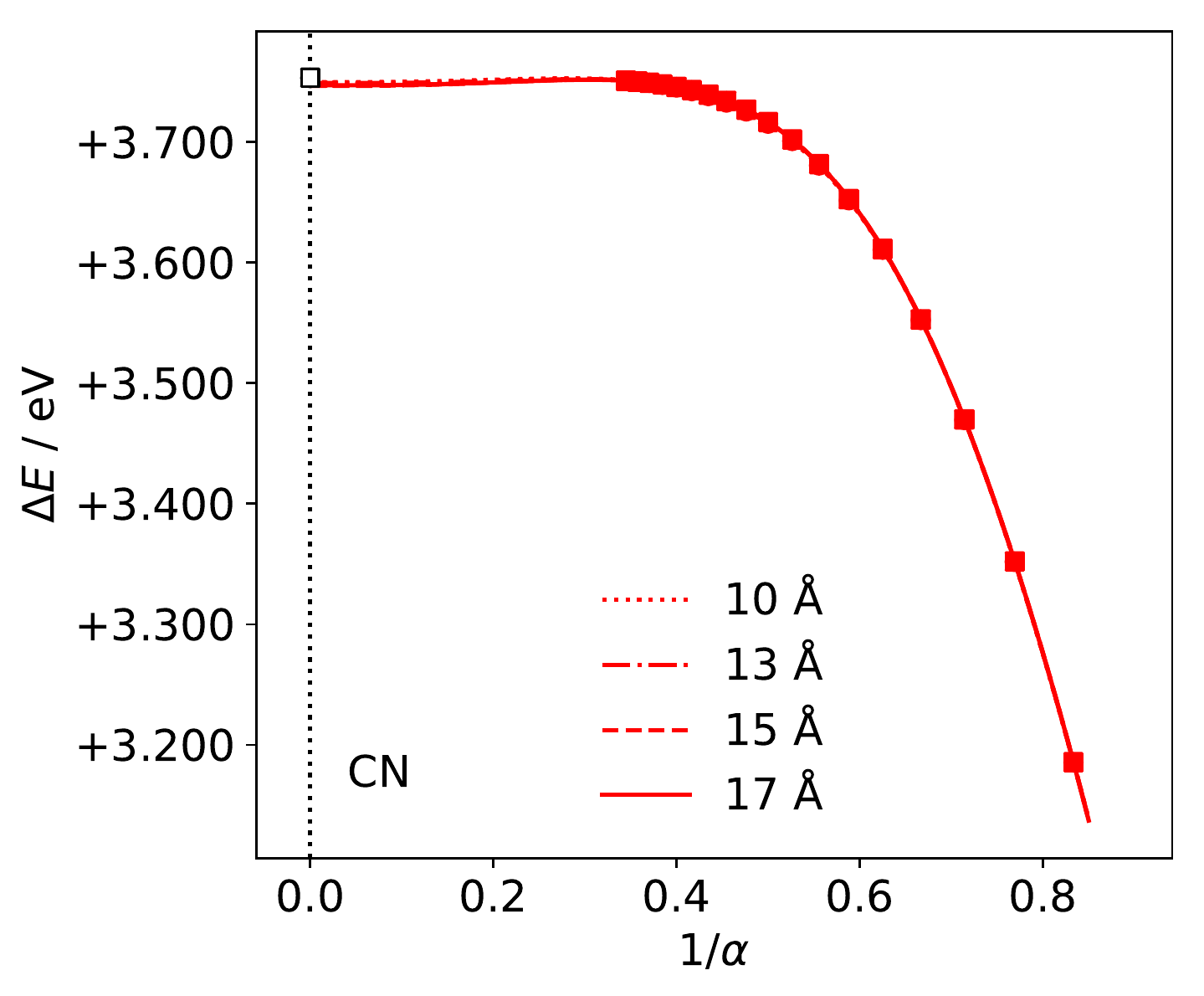} \\
\end{centering}
\caption{$\Delta E$ extrapolation for the species of the G2-1 dataset, using the confining-potential extrapolation towards large cavity sizes. Lines represent constrained 12th-degree polynomial functions (ridge-regression regularization has been employed to allow stable extrapolations).}
\label{fig:extrapolation-confining-radius}
\end{figure}

Figures \ref{fig:extrapolation-eps-sccs}--\ref{fig:extrapolation-confining-radius} illustrate how the vacuum total energy differences between the neutral and the anionic species of the G2-1 datasets have been extrapolated from the different types of embedding. In particular, Figure \ref{fig:extrapolation-eps-sccs} and Figure \ref{fig:extrapolation-eps-sscs} present results concerning the dielectric extrapolation, using the SCCS and the SSCS cavity, respectively. Polynomial functions have been used to fit and extrapolate the data to $\varepsilon_0 = 1$. Figure \ref{fig:extrapolation-confining-energy} and Figure \ref{fig:extrapolation-confining-radius} show instead results concerning the confining potential extrapolation, using the zero-potential and the large-cavity limit, respectively. Also here, polynomial functions have been used to fit the data and to extrapolate to vacuum conditions, i.e. $\kappa = 0$ Ry and $1/\alpha = 0$. For the large-cavity extrapolation, we have employed a higher-order polynomial where, however, we have constrained the values of the first- and second-order coefficients to be equal to and lower than zero, respectively, in order to force zero first derivative and negative second derivative of the function at $1/\alpha = 0$. These constraints guarantee that the extrapolating function approaches the large-cavity limit from below with a horizontal tangent at $1/\alpha = 0$. In addition, we have made use of ridge-regression regularization in order to avoid oscillations and to obtain smooth functions of $1/\alpha$ that could be reliably extrapolated to $1/\alpha = 0$.  

Figures \ref{fig:extrapolation-eps-sccs}--\ref{fig:extrapolation-confining-radius} also illustrate the role of the cavity size on the computed $\Delta E$ and $\Delta E'$ values. They show that converged values (within less than 15 meV) can be obtained using a cubic cell with a side length of 13 \AA . 

\begin{figure}
\begin{centering}
 \includegraphics[width=0.26\columnwidth]{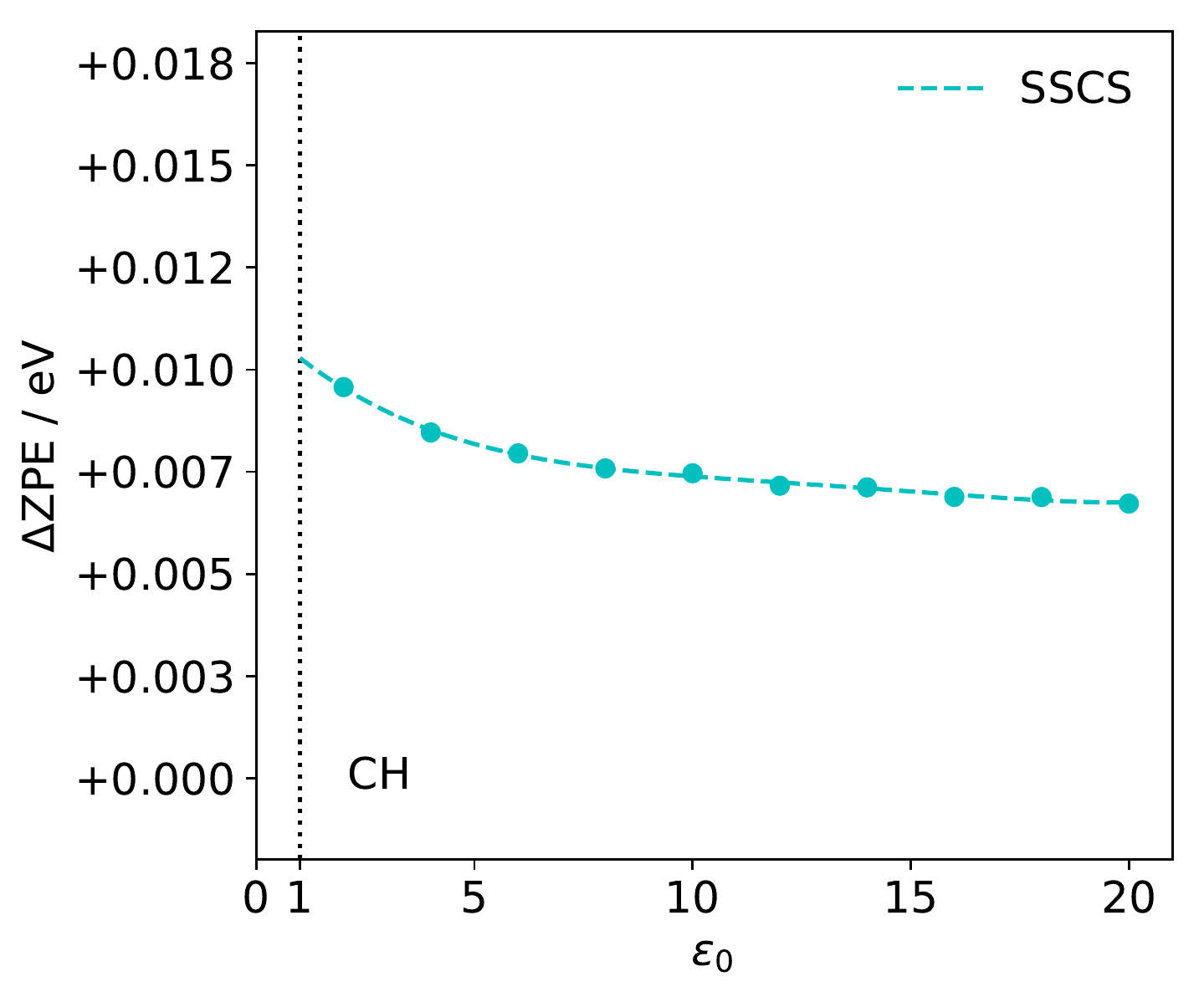}\includegraphics[width=0.26\columnwidth]{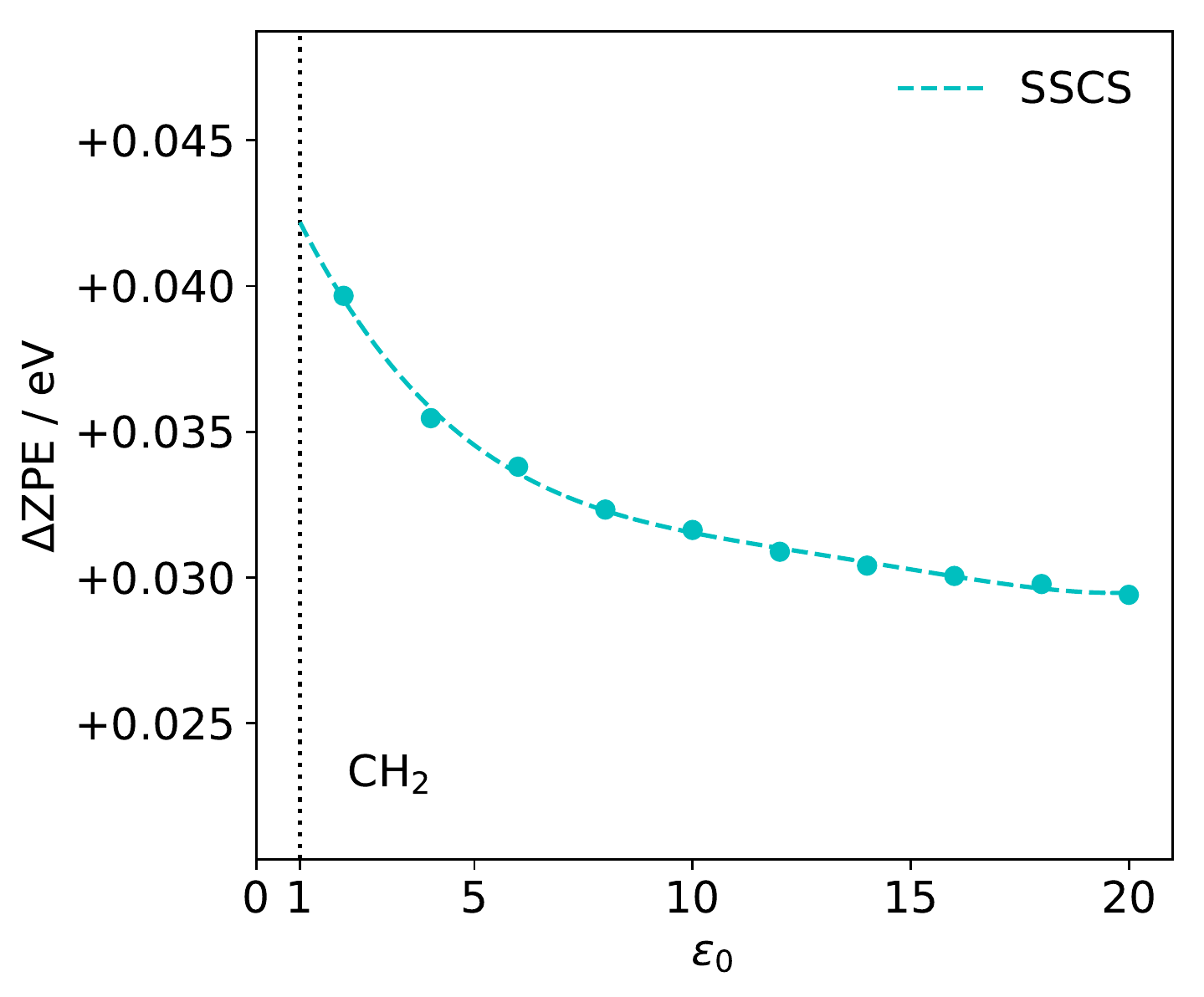}\includegraphics[width=0.26\columnwidth]{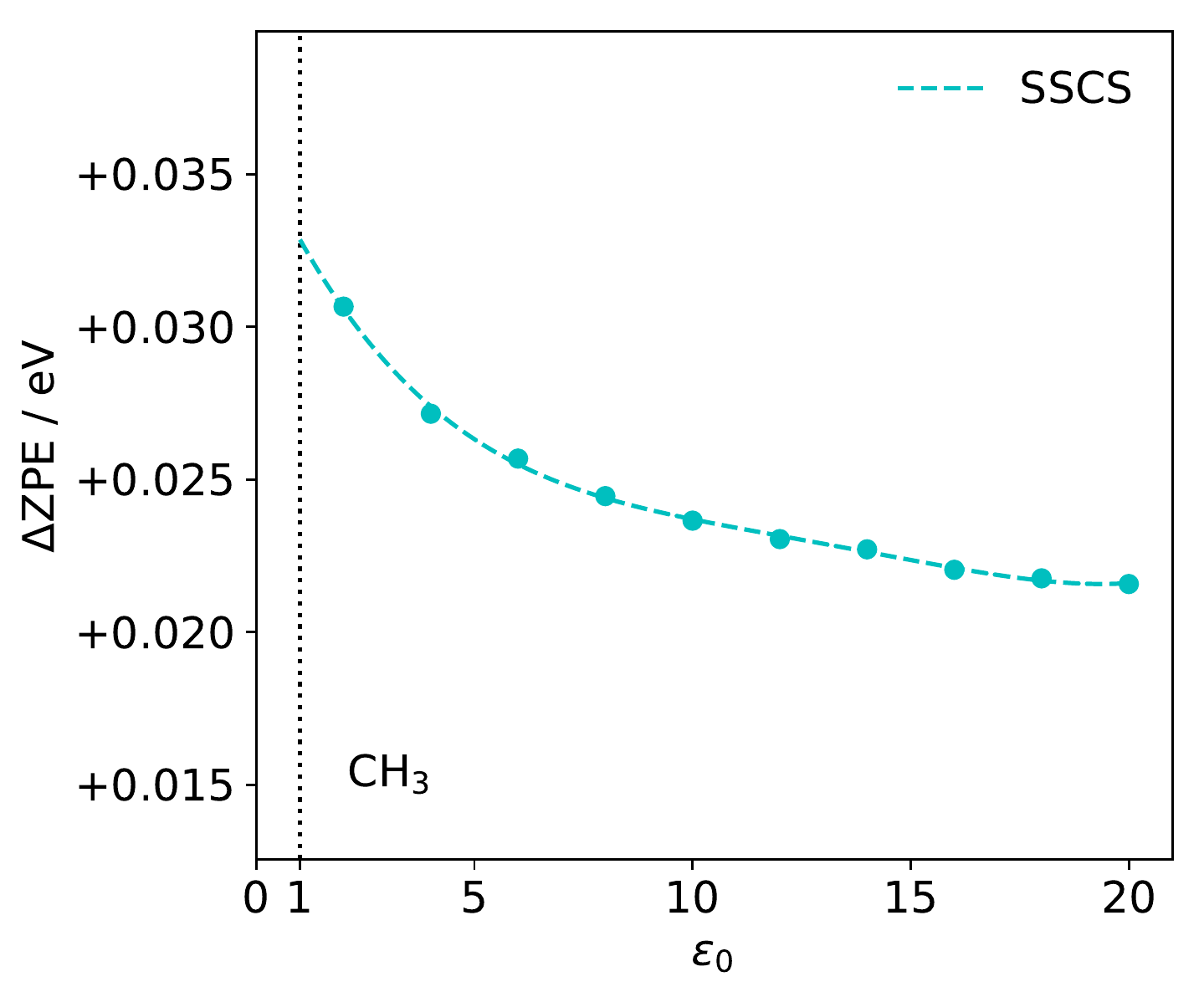} \\
\includegraphics[width=0.26\columnwidth]{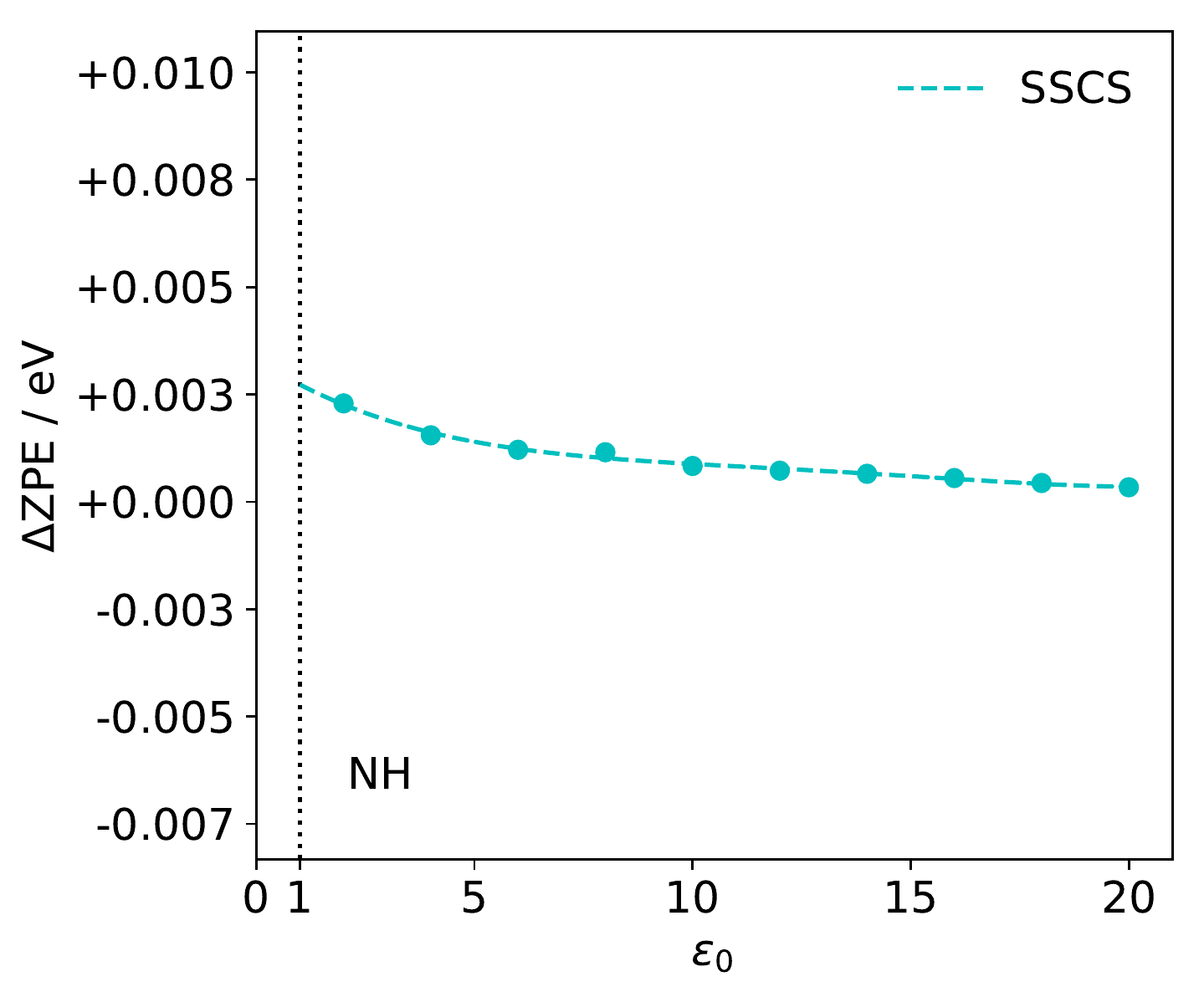}\includegraphics[width=0.26\columnwidth]{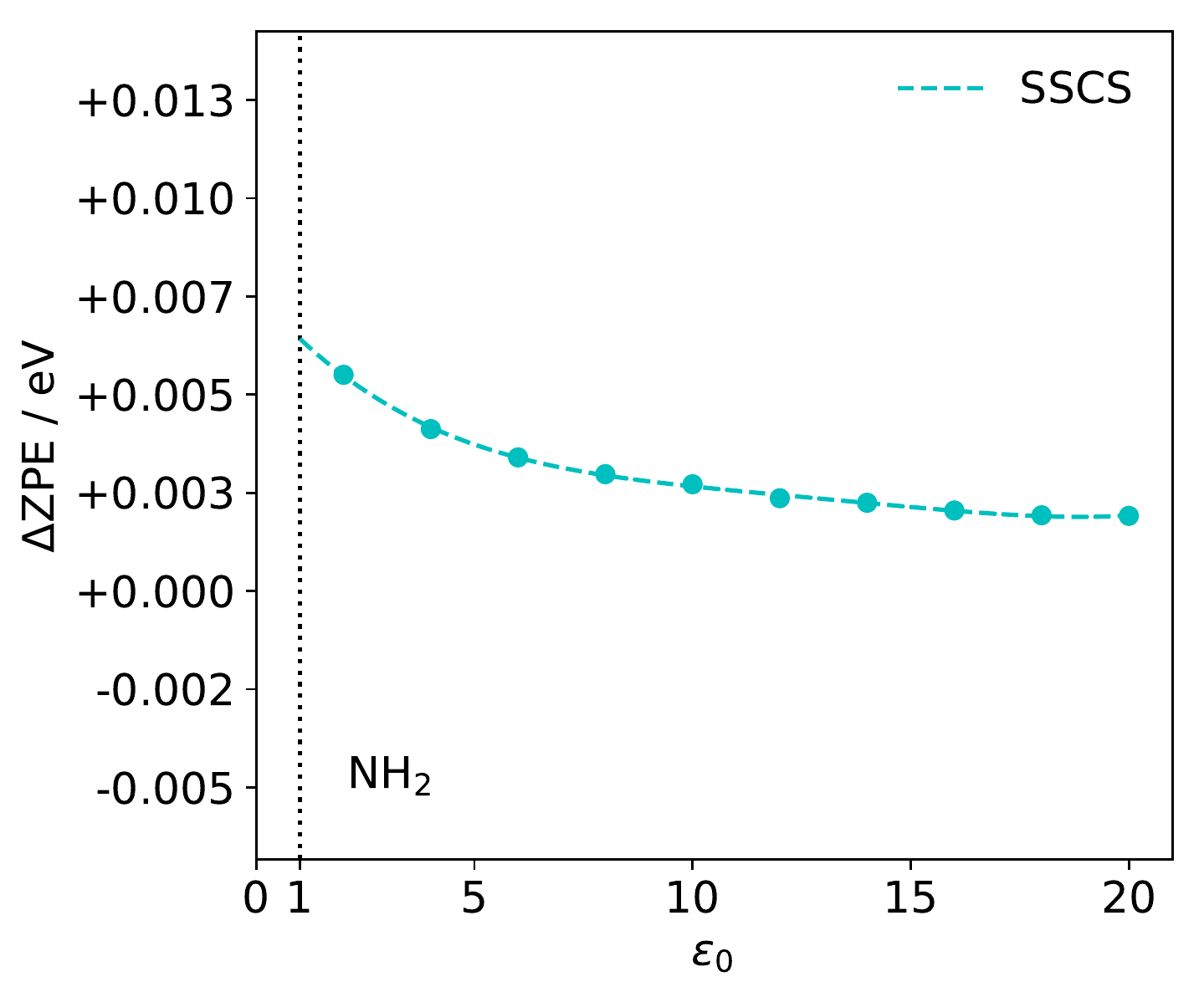}\includegraphics[width=0.26\columnwidth]{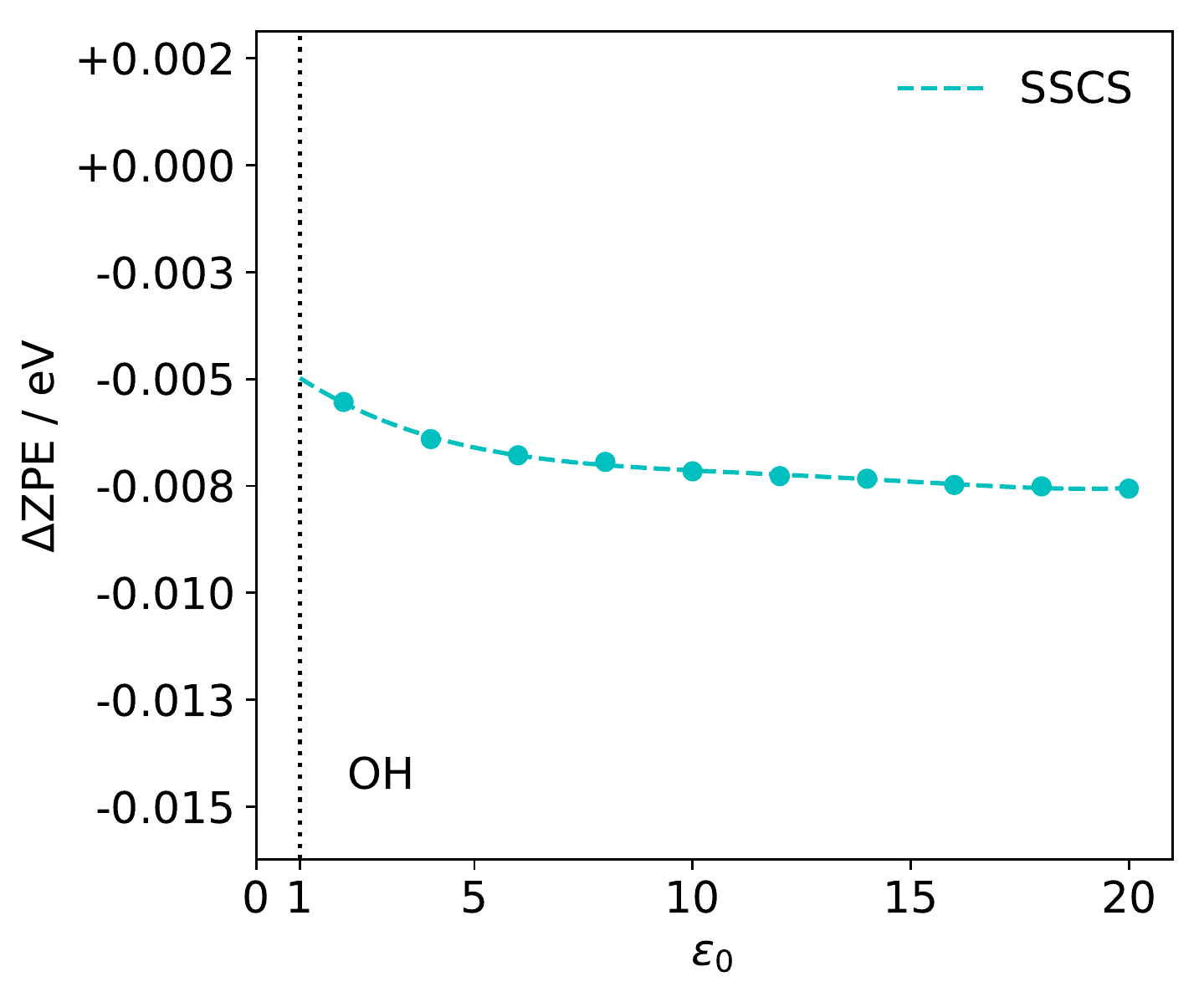} \\
\includegraphics[width=0.26\columnwidth]{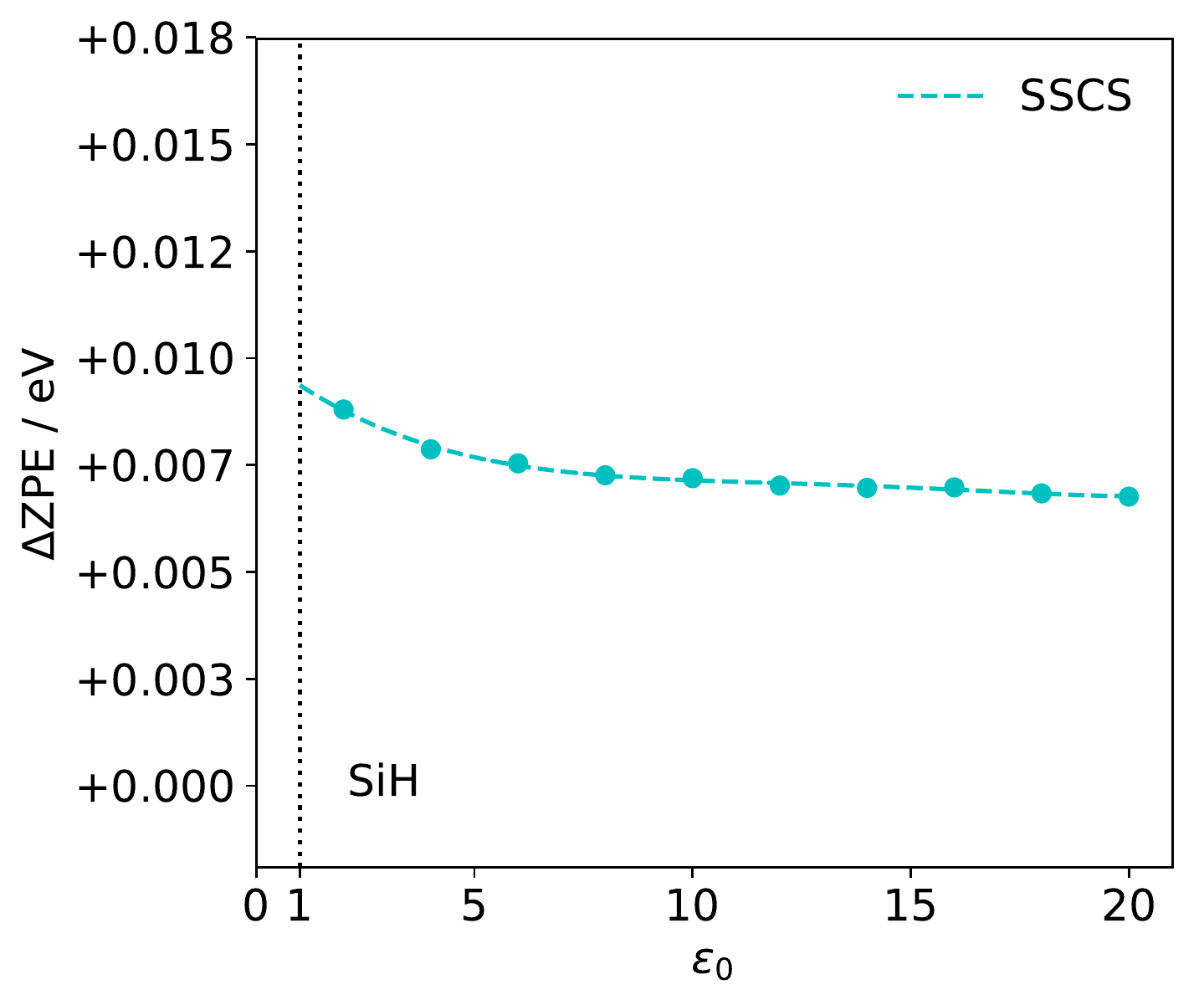}\includegraphics[width=0.26\columnwidth]{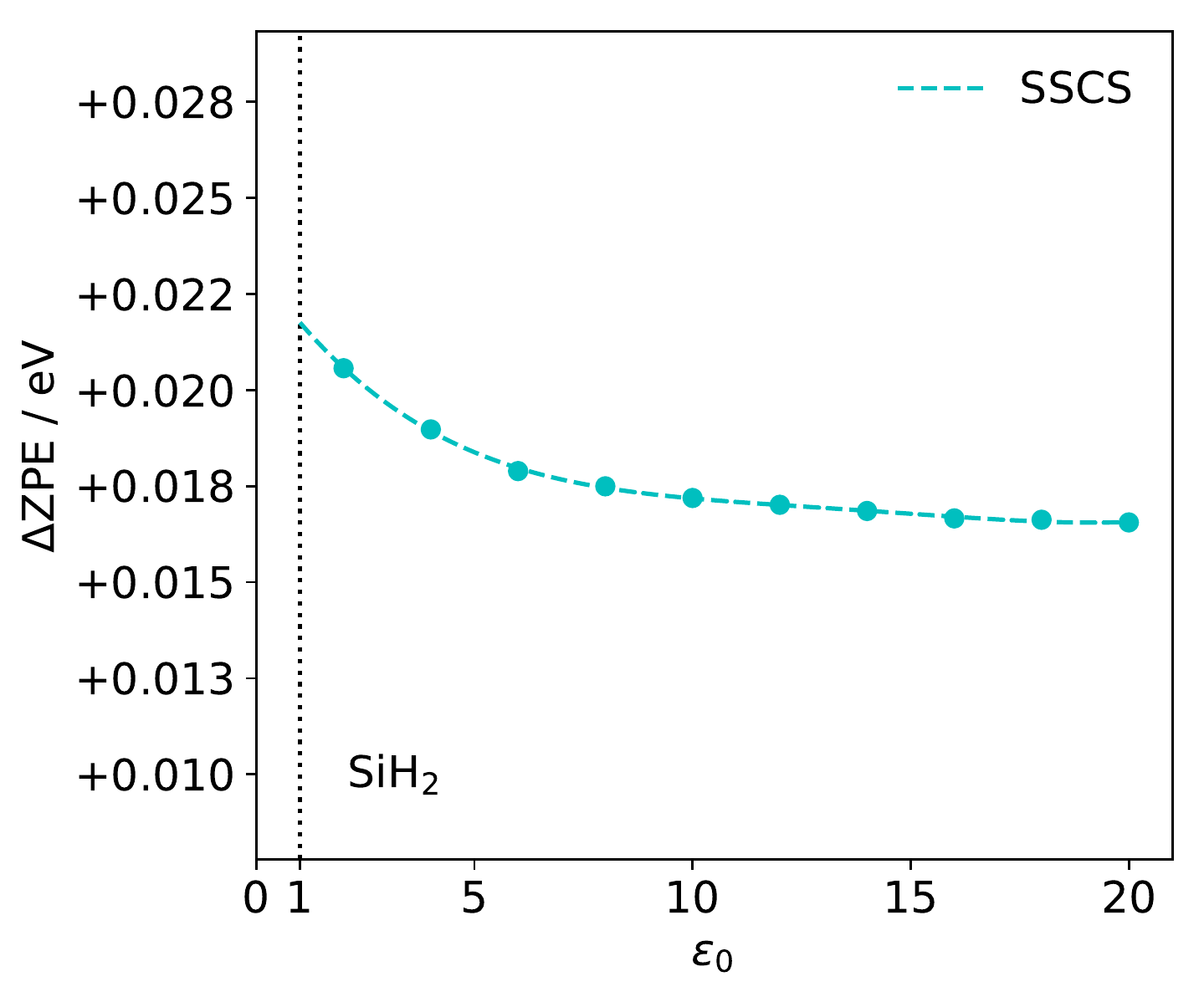}\includegraphics[width=0.26\columnwidth]{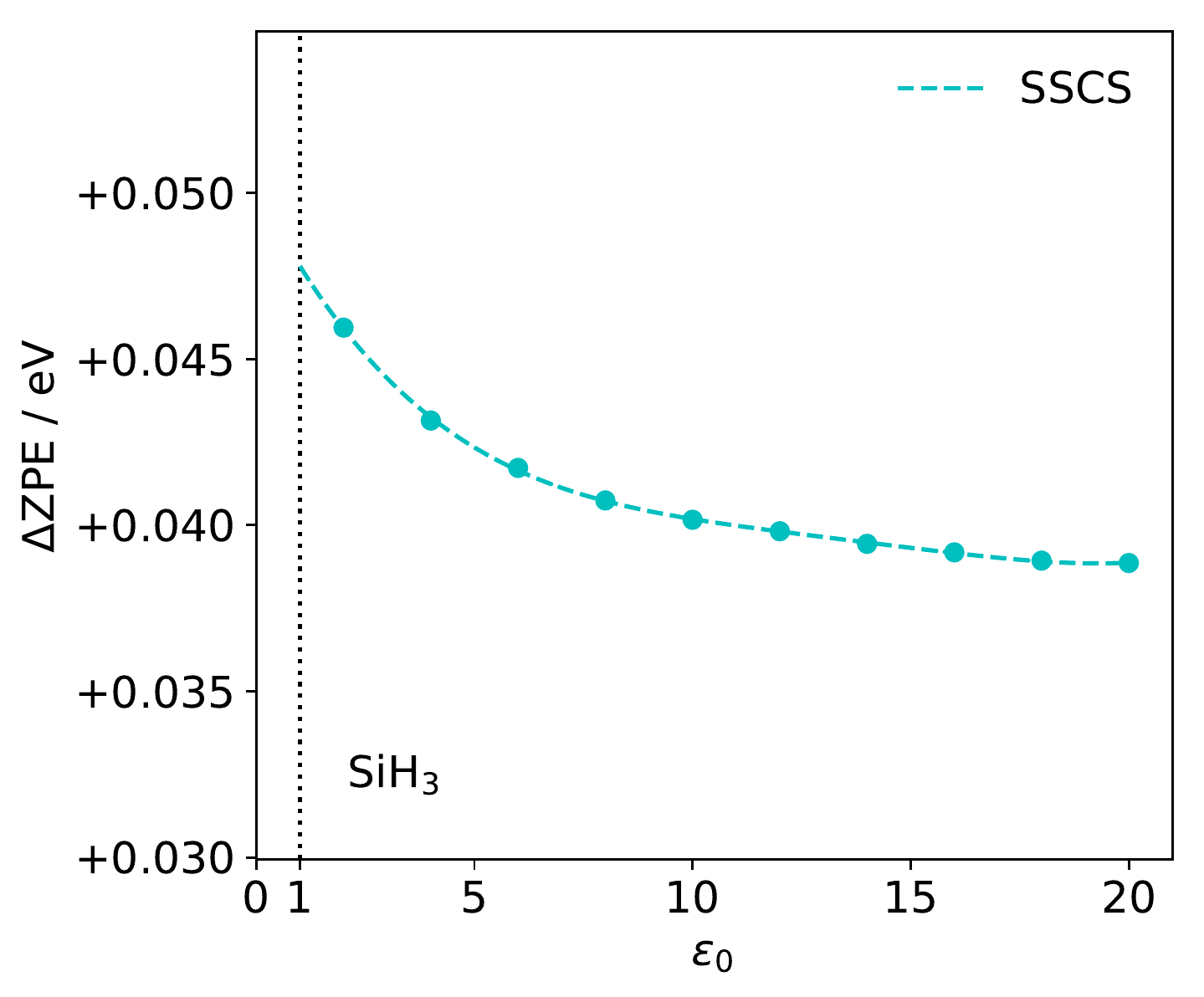} \\
\includegraphics[width=0.26\columnwidth]{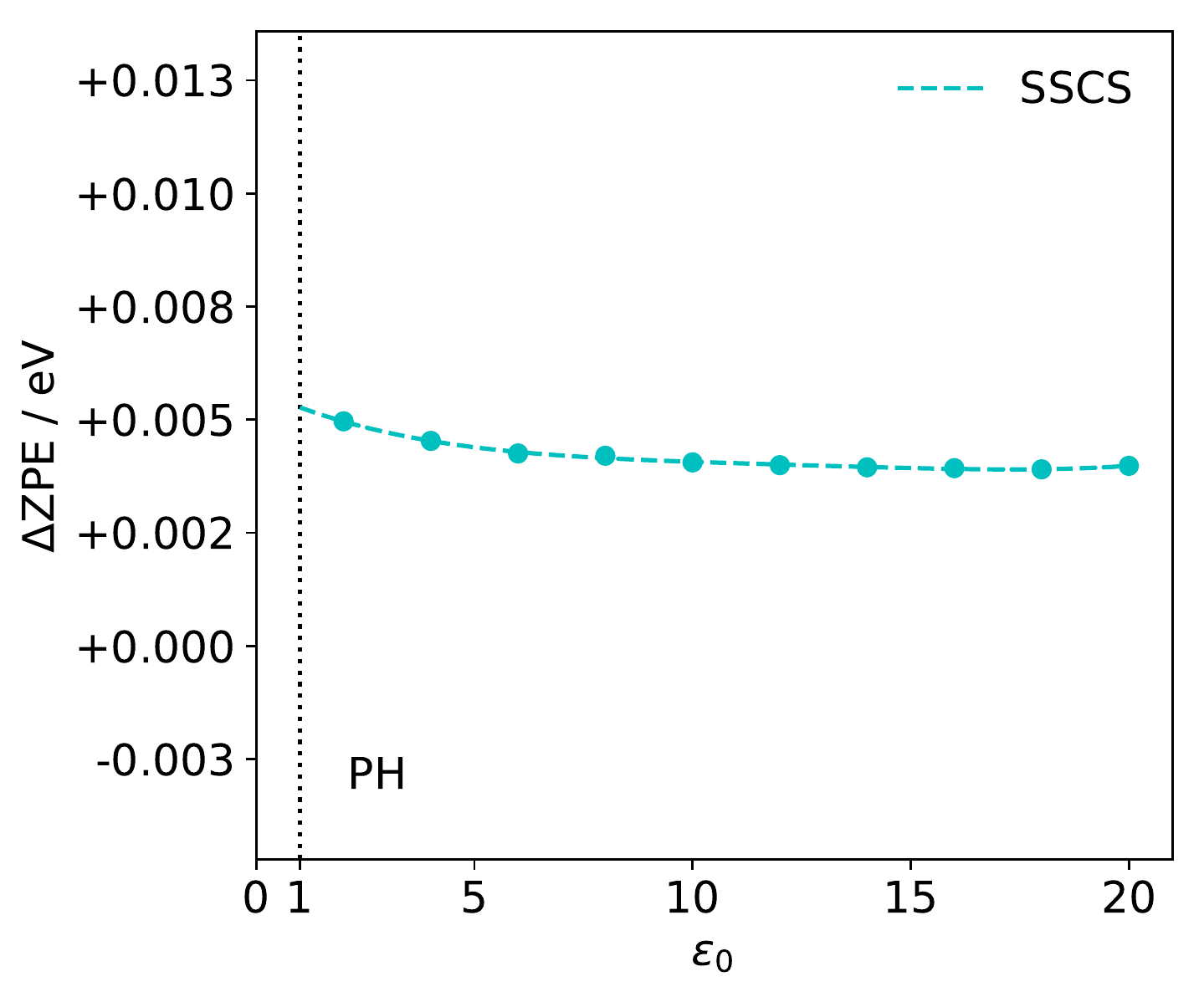}\includegraphics[width=0.26\columnwidth]{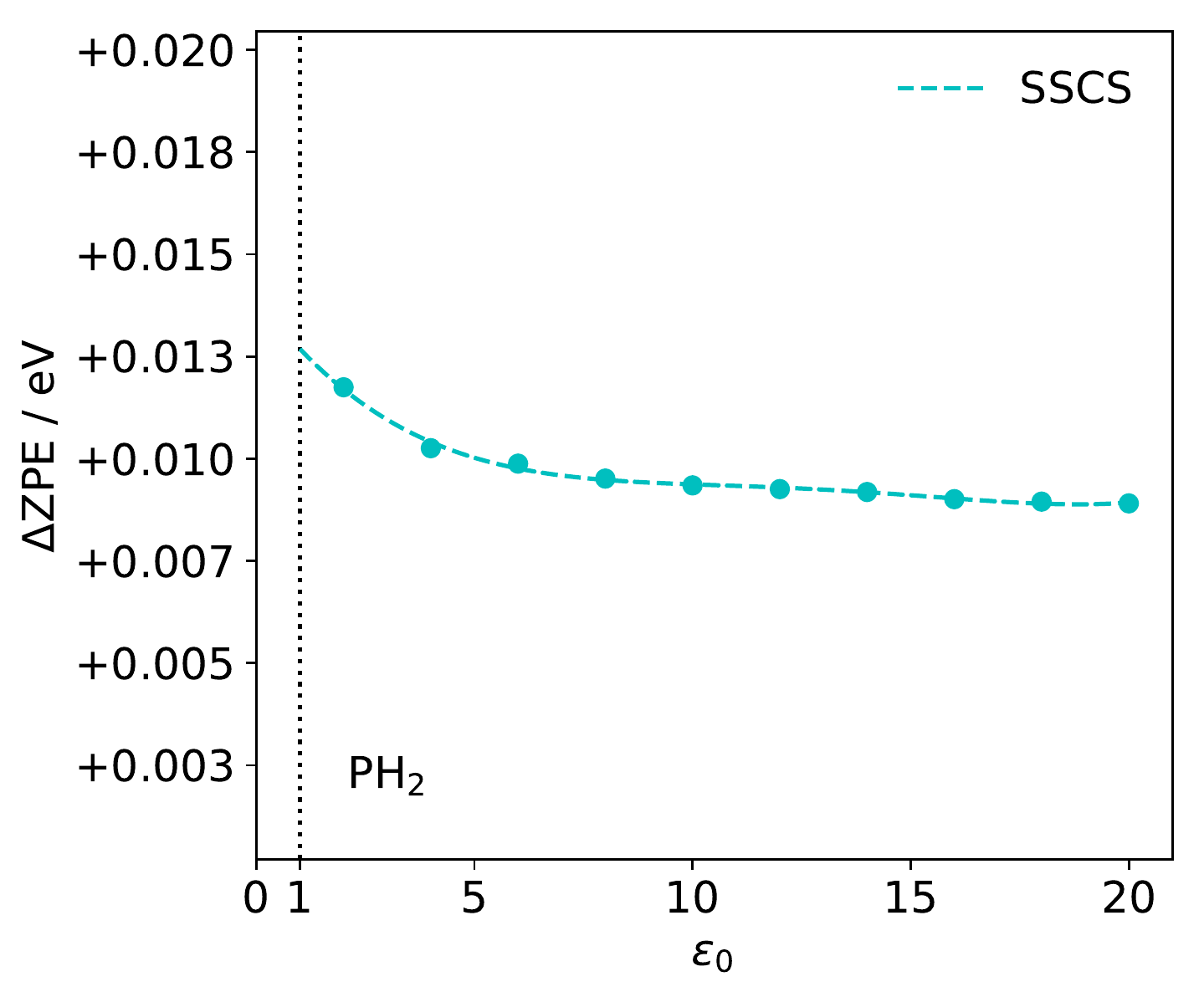}\includegraphics[width=0.26\columnwidth]{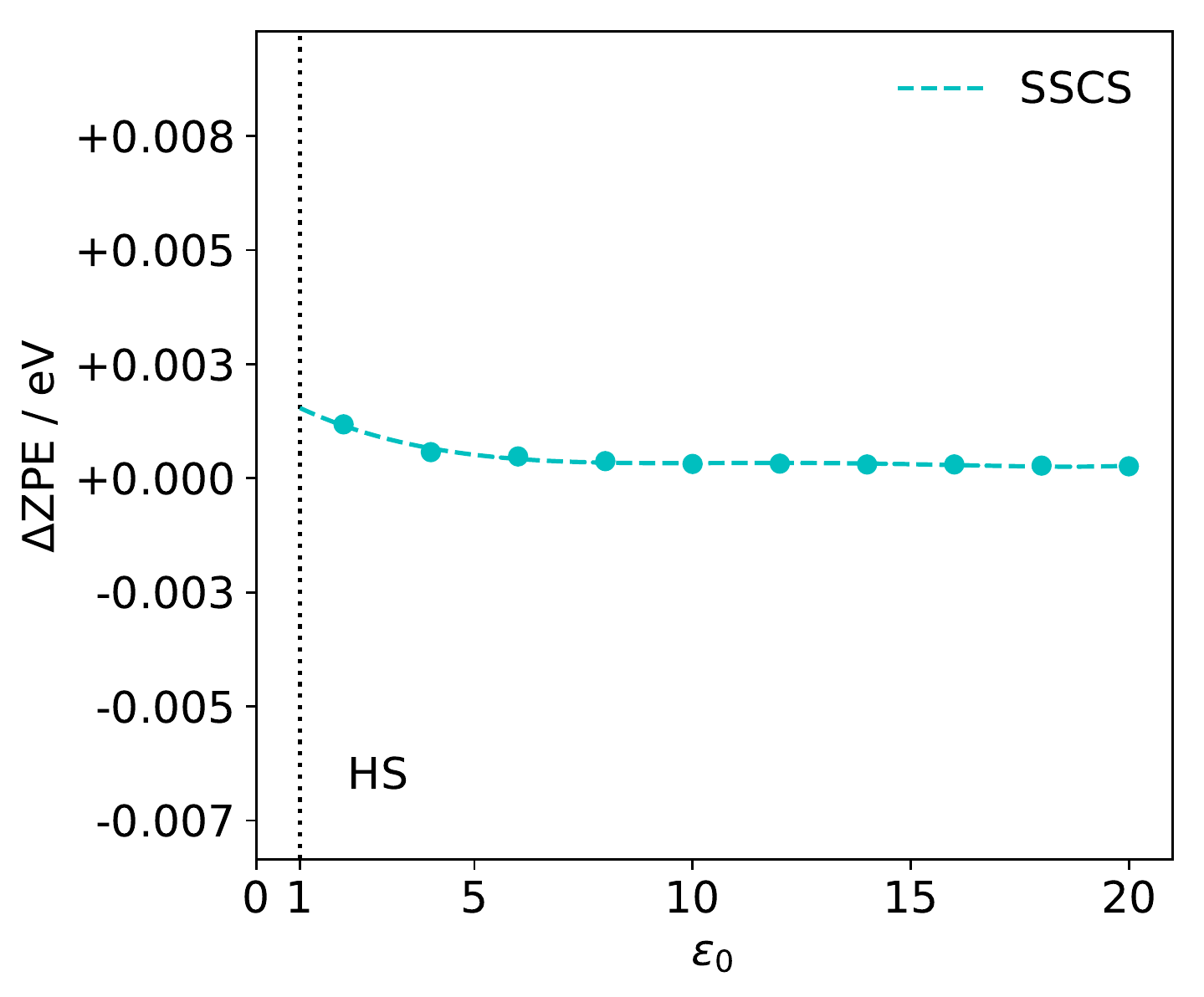} \\
\includegraphics[width=0.26\columnwidth]{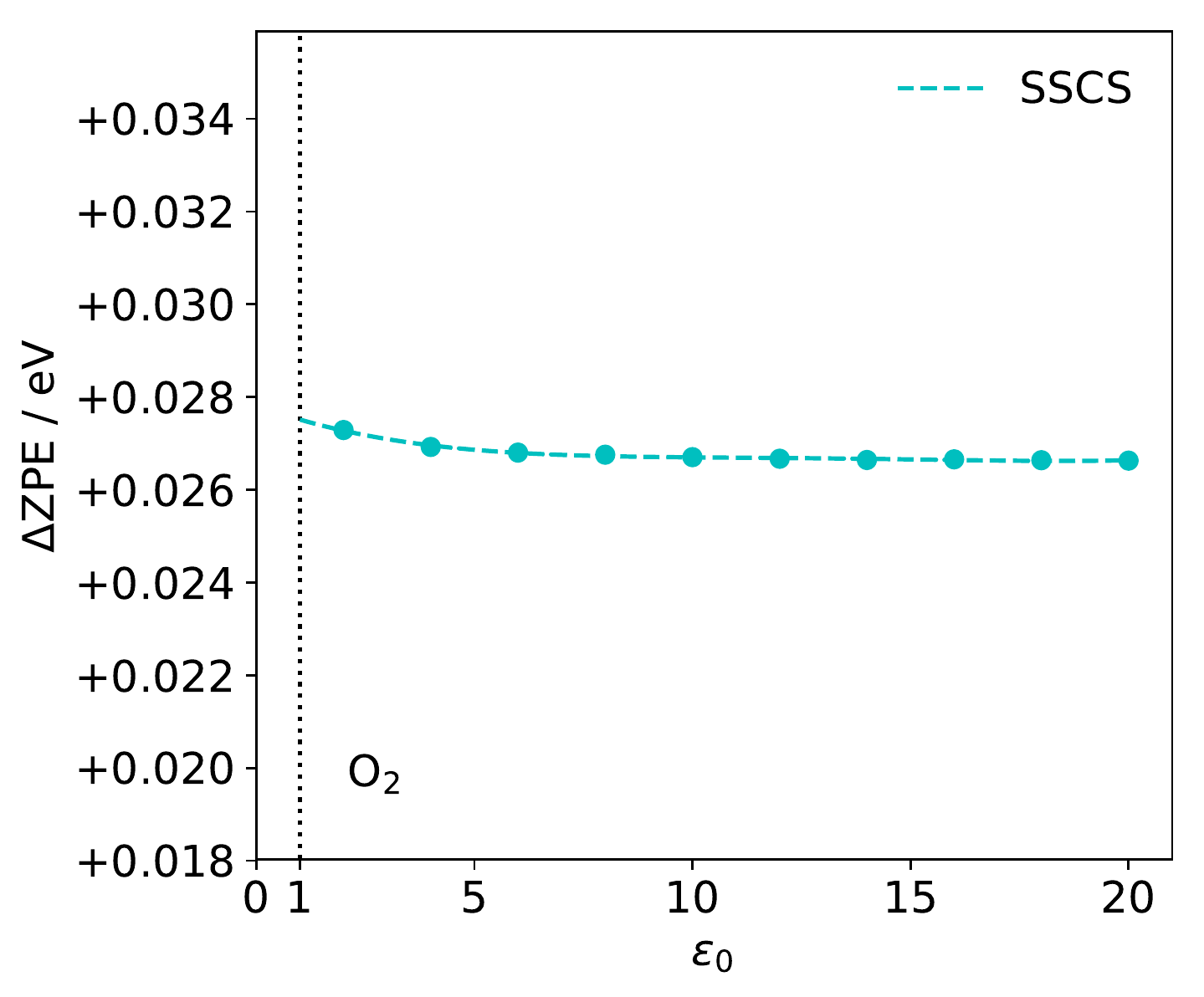}\includegraphics[width=0.26\columnwidth]{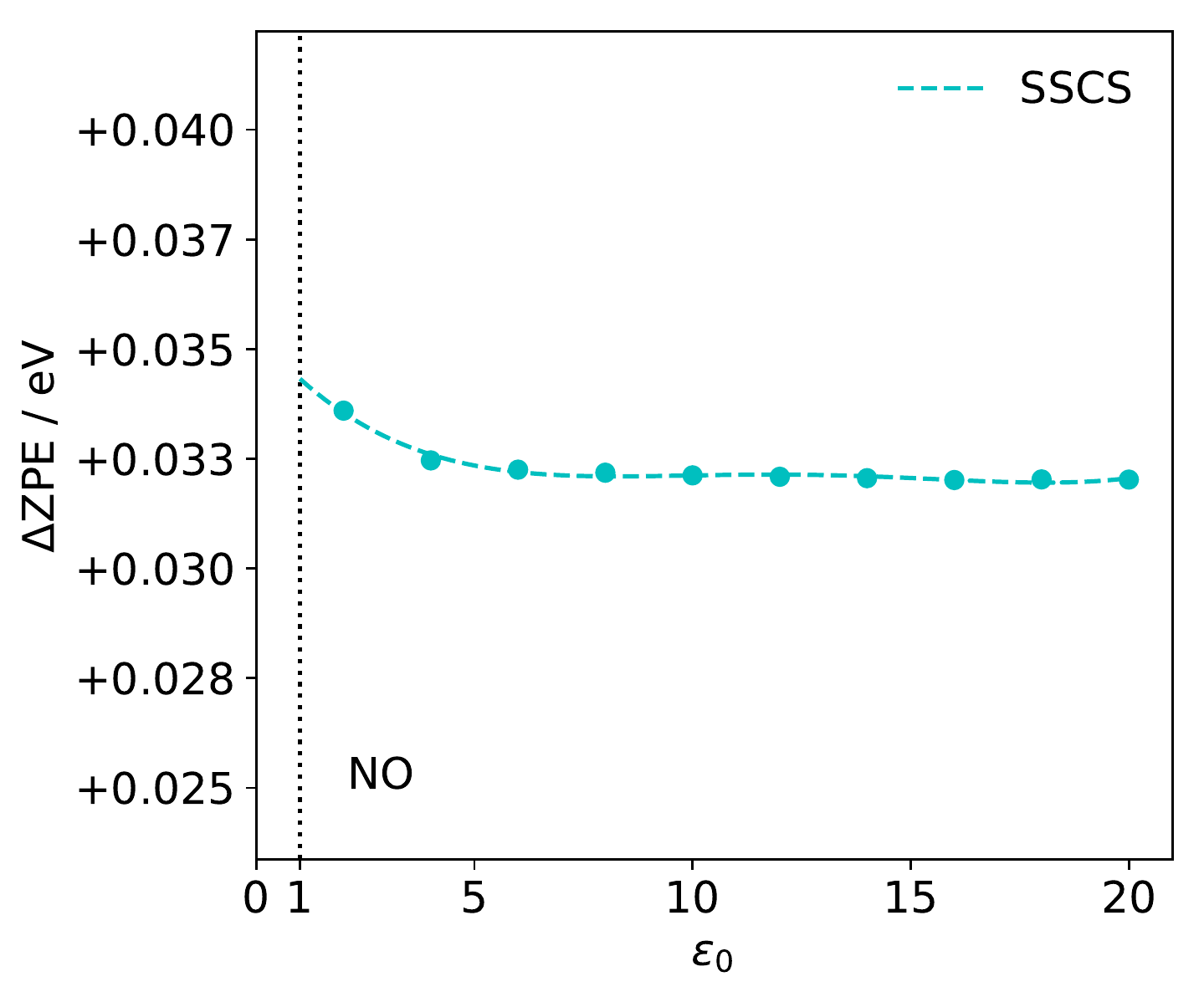}\includegraphics[width=0.26\columnwidth]{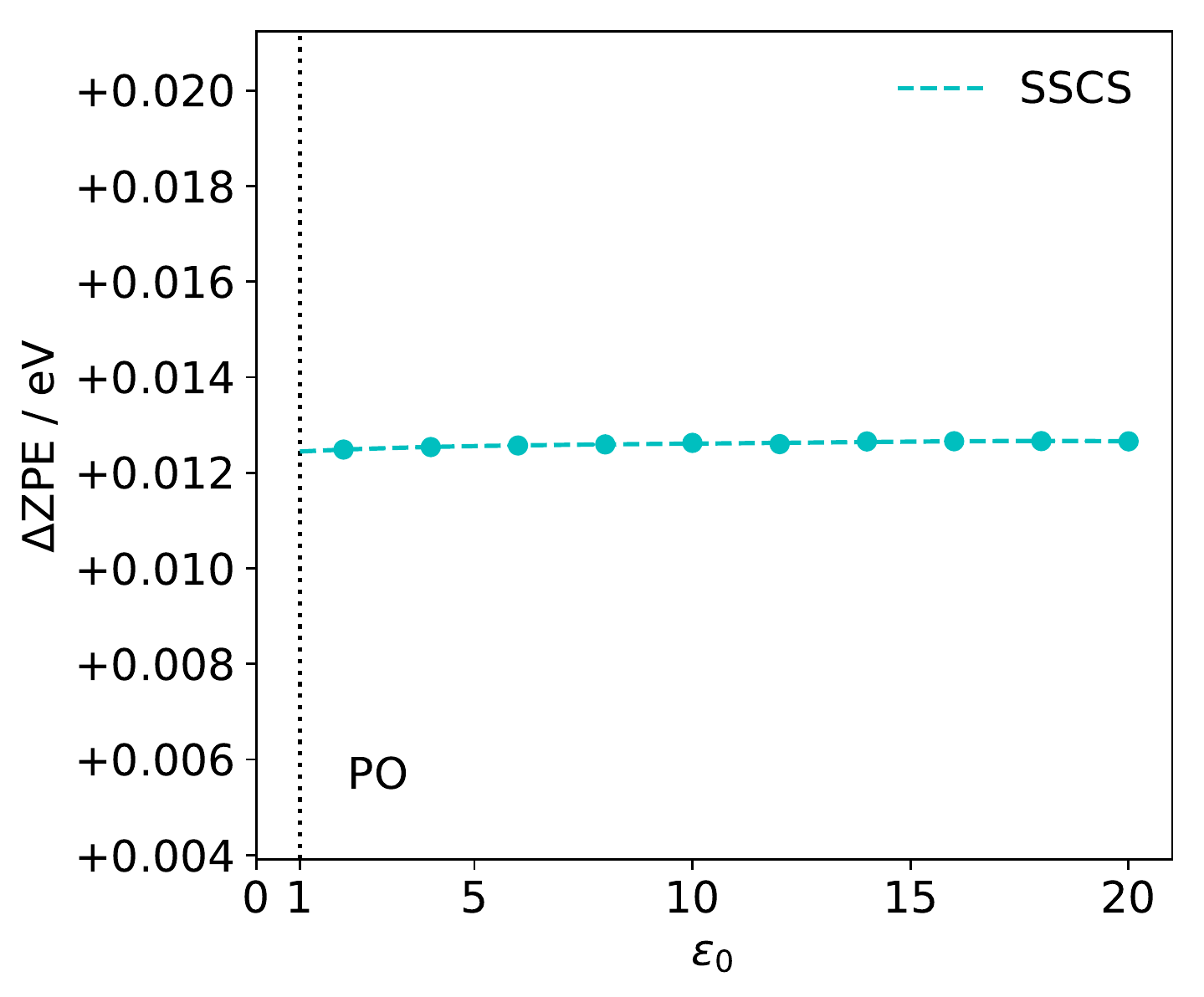} \\
\includegraphics[width=0.26\columnwidth]{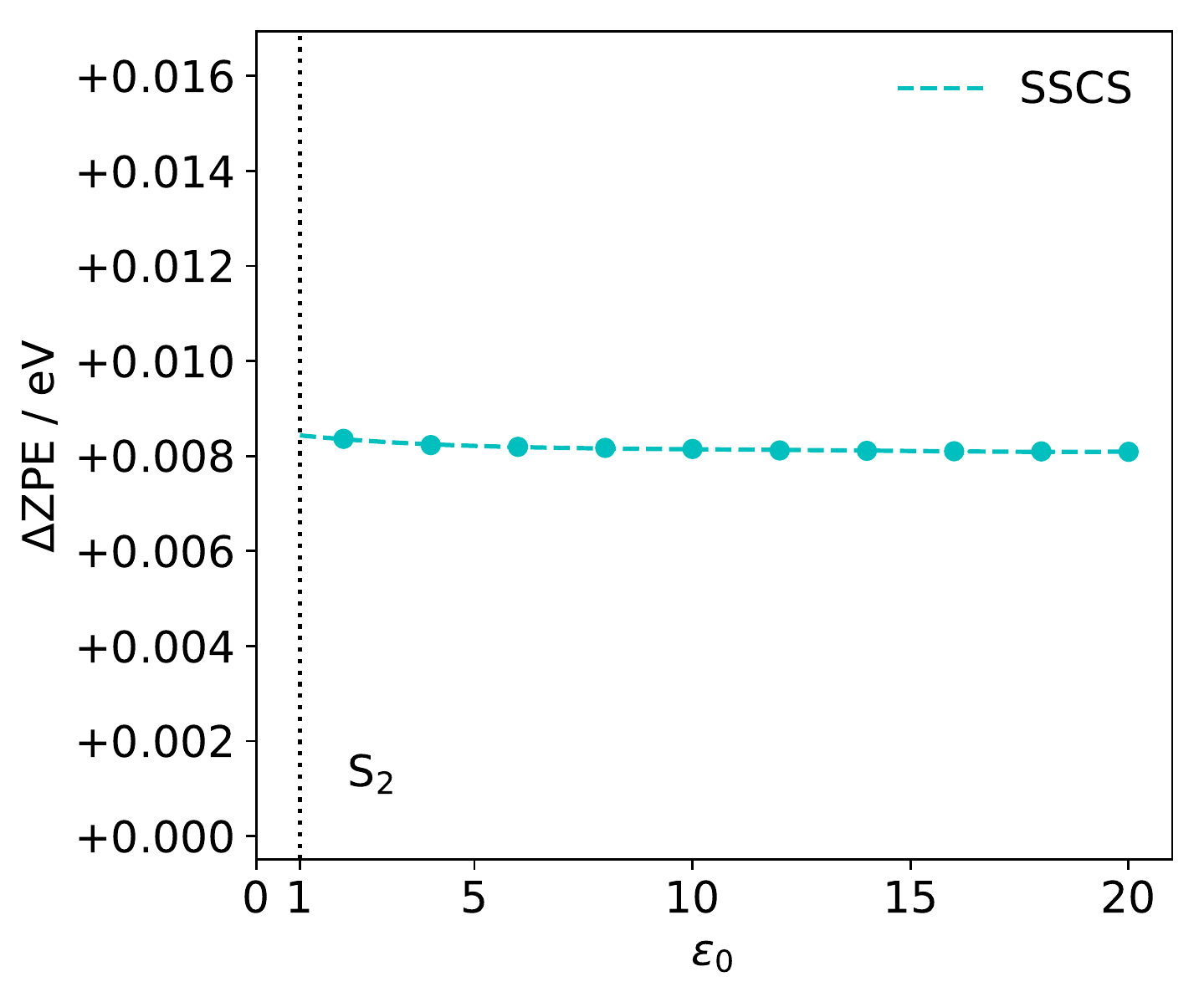}\includegraphics[width=0.26\columnwidth]{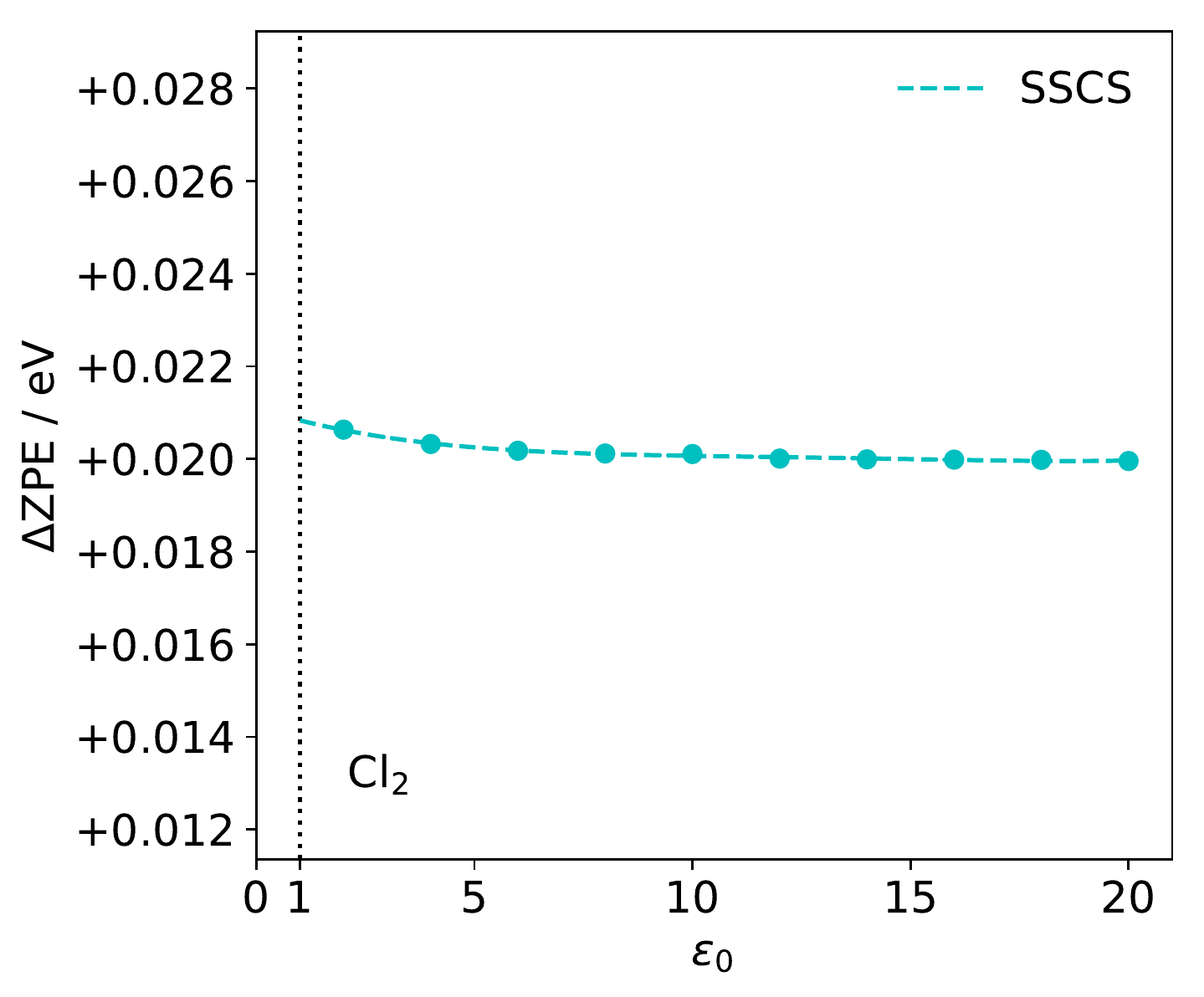}\includegraphics[width=0.26\columnwidth]{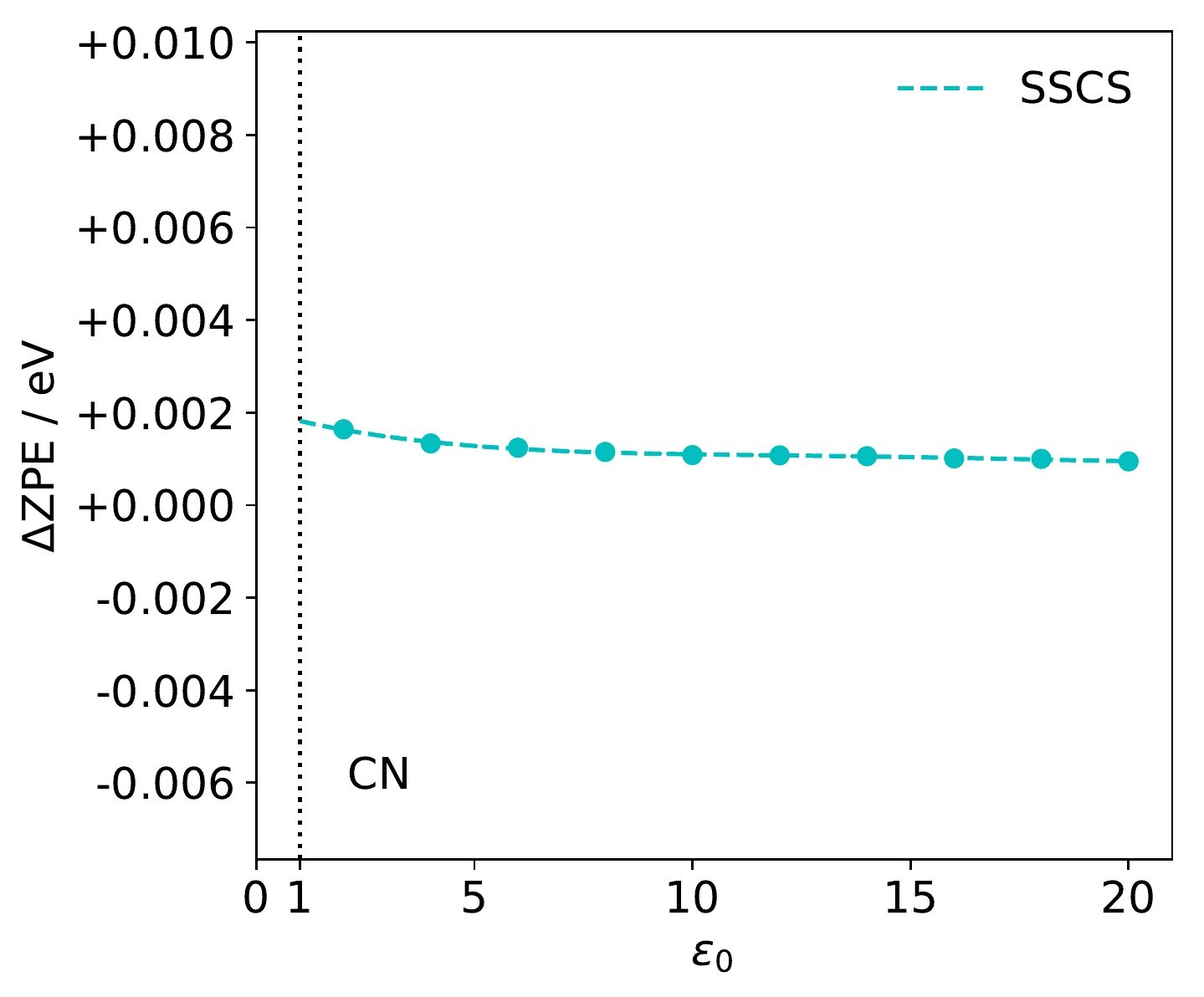} \\
\end{centering}
\caption{$\Delta$ZPE extrapolation for the species of the G2-1 dataset, using the dielectric extrapolation and the SSCS cavity. A cubic cell with a 13 \AA\ long side has been employed for all species. Lines represent 4th-degree polynomial functions.}
\label{fig:extrapolation-zpecorr}
\end{figure}

Figure \ref{fig:extrapolation-zpecorr} shows how the vacuum zero-point energy (ZPE) corrections $\Delta$ZPE have been obtained. The dielectric extrapolation technique, in combination with the SSCS cavity has been employed for this purpose . 

Finally, the computed electron affinity (EA) values for the various extrapolation methods, together with the corresponding ZPE corrections, are presented in Table \ref{tab:results}.

\begin{table}
\begin{centering}
{\renewcommand{\arraystretch}{1.5}
\begin{tabularx}{\textwidth}{@{}l *5{>{\centering\arraybackslash}X}@{}}
\hline
\hline
Molecule	&  $\varepsilon_0\rightarrow 1$, SCCS &  $\varepsilon_0\rightarrow 1$, SSCS & $\kappa\rightarrow 0$ & $1/\alpha\rightarrow 0$ & $\Delta$ZPE \tabularnewline
\hline 
CH         &  1.544  &  1.544  &  1.540  &  1.580  &  0.010  \tabularnewline
CH$_2$     &  0.780  &  0.782  &  0.773  &  0.815  &  0.042  \tabularnewline
CH$_3$     &  0.184  &  0.183  &  0.160  &  0.260  &  0.033  \tabularnewline
NH         &  0.547  &  0.542  &  0.511  &  0.596  &  0.003  \tabularnewline
NH$_2$     &  0.839  &  0.839  &  0.836  &  0.884  &  0.006  \tabularnewline
OH         &  1.987  &  1.990  &  1.991  &  2.019  & -0.005  \tabularnewline
SiH        &  1.374  &  1.375  &  1.368  &  1.402  &  0.009  \tabularnewline
SiH$_2$    &  1.257  &  1.259  &  1.252  &  1.274  &  0.022  \tabularnewline
SiH$_3$    &  1.398  &  1.399  &  1.392  &  1.400  &  0.048  \tabularnewline
PH         &  1.037  &  1.035  &  1.029  &  1.057  &  0.005  \tabularnewline
PH$_2$     &  1.225  &  1.225  &  1.217  &  1.244  &  0.013  \tabularnewline
HS         &  2.304  &  2.305  &  2.301  &  2.311  &  0.002  \tabularnewline
O$_2$      &  0.462  &  0.464  &  0.447  &  0.475  &  0.028  \tabularnewline
NO         &  0.367  &  0.363  &  0.314  &  0.423  &  0.034  \tabularnewline
PO         &  1.245  &  1.246  &  1.235  &  1.273  &  0.012  \tabularnewline
S$_2$      &  1.534  &  1.535  &  1.530  &  1.530  &  0.008  \tabularnewline
Cl$_2$     &  2.566  &  2.567  &  2.565  &  2.567  &  0.021  \tabularnewline
CN         &  3.750  &  3.753  &  3.752  &  3.751  &  0.002  \tabularnewline
\hline
\hline
\end{tabularx}
}
 \end{centering}
\caption{EA values (in eV) computed using the different extrapolation methods for the various species of the G2-1 dataset. The ZPE corrections ($\Delta$ZPE, in eV) are also reported.}
\label{tab:results}
\end{table}
 

